\documentclass[aps,prd,twocolumn,preprintnumbers,showpacs,nofootinbib,amssymb]{revtex4}
\usepackage{graphicx}
\usepackage{amsmath}
\usepackage{amssymb}
\usepackage{amsfonts}
\usepackage{bm}
\usepackage{color}

\def\be{\begin{equation}}
\def\ee{\end{equation}}
\def\bea{\begin{eqnarray}}
\def\eea{\end{eqnarray}}

\begin{document}

\title{A heuristic wave equation parameterizing BEC dark matter halos with \\
a quantum  core and an isothermal atmosphere}
\author{Pierre-Henri Chavanis}
\email{chavanis@irsamc.ups-tlse.fr}
\affiliation{Laboratoire de Physique Th\'eorique, Universit\'e de Toulouse,
CNRS, UPS, France}

\begin{abstract}

The Gross-Pitaevskii-Poisson equations that govern the evolution of
self-gravitating
Bose-Einstein condensates, possibly representing dark matter halos, experience a
process of gravitational cooling and violent relaxation. We propose a heuristic parametrization of
this complicated process in the spirit of Lynden-Bell's theory of violent
relaxation for collisionless stellar systems. We derive a generalized wave equation
(that was introduced phenomenologically in [P.H. Chavanis, Eur.
Phys. J. Plus {\bf
132}, 248 (2017)]) involving a
logarithmic nonlinearity associated with an effective temperature $T_{\rm eff}$
and a damping term associated with a friction $\xi$. These terms can be obtained
from a maximum entropy production principle and are linked by a form of Einstein
relation expressing the fluctuation-dissipation theorem. The wave equation
satisfies
an $H$-theorem for the Lynden-Bell entropy and relaxes towards a stable
equilibrium state which is a maximum of entropy at fixed mass and energy. This
equilibrium state represents the most probable state 
of a Bose-Einstein condensate dark matter halo. It generically has a
core-halo structure. The quantum core prevents gravitational collapse and
may solve the
core-cusp problem. The isothermal halo leads to flat rotation curves in
agreement with the observations.  These results are
consistent with the phenomenology of dark matter halos. Furthermore, as shown in a
previous paper [P.H. Chavanis, Phys. Rev. D {\bf 100}, 123506 (2019)], the maximization of entropy with respect to the core mass
at fixed total mass and total energy determines a core
mass--halo mass relation
which agrees with the relation obtained in direct numerical
simulations.  We stress the importance of using a microcanonical description
instead of a canonical one. We also explain how our formalism
can be applied to the case of
fermionic dark matter halos.

\end{abstract}

\pacs{95.30.Sf, 95.35.+d, 95.36.+x, 98.62.Gq, 98.80.-k}

\maketitle

\section{Introduction}
\label{sec_intro}

The nature of dark matter (DM) is still unknown and remains one of the greatest
mysteries of modern cosmology even after almost $100$ years of research.

The suggestion that DM may constitute a large part of the universe was
made by Zwicky \cite{zwicky} in 1933. Using the virial theorem to infer the
average mass of galaxies within the Coma cluster, he obtained a
much higher value than the mass of luminous material. He realized therefore
that some mass was missing in order to account for the observations (he
called this unseen mass {\it dunkle Materie}). This virial mass discrepancy in
galaxy clusters was confirmed later by more accurate measurements of velocity
dispersion showing that DM should represent about $90\%$ of the mass of the
cluster \cite{kravtsov}.

Another evidence of the  missing mass problem and of the existence of DM comes from the study of the rotation curves of disk galaxies \cite{flat1,flat2,flat3,flat4}. The rotational velocity of hydrogen
clouds in spiral galaxies measured from the Doppler effect is found first to increase near the galaxy center in agreement with Newtonian theory but then to saturate to an approximately constant value $v_\infty\sim 200\, {\rm km/s}$, even at large distances where very little baryonic matter can be detected, instead of decreasing according to the Keplerian law (like for the rotation of the planets around the Sun). This suggests that galaxies are surrounded by an
extended halo of DM, whose mass $M(r)\sim  v_\infty^2 r/G$
increases linearly with radius. This can be conveniently modeled by a
classical isothermal gas whose density decreases as $r^{-2}$ \cite{bt}.

On the cosmological side, the  observations of distant type Ia supernovae \cite{novae1,novae2,novae3,novae4} and the recent Planck satellite measurements of the Cosmic Microwave Background (CMB)  radiation \cite{planck2013,planck2016} have revealed that the content of the universe is made of about $70\%$
dark energy, $25\%$ DM, and $5\%$ baryonic (visible) matter. DM is also required
to interpret the data of
gravitational lensing \cite{lensing1,lensing2,lensing3} and the observations of
the Bullet Cluster,  resulting from the  collision of two clusters of galaxies,
in which the baryonic and the DM
components are clearly separated \cite{bullet}.

There have been some attempts to interpret the observations without assuming the existence of DM. For example Milgrom \cite{milgrom} proposed a modification of Newton's law (MOND theory) on galactic scales
to explain the rotation curves of spiral galaxies without invoking DM. Other
theories of modified gravity have been introduced as alternatives to DM
\cite{harkolobo} but  the DM hypothesis is favored by most astrophysicists. It
is likely that DM is made of a new type of particles, interacting only
gravitationally with ordinary matter, not yet included in the standard model of
particle physics. In the standard cold dark matter (CDM) model, DM is assumed to
be made of  (still hypothetical) weakly interacting massive particles (WIMPs)
with a mass in the GeV-TeV range. They may correspond to supersymmetric (SUSY)
particles \cite{susy}. These particles freeze out from thermal equilibrium in
the early universe and, as a consequence of this decoupling,  cool off rapidly
as the universe expands.
In the warm dark matter (WDM)  model
\cite{wdm,overduin}, DM is thought to be made of fermions, such as
massive neutrinos, with a mass in the keV range.  Other popular DM candidates
are bosons like the axion. The QCD axion \cite{kc} with a mass $m\sim 10^{-4}\,
{\rm eV}/c^2$ was initially proposed as a solution of the charge parity (CP)
problem of quantum chromodynamics (QCD) \cite{pq} but ultra light axions (ULA)
arising from string theory with a mass possibly as small as $m\sim  10^{-22}\,
{\rm eV/c^2}$   are also actively considered at present \cite{marshrevue}.

In the standard CDM model, DM is represented as a
classical pressureless gas at zero temperature ($T=0$) described by the
Euler-Poisson equations, or as a collisionless system of classical particles
described by
the Vlasov-Poisson equations \cite{peeblesbook}. The CDM model
works remarkably well at large (cosmological) scales and is consistent with ever
improving measurements of the CMB from WMAP and
Planck missions \cite{planck2013,planck2016}. However, it experiences serious
problems at small (galactic) scales. In particular, classical collisionless $N$-body
numerical simulations predict that DM halos
should be cuspy \cite{nfw}, with a density diverging as $r^{-1}$ for
$r\rightarrow 0$,  while observations reveal that they have a constant 
density core \cite{observations}.\footnote{Numerical simulations of CDM show
that DM halos have a  universal density profile called the Navarro-Frenk-White
(NFW) profile \cite{nfw}. The density diverges approximately as $r^{-1}$ (cusp)
for $r\rightarrow 0$ and decreases as $r^{-3}$ for $r\rightarrow +\infty$.  The
observational Burkert
\cite{observations} profile also decreases as $r^{-3}$ at large distances but tends to a constant (core) at the center.} On the other hand, the CDM model predicts
an over-abundance of small-scale structures (subhalos/satellites), much more
than what is observed around the Milky Way
\cite{satellites1,satellites2,satellites3}. Finally,
dissipationless CDM simulations predict that the
majority of the most massive subhaloes of the Milky Way are too dense to host
any of its bright satellites.  These problems are
referred to as the ``core-cusp'' problem \cite{moore}, the ``missing
satellites'' problem \cite{satellites1,satellites2,satellites3}, and the ``too
big to fail'' problem
\cite{boylan}. The
expression ``small-scale crisis of CDM'' has been coined \cite{crisis}. The small-scale problems of the CDM model are somehow related to the assumption that DM
is pressureless.

These problems may be relieved if we assume that DM is warm or if the DM
particles are self-interacting.\footnote{The possibility to solve the CDM crisis
without changing the basic assumptions of the CDM model invokes the feedback of
baryons that can
transform cusps into cores \cite{romano1,romano2,romano3}.}
In the WDM model the dispersion of the particles is responsible for a pressure force that can halt gravitational collapse and prevent the formation of cusps \cite{wdm}. Similarly, in the self-interacting dark matter (SIDM) model with a large scattering cross section but negligible annihilation or dissipation
\cite{spergelsteinhardt}, collisions can cause the relaxation of
the particles in the regions of high density and establish an isothermal
(Maxwellian) distribution function (DF). As a result, the system presents an
isothermal core (instead of a cusp) and a NFW halo. When this model is
confronted to observations it is found that the cross section  $\sigma/m$ of the
 DM particles depends on their velocity dispersion. This may be explained by a
dark photon model where self-interactions are described by a Yukawa potential
\cite{kaplinghat}, by a short-range interaction model with a large scattering
length \cite{bkl}, or by a dark fusion model \cite{mcdermott}. For dwarf and
low surface brightness (LSB) galaxies, the cross-section is approximately
constant with a value $\sigma/m\sim 1\, {\rm cm^2/g}$. Observations of the
Bullet Cluster leads to an upper limit $\sigma/m< 1.25\, {\rm cm^2/g}$.

Another possibility to solve the
small-scale crisis of CDM is to take into account the quantum nature of the
DM particle. In this paper, we shall assume that the DM particle  is a boson
like an axion \cite{marshrevue}.\footnote{Some authors have considered the case
where the DM particle is a fermion like a massive neutrino (see the Introduction
of Ref. \cite{gr1} for a short review and an exhaustive list of references).
In this model, gravitational collapse is prevented by the quantum pressure
arising
from the Pauli exclusion principle.} At very low temperatures, bosons form
self-gravitating Bose-Einstein condensates (BECs).\footnote{The
condensation of integer spin particles was theoretically predicted by Bose
\cite{bose} and
Einstein \cite{einstein1,einstein2} in 1924 and observed for the first time in
laboratory experiments
of dilute alkali gases in 1995 \cite{aemwc,dmaddkk,bradley1}. The condensation
occurs when the particles in
the gas become correlated quantum mechanically, i.e., when the de Broglie
thermal wavelength of a particle $\lambda_{\rm dB}=\sqrt{2\pi\hbar^2/mk_B T}$
turns out to be greater than the mean interparticle distance $l=n^{-1/3}$ (i.e.
$n\lambda_{\rm dB}^3>1$). The exact condensation temperature is given by
\begin{eqnarray}
\label{me5tc}
T_c=\frac{2\pi\hbar^2n^{2/3}}{m k_B \zeta(3/2)^{2/3}}
\end{eqnarray}
with $\zeta(3/2)=2.612...$. If we apply these results to ULAs
with a mass $m=2.92\times 10^{-22}\, {\rm eV/c^2}$ and consider a typical
DM halo density
$\rho=7.02\times 10^{-3}\, M_{\odot}/{\rm pc}^3$ (medium spiral), we get
$T_c=4.82\times 10^{36}\, {\rm K}$. On the other hand, the typical temperature
of the halo (obtained from the virial relation $k_B T_{\rm eff}/m\sim
v_h^2=GM_h/r_h$) is $T_{\rm eff}\sim 4.41\times 10^{-25}\, {\rm K}$ where we
have taken $(k_B
T_{\rm eff}/m)^{1/2}=108\, {\rm km/s}$.
Therefore, we find
that $T_{\rm eff}\ll T_c$.
This inequality shows that $T_{\rm eff}$ is necessarily an out-of-equilibrium
(effective)
temperature otherwise the
halo would be completely condensed.
For $m=1.10\times
10^{-3}\, {\rm eV/c^2}$  we get
$T_c=5.29\times 10^5\, {\rm K}$ and  $T=1.66\times
10^{-6}\, {\rm K}$. If we apply the
self-gravitating BEC model to neutron stars with the idea that neutrons could
form Cooper pairs and
behave as bosons of mass $m=2m_n$ \cite{chavharko} we find $T_c=8.26\times
10^{11}\, {\rm
K}$, where we have taken $m_n=1.675\times 10^{-24}\, {\rm g}$ and
$\rho_0=3.54\times 10^{21}\, {\rm g/m^3}$.} In that case, DM halos can
be viewed as gigantic bosonic atoms  at $T=0$ where
the bosonic particles are condensed in a
single
macroscopic quantum state. They are described by a scalar field (SF) that can be
interpreted as the wavefunction $\psi({\bf r},t)$ of the condensate. The
evolution of the wave function of the condensate is governed by the
Schr\"odinger-Poisson equations when the bosons are noninteracting or by the
Gross-Pitaevskii-Poisson (GPP) equations when the bosons are self-interacting.
By using the Madelung \cite{madelung} transformation, these wave equations may
be written in the form of hydrodynamic equations including a quantum potential.
The wave properties of the SF are negligible at large (cosmological) scales
where the SF behaves as CDM, but they become relevant at small (galactic) scales
and can prevent gravitational collapse.
This model has been given several names such as wave DM, fuzzy dark matter
(FDM), BECDM, $\psi$DM,
or SFDM.\footnote{See the Introduction of Ref. \cite{tunnel} for a short review
and an exhaustive list of references. See also  the
Introduction of Ref. \cite{prd1} and Ref. \cite{leerevue} for an early
history of
this model, and Refs.
\cite{srm,rds,chavanisbook,marshrevue,braatenrevue,niemeyer,ferreira}
for recent reviews on the subject.} We shall
refer to this model as BECDM. In the BECDM model, gravitational
collapse is prevented by the quantum pressure arising from the Heisenberg
uncertainty principle or from the scattering of the bosons (when the
self-interaction is repulsive).  This leads to DM halos presenting a central
core instead of a cusp. Since the quantum Jeans length is nonzero
\cite{khlopov,bianchi,hu,sikivie,prd1,abrilmnras,aacosmo,abrilph,
suarezchavanisprd3,harkoj,jeansunivers}, the
formation of small-scale structures is
suppressed even
at $T=0$. Therefore, quantum mechanics may be a way to solve the small-scale
problems of the CDM model such
as the core-cusp problem and the missing satellite problem.

The GPP equations have a very complicated dynamics.
A self-gravitating BEC at $T=0$ that is not initially in a steady state
undergoes Jeans instability, gravitational collapse (free fall), displays damped
oscillations, and finally settles down on a quasistationary state
(virialization) by radiating part
of the scalar field \cite{seidel94,gul0,gul}. This is the process of
gravitational cooling initially introduced by Seidel and Suen \cite{seidel94} in
the context of boson stars.

As a result of gravitational cooling, the system
reaches an equilibrium configuration with a core-halo structure. The condensed
core (soliton/BEC) is stabilized by quantum mechanics and has a smooth (finite)
density. This is a stable
stationary solution of the GPP equations at
$T=0$ (ground state). Gravitational collapse is
prevented by the quantum potential arising from the Heisenberg principle or by
the pressure $P=2\pi a_s\hbar^2\rho^{2}/m^3$ arising  from the self-interaction
of the bosons which is measured by their scattering length $a_s$.\footnote{A
repulsive self-interaction ($a_s>0$) stabilizes the quantum
core. By contrast, an attractive  self-interaction destabilizes the quantum core
above a maximum mass $M_{\rm max}=1.012\, \hbar/\sqrt{Gm|a_s|}$
first identified in \cite{prd1}.}
This quantum core (ground state) is surrounded by a halo of scalar radiation
corresponding to the quantum
interferences of excited states. As shown by Schive {\it et al.}
\cite{ch2,ch3}, and further discussed in \cite{hui,bft,bft2,meff}, these
interferences
produce time-dependent small-scale
density granules -- or quasiparticles -- of the size of the solitonic
core $\lambda_{\rm dB}\sim\hbar/m v\sim 1\, {\rm kpc}$ (de Broglie wavelength)
and effective mass $m_{\rm eff}\sim\rho\lambda_{\rm dB}^3\sim 10^7\,
M_{\odot}\gg m$
that counter self-gravity and create an effective thermal pressure. These
noninteracting excited states are analogous to collisionless particles in
classical mechanics. As a result, the halo behaves essentially as CDM and  is
approximately isothermal with an equation of state $P=\rho T_{\rm eff}$
involving an effective temperature $T_{\rm eff}$ resulting from a process of
collisionless violent relaxation (see below). The quantum core
(soliton)
may solve the core-cusp problem of the CDM model and the
isothermal halo where the density decreases as $r^{-2}$ yields flat rotation
curves  in agreement with the observations.\footnote{The halo cannot
be exactly
isothermal otherwise it would have an infinite
mass \cite{bt}. In reality, the density in the
halo decreases as $r^{-3}$, similarly to the NFW
\cite{nfw} and Burkert
\cite{observations} profiles, or even as $r^{-4}$ (see
Appendix D of \cite{clm1} and Appendix I of \cite{clm2}),
instead of $r^{-2}$ corresponding to the
isothermal
sphere \cite{bt}. This extra-confinement may be due to incomplete
relaxation, tidal
effects, and stochastic
perturbations as discussed in Appendix B of \cite{modeldm}. We
stress that the halo is in a dynamical  equilibrium (virialized) state but in an
out-of-equilibrium thermodynamical equilibrium state (see footnote 4). As
discussed in \cite{hui,bft,bft2,meff}, the
quasiparticles are responsible for a slow (secular) collisional evolution of
the halo towards thermodynamical equilibrium on a timescale of
the order of the Hubble time. By this process part of the halo
condense  (since $T\ll T_c$) and feeds the soliton. We shall not consider
this collisional regime in the present paper.}
This core-halo
structure (and the presence of granules) has been clearly
evidenced in numerical simulations of the Schr\"odinger-Poisson equations
\cite{ch2,ch3,schwabe,mocz,moczSV,veltmaat,moczprl,moczmnras,veltmaat2}. The
core
mass-halo mass
relation
$M_c(M_v)$ has been obtained numerically (and explained heuristically from a
nonlocal uncertainty principle) in Ref. \cite{ch3}. For
noninteracting bosons, the core mass scales as $M_c\propto M_v^{1/3}$.

Gravitational cooling  is a dissipationless relaxation mechanism similar in some
respect to the concept of violent relaxation introduced by  Lynden-Bell \cite{lb}  in the context of
collisionless stellar systems described by the
Vlasov-Poisson equations.
A collisionless stellar 
system  that is not initially in a dynamically stable steady
state undergoes Jeans instability, gravitational collapse (free fall), displays damped
oscillations, and finally settles down on a quasistationary (virialized)
state
 by sending some of the particles at large distances. This
process, which takes place on a dynamical timescale,  is related to phase mixing
and
nonlinear Landau damping.

Lynden-Bell \cite{lb} developed the statistical mechanics of violent relaxation
and determined the coarse-grained DF at statistical equilibrium (most
probable state) from a maximum
entropy principle (MEP) taking into account all the constraints of the
Vlasov-Poisson equations. The quasistationary state reached by the system as a
result of violent relaxation is expected to be in this most
probable state.
The Lynden-Bell DF is similar to the Fermi-Dirac distribution
but with, of course, a completely different interpretation. It takes into
account a sort of exclusion principle implied by the Vlasov equation, similar to
the Pauli exclusion principle for fermions, but of nonquantum origin. In
Lynden-Bell's theory, further developed by Chavanis and Sommeria \cite{csmnras},
the quasistationary state generically has a core-halo structure with
a  completely degenerate core at $T=0$ (effective fermion ball)  and an
isothermal atmosphere with an effective temperature $T_{\rm eff}$. The core
has a polytropic  equation of state $P=(1/5)[3/(4\pi\eta_0)]^{2/3}\rho^{5/3}$
and the halo has an isothermal equation of state $P=\rho T_{\rm
eff}$.\footnote{Lynden-Bell \cite{lb}, who was concerned with the study of
elliptical galaxies, argued that these objects are described by the
nondegenerate limit of his theory where his DF is similar to
the Boltzmann distribution. However, his theory may also apply to fermionic
DM halos where
degeneracy effects may be important as suggested in \cite{csmnras,clm1,clm2}.
The theory of violent relaxation explains how a collisionless self-gravitating
system may reach an isothermal distribution on a very short timescale (of the
order of the dynamical time $t_D$) without recourse to collisions or
gravitational
encounters that operate on a relaxation timescale $t_R\sim (N/\ln N)t_D$ much
larger than the age of
the universe \cite{bt}. The theory of violent relaxation thus solves a
notoriously
important timescale problem in astrophysics \cite{lb,csmnras,clm1,clm2}.}  The
degenerate core (in the sense
of Lynden-Bell) may solve the core-cusp problem of the CDM model. On the other
hand,
the density decreases as $r^{-2}$ in the isothermal halo, yielding flat rotation
curves in agreement
with the observations \cite{flat1,flat2,flat3,flat4}. This core-halo structure
has been studied in detail in models of self-gravitating fermions and in
relation to Lynden-Bell's theory of violent relaxation (see Sec. V.A of
\cite{modeldm} for an exhaustive list of references).
In the analogy between the gravitational
cooling of self-gravitating BECs and the violent
relaxation of collisionless self-gravitating  systems, the bosonic core
(BEC/soliton) corresponds to the effective
fermion ball and the halo made of scalar radiation corresponds to the isothermal
halo predicted by Lynden-Bell.  Actually, since a collisionless system of bosons
is described by the Vlasov-Poisson equations at large scales where quantum
effects are negligible (see \cite{moczSV} for the
Schr\"odinger-Poisson--Vlasov-Poisson correspondence) it is very likely that
both processes -- gravitational
cooling and violent relaxation -- occur  in self-gravitating BECs and may
even correspond to the same phenomenon.\footnote{The connection between
self-gravitating BECs and the Lynden-Bell theory of violent relaxation was
mentioned in \cite{modeldm,chavtotal} and in the Appendix
of \cite{moczSV}.} As a
result, self-gravitating
BECs should have a core that is partly bosonic (soliton) and partly fermionic
(in the sense of Lynden-Bell), surrounded by an effective isothermal halo.  Gravitational cooling and violent relaxation may be at work during hierarchical clustering, a process
 by which small DM halos merge and form larger halos in a bottom-up
structure formation scenario.
It is believed that DM halos acquire an approximately isothermal
profile, or more realistically a NFW or Burkert profile (see footnote 7),
as a result of successive mergings. Gravitational cooling and violent relaxation explain
how collisionless self-gravitating systems can rapidly thermalize and acquire a
large effective temperature $T_{\rm eff}$ even if $T=0$ fundamentally.

In the context of the violent relaxation of collisionless stellar systems, we
have derived 
a relaxation equation for the coarse-grained DF
$\overline{f}({\bf r},{\bf v},t)$ by using a maximum entropy production
principle (MEPP) \cite{csr,chavmnras,dubrovnik}. This coarse-grained Vlasov
equation has the form of a
fermionic Kramers equation involving a diffusion term and a friction
term.\footnote{Alternatively, we can describe the effective
collision term by a fermionic Landau operator
\cite{kp,sl,chavmnras,kingen,dubrovnik}.} The
diffusion term accounts for fluctuations and effective thermal effects while the
friction term accounts for collisionless dissipation (nonlinear Landau damping).
The diffusion and friction coefficients are linked by a form of Einstein
relation expressing the fluctuation-dissipation theorem. The coarse-grained
Vlasov equation respects the Lynden-Bell exclusion principle and satisfies an
$H$-theorem for the Lynden-Bell entropy. As a result, it relaxes towards the
Lynden-Bell distribution. Starting from the coarse-grained Vlasov equation, we
can derive a set of hydrodynamic equations by closing the hierarchy of Jeans
equations with a local thermodynamic equilibrium (LTE) approximation based on
the Lynden-Bell distribution \cite{csr}. These damped Euler-Poisson equations
provide a heuristic parametrization of violent relaxation for classical
collisionless self-gravitating systems described by the Vlasov-Poisson
equations.

In this paper, we develop a similar heuristic parametrization for
self-gravitating BECs at $T=0$.  Our parametrization accounts for the
complicated processes of  gravitational cooling and violent relaxation. We first
reformulate the Schr\"odinger equation as a  Wigner (or quantum Vlasov)
equation. We then introduce a coarse-grained Wigner equation including an 
effective fermionic Kramers collision term taking into account the Lynden-Bell
exclusion principle like in the classical coarse-grained Vlasov equation.  We
then take the hydrodynamic moments of the coarse-grained Wigner equation and
close the hierarchy by making a LTE approximation based on the Lynden-Bell
distribution.  The quantum potential is automatically taken into account
in our procedure. This leads to the quantum damped Euler-Poisson equations.
Finally, we use  the inverse Madelung transformation to derive a wave equation
associated with these hydrodynamic equations. This procedure provides a more
precise justification of the wave equation introduced phenomenologically in our
previous papers \cite{chavtotal,modeldm}. This wave equation involves a
logarithmic nonlinearity associated with an effective temperature $T_{\rm eff}$
and a damping term associated with a friction  $\xi$. These terms arise from the
MEPP and are linked by a form of Einstein relation expressing the
fluctuation-dissipation theorem. The wave equation satisfies an $H$-theorem for
the Lynden-Bell entropy and relaxes towards a stable
equilibrium state which is a maximum of entropy at fixed mass and energy.  This
equilibrium state represents the most probable state of a BECDM halo. It
generically has a core-halo structure. The quantum core prevents
gravitational collapse and may solve the
core-cusp problem. The isothermal halo leads to flat rotation curves in
agreement with the observations.  These results are
consistent with the phenomenology of DM halos. Furthermore, as shown in a
previous paper \cite{mcmh}, the maximization of the entropy $S(M_c)$ with
respect to the core mass $M_c$ at fixed total mass $M_h$  and
total energy $E_h$ determines a core mass -- halo mass relation $M_c(M_h)$
which agrees with the relation obtained in direct numerical
simulations of noninteracting bosons \cite{ch3}, giving further support to our
effective thermodynamical
approach. This thermodynamical approach also allowed us to
derive the general expression of the core mass -- halo mass
relation $M_c(M_h)$ for
self-interacting bosons and fermions \cite{mcmh,mcmhbh}, making 
new predictions that still have to be confirmed numerically.

The paper is organized as follows. In Sec. \ref{sec_wfa}, we
recall the wave function approach of BECDM halos based on the
Schr\"odinger-Poisson equations. We discuss the hydrodynamic representation of
these equations based on the Madelung transformation and introduce the concepts
of gravitational cooling and violent relaxation. In Sec. \ref{sec_dfaw}, we
recall the DF approach of BECDM halos based on the Wigner-Poisson equations. We
propose a heuristic coarse-graining of these equations based on a MEPP and
derive the corresponding hydrodynamic equations. In Sec. \ref{sec_pcd}, we
obtain a generalized wave equation which is equivalent to the
hydrodynamic equations associated with the coarse-grained Wigner-Poisson
equations. This generalized wave equation provides a heuristic parametrization
of the complex dynamics of BECDM halos. In Sec. \ref{sec_es}, we study the
equilibrium states of this equation and show that their structure agrees with
the phenomenology of BECDM halos. In Sec. \ref{sec_cmm}, we stress the
importance of developing a microcanonical description of the process of violent
relaxation and show how it can be implemented in our parametrization. In Sec.
\ref{sec_fermidm}, we explain how our formalism can be applied
to the case of
fermionic DM halos. The Appendices provide additional results that are needed
in
our study. In Appendix \ref{sec_cw}, we derive the basic properties of the
Wigner distribution. In Appendix \ref{sec_gwe}, we recall the basic properties
of the generalized GPP equations introduced in \cite{chavtotal}. We mention
that, in addition to providing a parametrization of the processes of 
gravitational cooling and violent relaxation, these equations can
serve as a numerical algorithm to construct stable stationary solutions of
the GPP equations. In Appendix
\ref{sec_ccsgs}, we discuss certain aspects of the dynamical evolution of
classical collisionless self-gravitating systems. We point out in particular
the limitation of the single-speed solution and the need to introduce more
sophisticated parametrizations. In Appendix \ref{sec_vrx}, we
discuss the violent relaxation of classical collisionless self-gravitating
systems towards a quasistationary state in terms of
statistical mechanics. In Appendix \ref{sec_cl}, we discuss the classical limit
$\hbar\rightarrow 0$ of the quantum equations derived in this
the paper. In Appendix \ref{sec_vb}, we briefly consider the
Vlasov-Bohm equation instead of the Wigner equation. In Appendix
\ref{sec_multistate}, we explain how our results can be extended to multistate
systems like fermions.

\section{Wave function approach}
\label{sec_wfa}

\subsection{Schr\"odinger-Poisson equations}
\label{sec_he}

Let us consider a system of $N$ bosons of individual mass $m$ interacting 
via a binary potential $u(|{\bf r}-{\bf r}'|)$. Their Hamiltonian is given, in
the second quantization, by
\begin{eqnarray}
\label{he1}
{\hat H}&=&\int d{\bf r}\, {\hat \psi}^{\dag}({\bf r})\left (-\frac{\hbar^2}{2m^2}\Delta\right ){\hat \psi}({\bf r})\nonumber\\
&+&\frac{1}{2}\int d{\bf r}d{\bf r}' {\hat \psi}^{\dag}({\bf r}){\hat \psi}^{\dag}({\bf r}')u(|{\bf r}-{\bf r}'|){\hat \psi}({\bf r}){\hat \psi}({\bf r}'),
\end{eqnarray}
where ${\hat \psi}({\bf r})$ and ${\hat \psi}^{\dag}({\bf r})$ are the bosonic field operators that annihilate and create a particle at the position ${\bf r}$, respectively. We write the field operator in the Heisenberg representation as
\begin{eqnarray}
\label{he2}
{\hat \psi}({\bf r},t)={\psi}({\bf r},t)+{\tilde \psi}({\bf r},t),
\end{eqnarray}
where the expectation value of the field 
operator ${\psi}({\bf r},t)=\langle {\hat \psi}({\bf r},t) \rangle$ is the
condensate wavefunction  and ${\tilde \psi}({\bf r},t)$  is the noncondensate
field operator whose average is zero by construction: $\langle {\tilde
\psi}({\bf r},t)\rangle =0$. In this manner, the BEC contribution has been
separated from the total bosonic field operator. The condensate wavefunction 
${\psi}({\bf r},t)$ is a classical field and its squared modulus determines the
condensate mass density  $\rho({\bf r},t)=| {\psi}({\bf r},t)|^2$. It is
normalized such that $M=\int \rho({\bf r},t)\, d{\bf r}$ represents the total
mass of the condensate.

The Heisenberg equation of motion for the quantum field operator ${\hat
\psi}({\bf r},t)$  
corresponding to the many-body Hamiltonian given by Eq. (\ref{he1}) is
\begin{eqnarray}
\label{he3}
&&i\hbar \frac{\partial}{\partial
t}{\hat \psi}({\bf r},t)=\lbrack {\hat \psi}, {\hat H}\rbrack\nonumber\\
&=&\left\lbrack -\frac{\hbar^2}{2m}\Delta+m\int d{\bf r}'\,  {\hat \psi}^{\dag}({\bf r},t)u(|{\bf r}-{\bf r}'|){\hat \psi}({\bf r}',t)\right\rbrack{\hat \psi}({\bf r},t).\nonumber\\
\end{eqnarray}
This equation is exact and describes the dynamics of a system of interacting
bosons at arbitrary temperature $T$. If we take the average of this equation, we
obtain an exact equation for the condensate wavefunction   ${\psi}({\bf
r},t)$. This equation contains correlation terms which describe the interaction
between the condensate and the thermal cloud of noncondensed bosons. This
equation must be supplemented by a kinetic equation for the distribution
function $f({\bf r},{\bf p},t)$ of the noncondensed bosons which itself depends
on the interaction with the condensate. This yields a system of two coupled
differential equations for the condensate and the thermal cloud. These equations
describe the collisional evolution of the system of interacting bosons. The
corresponding kinetic theory is studied in detail in Ref. \cite{gnz}. However,
this is not the regime that we consider in the present paper.

In the following, we assume that $T\ll T_c$ and $N\gg 1$, where $T_c$ is the
condensation temperature and $N$ is the number of bosons. In that case, most of
the bosons lie in the same single-particle quantum state and we can make the
mean field approximation which consists in replacing the field operator ${\hat
\psi}({\bf r},t)$ by the condensate wavefunction ${\psi}({\bf r},t)$. This
yields the mean field equation
\begin{eqnarray}
\label{he4}
&&i\hbar \frac{\partial}{\partial
t}{\psi}({\bf r},t)\nonumber\\
&=&\left\lbrack -\frac{\hbar^2}{2m}\Delta+m\int d{\bf r}'\, u(|{\bf r}-{\bf r}'|)|\psi({\bf r}',t)|^2 \right\rbrack {\psi}({\bf r},t).\quad
\end{eqnarray}
This equation gives a very good description of the condensate when $T\ll T_c$
and $N\gg 1$. It can be written as a Schr\"odinger equation
\begin{eqnarray}
\label{he5}
i\hbar \frac{\partial\psi}{\partial
t}=-\frac{\hbar^2}{2m}\Delta\psi+m\Phi_{\rm tot}\psi
\end{eqnarray}
with a mean field potential
\begin{eqnarray}
\label{he6}
\Phi_{\rm tot}({\bf r},t)=\int  u(|{\bf r}-{\bf r}'|)\rho({\bf r}',t)\, d{\bf
r}'
\end{eqnarray}
produced by the particles self-consistently.
We now assume that the potential of interaction $u(|{\bf r}-{\bf r}'|)$ can be
written as the sum of a long-range
potential $u_{\rm LR}(|{\bf r}-{\bf r}'|)$ and a  short-range potential $u_{\rm
SR}(|{\bf r}-{\bf r}'|)$ so 
that $u=u_{\rm LR}+u_{\rm SR}$. The long-range potential corresponds to the
gravitational interaction and is given by $u_{\rm LR}=-G/|{\bf r}-{\bf r}'|$.
The short-range potential takes into account binary collisions. In a dilute cold
gas they can be modeled by a zero-range pseudo-potential $u_{\rm
SR}=g\delta({\bf r}-{\bf r}')$ of strength  $g=4\pi a_s\hbar^2/m^3$, where $a_s$
is the $s$-wave scattering length of the bosons. The scattering length $a_s$ can
be positive (corresponding to a repulsive self-interaction) or negative
(corresponding to an attractive self-interaction). Under these conditions, the
total mean field potential can be written as $\Phi_{\rm tot}({\bf
r},t)=\Phi({\bf r},t)+h(\rho)$, where $\Phi({\bf r},t)=-G\int \rho({\bf
r}',t)|{\bf r}-{\bf r}'|^{-1}\, d{\bf r}'$ is the gravitational potential and
$h(\rho)=g\rho$ is an effective potential modeling short-range interactions.
When this form of potential is substituted into Eq. (\ref{he4}), we obtain
the GPP equations
\begin{eqnarray}
\label{he7}
i\hbar \frac{\partial\psi}{\partial
t}=-\frac{\hbar^2}{2m}\Delta\psi+m\Phi\psi+\frac{4\pi a_s\hbar^2}{m^2}|\psi|^2\psi,
\end{eqnarray}
\begin{eqnarray}
\label{he8}
\Delta\Phi=4\pi G |\psi|^2.
\end{eqnarray}
When the self-interaction of the bosons can be neglected  they reduce to the
Schr\"odinger-Poisson equations
\begin{eqnarray}
\label{h1}
i\hbar \frac{\partial\psi}{\partial
t}=-\frac{\hbar^2}{2m}\Delta\psi+m\Phi\psi,
\end{eqnarray}
\begin{eqnarray}
\label{h2}
\Delta\Phi=4\pi G |\psi|^2.
\end{eqnarray}
The Schr\"odinger-Poisson equations can be combined into a 
single
equation of the form  
\begin{eqnarray}
\label{h1b}
i\hbar \frac{\partial\psi}{\partial
t}=\left\lbrack -\frac{\hbar^2}{2m}\Delta-\int \frac{Gm}{|{\bf r}-{\bf
r}'|}|\psi|^2({\bf r}',t)\, d{\bf r}'\right\rbrack\psi,
\end{eqnarray}
or
\begin{eqnarray}
\label{h1c}
i\hbar \frac{\partial\Delta\ln\psi}{\partial
t}=-\frac{\hbar^2}{2m}\Delta\left (\frac{\Delta\psi}{\psi}\right
)+4\pi G m |\psi|^2.
\end{eqnarray}

{\it Remark:} The mean field approximation is justified for two reasons. 
First, the temperature of our system is much smaller than the condensation
temperature $T_c$. Therefore, we are in a situation where most of the bosons are
condensed in the same quantum state. As a result, we can treat
the wavefunction as a classical field (the condition $T\ll T_c$ is equivalent to
a large occupation number ${\cal N}=n\lambda_{\rm dB}^3\sim 10^{94}\gg
1$ \cite{meff}). In that case, the Schr\"odinger equation can be interpreted as
a
classical wave equation.\footnote{
To describe a system of bosons in interaction we should in principle
``second-quantize''
the Schr\"odinger equation, which then becomes an equation for
the evolution of field operators with appropriate commutation rules [see Eq.
(\ref{he3})]. However, in the limit of very large occupation numbers ${\cal
N}\gg 1$, the
classical description of the Schr\"odinger wave equation
equation is adequate.} On the other hand, in the
case of systems with long-range
interactions such as self-gravitating systems, the collisional
relaxation time scales like  $(N/\ln N)t_D$ and is generally extremely large
(see,
e.g., Ref. \cite{bt}).\footnote{In the case of spiral or
elliptical galaxies where $N\sim 10^{12}$, the relaxation time is much larger
than
the age of the universe \cite{bt}. In the case of bosonic DM halos,  the
relaxation time
is reduced by Bose stimulation (the number of quasiparticles is $N_{\rm eff}\sim
10^5$ \cite{meff}) but it remains relatively long, of the order of the Hubble
time, like in the
case of globular clusters where $N\sim 10^5$ \cite{bt}.} In a first regime, the
evolution
of the system is
essentially
collisionless and a mean field approximation can be implemented. In the analogy
with the kinetic theory of classical self-gravitating systems (see, e.g.,
\cite{physicaA}), Eq. (\ref{he3})
is the counterpart of the Klimontovich  equation which is an exact
equation equivalent to the  Liouville equation or to the Hamilton equations of
motion, and Eq. (\ref{h1}) is the
counterpart of the Vlasov equation which is a mean field equation valid in the
collisionless regime. It describes the evolution of a smooth
wavefunction $\psi({\bf r},t)$.\footnote{We
can also base the (collisional) kinetic theory of self-gravitating bosons on an
equation similar to Eq.
(\ref{h1}) for a classical field $\psi_d({\bf r},t)$ provided that we take 
fluctuations into account (i.e., $\psi_d$ can be decomposed into 
$\psi_d=\psi+\delta\psi$ where $\psi$ is a smooth component and $\delta\psi$
a stochastic component).}

\subsection{Madelung transformation}
\label{sec_mad}

We can use the Madelung \cite{madelung} transformation to write the
Schr\"odinger-Poisson
equations (\ref{h1}) and (\ref{h2}) under the form of hydrodynamic equations.
To that purpose, we  write the wave function as
\begin{equation}
\label{mad1}
\psi({\bf r},t)=\sqrt{{\rho({\bf r},t)}} e^{iS({\bf
r},t)/\hbar},
\end{equation}
where  $\rho({\bf r},t)$ is the mass density and $S({\bf r},t)$ is
the action given by
\begin{equation}
\label{mad2}
\rho=|\psi|^2\qquad {\rm and}\qquad
S=-i\frac{\hbar}{2}\ln\left
(\frac{\psi}{\psi^*}\right ).
\end{equation}
Following Madelung \cite{madelung}, we introduce the velocity field
\begin{equation}
\label{mad3}
{\bf u}=\frac{\nabla S}{m}.
\end{equation}
Since the velocity is potential, the flow is irrotational: $\nabla\times {\bf
u}={\bf 0}$. Substituting
Eq. (\ref{mad1}) into Eqs. (\ref{h1}) and (\ref{h2}) and separating the real and
the imaginary parts, we find that the Schr\"odinger-Poisson
equations are equivalent to
hydrodynamic
equations of the form
\begin{equation}
\label{mad4}
\frac{\partial\rho}{\partial t}+\nabla\cdot (\rho {\bf
u})=0,
\end{equation}
\begin{equation}
\label{mad5}
\frac{\partial S}{\partial t}+\frac{1}{2m}(\nabla
S)^2+m\Phi+Q=0,
\end{equation}
\begin{equation}
\label{mad6}
\frac{\partial {\bf u}}{\partial t}+({\bf u}\cdot \nabla){\bf
u}=-\frac{1}{m}\nabla
Q-\nabla\Phi,
\end{equation}
\begin{eqnarray}
\label{mad7}
\Delta\Phi=4\pi G\rho,
\end{eqnarray}
where
\begin{equation}
\label{mad8}
Q=-\frac{\hbar^2}{2m}\frac{\Delta\sqrt{\rho}}{\sqrt{\rho}}=-\frac{\hbar^2}{4m}
\left\lbrack
\frac{\Delta\rho}{\rho}-\frac{1}{2}\frac{(\nabla\rho
)^2}{ \rho^2} \right\rbrack
\end{equation}
is the quantum (Bohm) potential taking into account the Heisenberg uncertainty
principle. Eq. (\ref{mad4}) is the continuity equation,
Eq. (\ref{mad5}) is the quantum Hamilton-Jacobi (or Bernoulli) equation and Eq.
(\ref{mad6})
is the quantum Euler
equation (it is obtained by taking the gradient of the quantum Hamilton-Jacobi
equation). The quantum Euler-Poisson equations (\ref{mad4})-(\ref{mad7}) are
equivalent to the
Schr\"odinger-Poisson
equations (\ref{h1}) and (\ref{h2}). Since there is no viscosity, these
hydrodynamic equations
describe a
superfluid. When $\hbar\rightarrow 0$, we recover the classical Euler-Poisson
equations. We note that quantum effects arise through the
ratio $\hbar/m$.

{\it Remark:} We note that the quantum force in Eq. (\ref{mad6}) can be written
as
\begin{equation}
\label{mad9}
-\frac{1}{m}
\nabla Q=-\frac{1}{\rho}\partial_jP_{ij}^{Q},
\end{equation}
where
\begin{equation}
\label{mad10}
P_{ij}^{Q}=-\frac{\hbar^2}{4m^2}\rho\,
\partial_i\partial_j\ln\rho=\frac{\hbar^2}{4m^2}\left (\frac{1}{\rho}
\partial_i\rho\partial_j\rho-\partial_i\partial_j\rho\right )
\end{equation}
is an anisotropic quantum pressure tensor. Therefore, the quantum Euler equation (\ref{mad6}) may be
rewritten as
\begin{equation}
\label{mad6b}
\frac{\partial {\bf u}}{\partial t}+({\bf u}\cdot \nabla){\bf
u}=-\frac{1}{\rho}\partial_jP_{ij}^{Q}-\nabla\Phi.
\end{equation}

\subsection{Gravitational cooling and violent relaxation}
\label{sec_gc}

The Schr\"odinger-Poisson equations (\ref{h1}) and (\ref{h2}) conserve the
mass
\begin{eqnarray}
\label{pme1}
M=\int |\psi|^2\, d{\bf r}
\end{eqnarray}
and the energy
\begin{eqnarray}
\label{pme2}
E_{\rm tot}=\frac{\hbar^2}{2m^2}\int |\nabla\psi|^2\, d{\bf r}+\frac{1}{2}\int |\psi|^2 \Phi\, d{\bf r},
\end{eqnarray}
which is the sum of the kinetic energy $\Theta$ and the gravitational energy $W$.

Equivalently, the quantum Euler-Poisson equations (\ref{mad4})-(\ref{mad7}) conserve the
mass
\begin{eqnarray}
\label{hme1}
M=\int \rho\, d{\bf r}
\end{eqnarray}
and the energy 
\begin{eqnarray}
\label{hme2}
E_{\rm tot}=\Theta_c+\Theta_Q+W,
\end{eqnarray}
where the first term is the  classical kinetic energy
\begin{eqnarray}
\label{hme3}
\Theta_c=\int\rho  \frac{{\bf u}^2}{2}\, d{\bf r},
\end{eqnarray}
the second term is the quantum kinetic energy
\begin{equation}
\label{hme4}
\Theta_Q=\frac{1}{m}\int \rho Q\, d{\bf r},
\end{equation}
and the third term is
the gravitational energy
\begin{eqnarray}
\label{hme5}
W=\frac{1}{2}\int\rho\Phi\, d{\bf r}.
\end{eqnarray}
The decomposition $\Theta=\Theta_c+\Theta_Q$ arises naturally from the Madelung
transformation.

Since the Schr\"odinger-Poisson equations (and the corresponding hydrodynamic
equations) are
reversible, they do not satisfy an $H$-theorem. As a
result, their relaxation
towards a quasistationary state (if the system is not initially in such a
state) is not
trivial.\footnote{This is a notoriously difficult mathematical problem even at
the classical level, i.e., for the Vlasov-Poisson equations. The relaxation of
the Vlasov-Poisson equations towards a 
quasistationary state on the coarse-grained scale 
is related to the concepts of violent relaxation, phase mixing and nonlinear
Landau damping \cite{villani}.} 
Numerical simulations show that
the Schr\"odinger-Poisson equations experience a process
of gravitational cooling \cite{seidel94,gul0,gul} which is similar to the process of violent
relaxation of collisionless stellar systems described by the Vlasov-Poisson
equations \cite{lb}. This process takes place on a very short timescale, of the order of the dynamical -- or free fall -- time. As a result of gravitational cooling  and violent
relaxation, the system displays damped oscillations and achieves a 
quasistationary state with a  core-halo
structure which is in virial equilibrium. The
virial theorem writes\footnote{See Appendix G of
\cite{chavtotal} for a general derivation of the quantum virial theorem of
self-gravitating BECs from the hydrodynamic representation of the
GPP equations.}
\begin{eqnarray}
\label{vireq}
2\langle\Theta\rangle+\langle W\rangle=0,
\end{eqnarray}
where $E_{\rm tot}=\Theta+W$ is constant. This core-halo structure has
been evidenced in direct numerical simulations of the Schr\"odinger-Poisson
equations
\cite{ch2,ch3,schwabe,mocz,moczSV,veltmaat,moczprl,moczmnras,veltmaat2}.

The
quantum
core corresponds to the ground state of the Schr\"odinger-Poisson equations. It
is described by a stationary
wavefunction of the form
\begin{eqnarray}
\label{nwh1}
\psi({\bf r},t)=\phi({\bf r})e^{-iEt/\hbar},
\end{eqnarray}
where $\phi({\bf r})$ and $E$  are real. They are determined by the eigenvalue
problem
\begin{eqnarray}
\label{wh1}
-\frac{\hbar^2}{2m}\Delta\phi+m\Phi\phi=E\phi,
\end{eqnarray}
\begin{eqnarray}
\label{wh2}
\Delta\Phi=4\pi G \phi^2.
\end{eqnarray}
The ground state is the solution of these equations with the lowest eigenenergy
$E$. This solution is spherically symmetric and has no node so that the density
profile decreases monotonically (see, e.g., \cite{prd1,prd2}).

In the hydrodynamic
representation, the core is determined by the condition of quantum hydrostatic
equilibrium
\begin{equation}
\label{gc2}
\frac{\rho}{m}\nabla
Q+\rho\nabla\Phi={\bf 0}
\end{equation}
coupled to the Poisson equation (\ref{mad7}). Eqs. (\ref{wh1}) and
(\ref{gc2}) are equivalent. Indeed, since  $\phi=\sqrt{\rho}$, Eq. (\ref{wh1}) 
can be rewritten as
\begin{eqnarray}
\label{wh1b}
Q+m\Phi=E.
\end{eqnarray}
Taking the gradient of this equation and multiplying by $\rho/m$, we obtain
Eq. (\ref{gc2}).

The quantum core (ground state) can also be obtained by
minimizing the energy at fixed mass \cite{prd1}:
\begin{equation}
\min\ \lbrace {E_{\rm tot}}\, |\,  M \,\, {\rm fixed} \rbrace.
\label{vpurt}
\end{equation}
First, one can show that an extremum of energy at fixed mass is a steady state
of the
Schr\"odinger-Poisson equations. Indeed, writing  the
variational principle as
\begin{equation}
\label{nw35qr}
\delta E_{\rm tot}-\frac{\mu}{m}\delta M=0,
\end{equation}
where $\mu/m$ is a Lagrange multiplier (global chemical potential) taking into
account the mass constraint, we get \cite{prd1}
\begin{equation}
\label{hyeb}
Q+m\Phi=\mu,
\end{equation}
which is equivalent to Eq. (\ref{wh1b}) with $E=\mu$. This establishes that the
eigenenergy is equal to the global chemical potential. Furthermore, one can
show   that an equilibrium state of the Schr\"odinger-Poisson
equations is stable if,
and only if, it is
a minimum of energy at fixed mass ($\delta^2E_{\rm tot}>0$ for all
perturbations that conserve mass).\footnote{See Appendix B of \cite{jeansMR}
for a detailed proof of this result.} These results are very general
for dynamical systems \cite{holm}. They are basically due to the fact that
$E_{\rm tot}$ and $M$ are individually
conserved by the Schr\"odinger-Poisson equations.

The equations for the ground state can be solved
numerically \cite{rb,membrado,gul0,gul,prd2,ch2,ch3,pop,hui} leading to
what is generally called a soliton. In
general, the soliton (quantum core) has a mass $M_c$ and an energy $E_c$ that
differ from the initial mass $M$ and initial energy $E$ of the system. Therefore,
the excess mass and the
excess energy must be redistributed in a halo made of
scalar waves (radiation) resulting from the interference of excited states.
Numerical
simulations of the Schr\"odinger-Poisson equations
\cite{ch2,ch3,schwabe,mocz,moczSV,veltmaat,moczprl,moczmnras,veltmaat2}  show
that this
halo has a 
density profile that is consistent with
the NFW profile obtained in CDM
simulations \cite{nfw}.\footnote{This is true in a 
``smoothed-out'' sense where the density profile is averaged on a scale larger
than the de Broglie length. Indeed, the small-scale interferences produce
time-dependent granules (or quasiparticles) of size $\lambda_{\rm
dB}$ that can induce a slow collisional evolution of the halo
\cite{hui,bft,bft2,meff}. We
shall not consider this collisional evolution here and remain at a purely
collisionless level.} The halo profile is also relatively close to an
isothermal profile with an
effective temperature $T_{\rm eff}$ (see, e.g., the Appendix of \cite{moczSV}). 
The analogy 
between the process of
gravitational cooling and the process of violent relaxation will help us
determining the structure of the halo. To that purpose, 
we have to introduce a DF approach as done in Sec. \ref{sec_dfaw}.

\subsection{Interferences}
\label{sec_inter}

In order to better undertand the structure of the halo, it
is useful to follow Ref. \cite{lin} and decompose the wavefunction under the
form
\begin{equation}
\label{inter1}
\psi({\bf r},t)=\sum_n c_n \psi_n({\bf r})e^{-i E_n t/\hbar},
\end{equation}
where $\psi_n({\bf r})$ are the eigenfunctions of the time-independent
Schr\"odinger equation constructed with the locally
averaged gravitational potential $\Phi({\bf r})$ and
$c_n$ is a random complex
amplitude for mode $n$ (the gravitational potential  is obtained
self-consistently from
the Poisson equation $\Delta\Phi=4\pi G \sum_n |c_n|^2 |\psi_n({\bf r})|^2$).
Instead of many particles, one deals here with a single
wave
that has many noninteracting eigenstates. The exact density profile
$\rho=|\psi|^2$ is given by 
\begin{eqnarray}
\label{inter2}
\rho({\bf r},t)&=&\sum_n |c_n|^2 |\psi_n({\bf r})|^2\nonumber\\
&+&\sum_{n\neq m} c_n
c_m^* \psi_n({\bf r})  \psi_m({\bf r})^*  e^{i (E_m-E_n) t/\hbar}.
\end{eqnarray}
The first term  in Eq. (\ref{inter2}) determines the smooth, stationary, profile
of the DM halo (the one
that can be compared to the NFW profile for example). Here, $p_n=|c_n|^2$ is a
weighting factor
which is proportional to the probability of the $n$-th
state. The fundamental mode $n=0$ (ground state) corresponds to
the condensate  (soliton) forming the quantum core and the excited states $n>0$
give rise to the halo. The second term in Eq. (\ref{inter2}),
which is time-dependent, represents the interferences of the different
eigenstates. It gives rise to density granules -- or quasiparticles -- of
size $\lambda_{\rm dB}$ that
provide pressure support against self-gravity and that can cause a slow
(secular) collisional evolution of the halo \cite{hui,bft,bft2,meff}.
Developing this  approach, Lin {\it et al.} \cite{lin} derived the classical
particle DF
$f({\bf r},{\bf v})$ of the halo using analytical and numerical technics
and found that it is well-fitted by the fermionic King model
\cite{chavmnras,clm2}.\footnote{In principle, for a system of
bosons, we would expect that
$f({\bf r},{\bf v})$ is given by the Bose-Einstein or Rayleigh-Jeans DF. That
would be the case if the statistical equilibrium state resulted from a
collisional relaxation \cite{meff}. However, in the present context, the
evolution of the
system is collisionless and the DF is given by the Lynden-Bell DF which is
similar to the Fermi-Dirac DF (see below). This is why Lin {\it et al.} 
\cite{lin} find that the DF of the halo is well-described by the Fermi-Dirac DF
\cite{lb,csmnras} or by the fermionic King model \cite{chavmnras,clm2}. In the
nondegenerate limit, these DFs reduce to the Boltzmann DF and to the classical
King
model.} By
contrast, the ground state (soliton) is a
highly nonlinear object that cannot be described by this DF. In the following
sections, we shall
develop a complementary description of the core-halo structure of DM halos based
on a
kinetic approach.

{\it Remark:} We note that the
Schr\"odinger-Poisson equations (\ref{h1}) and (\ref{h2}), or the corresponding
quantum Euler-Poisson equations (\ref{mad4})-(\ref{mad7}) do not relax
towards an equilibrium state. Indeed, the density $\rho({\bf r},t)$ from Eq.
(\ref{inter2}) is always time-dependent due to small-scale interferences.
However,
if we locally average over these fluctuations, we get a smooth density that
tends to an equilibrium state [first term in Eq. (\ref{inter2})]. This is very
similar to the process of violent relaxation based on the Vlasov-Poisson
equations (\ref{vlasov1}) and (\ref{vlasov2}) \cite{lb}. In that case, the
fine-grained DF
$f({\bf r},{\bf v},t)$ never reaches an equilibrium state but develops filaments
at smaller and smaller scales. However, if we locally
average over this filamentation, we obtain a coarse-grained DF
$\overline{f}({\bf r},{\bf v},t)$  that rapidly relaxes towards a
quasistationary state $\overline{f}_{\rm QSS}({\bf r},{\bf v})$
(see Appendix \ref{sec_vrx}).

\section{Distribution function approach}
\label{sec_dfaw}

\subsection{Wigner equation}
\label{sec_w}

From the wavefunction $\psi({\bf r},t)$  we
can define the Wigner DF
$f({\bf r},{\bf v},t)$ by \cite{wigner}
\begin{eqnarray}
\label{qw1}
f({\bf r},{\bf v},t)=\frac{m^3}{(2\pi\hbar )^3} \int d{\bf
y}\,  e^{im{\bf v}\cdot {\bf y}/\hbar} \nonumber\\
\times\psi^*\left ({\bf
r}+\frac{{\bf y}}{2},t\right )\psi\left ({\bf r}-\frac{{\bf
y}}{2},t\right ).
\end{eqnarray}
The term in the second line represents the density matrix. One
can check that $\int f\, d{\bf v}=|\psi|^2=\rho$ (see Appendix \ref{sec_cw}). On
the other hand, using the
Schr\"odinger equation (\ref{h1}), one can show
that $f({\bf r},{\bf
v},t)$ satisfies an equation of the form 
\begin{eqnarray}
\label{qw2}
\frac{\partial f}{\partial t}+{\bf v}\cdot \frac{\partial f}{\partial {\bf
r}}-\frac{im^4}{(2\pi\hbar)^3\hbar}\int e^{im({\bf v}-{\bf v}')\cdot
{\bf y}/\hbar}\nonumber\\
\times \left\lbrack\Phi\left ({\bf r}+\frac{{\bf
y}}{2},t\right )-\Phi\left ({\bf r}-\frac{{\bf
y}}{2},t\right )\right\rbrack f({\bf r},{\bf v}',t)\, d{\bf
y}d{\bf v}'=0,\nonumber\\
\end{eqnarray}
which is called the Wigner (or Wigner-Moyal) equation
\cite{wigner,moyal}. It can be viewed as the quantum
generalization of the Vlasov equation to which it reduces in the classical
limit $\hbar\rightarrow 0$ (see Appendix \ref{sec_cw}). For self-gravitating
BECs, the Wigner
equation must be coupled to the Poisson equation
\begin{eqnarray}
\label{qw2b}
\Delta\Phi=4\pi G\int f\, d{\bf v}.
\end{eqnarray}
Eqs. (\ref{qw2}) and (\ref{qw2b}) form 
the Wigner-Poisson
equations. The Wigner-Poisson equations conserve the
energy $E=(1/2)\int f v^2\, d{\bf
r}d{\bf v}+(1/2)\int \rho\Phi\, d{\bf r}$ and the mass $M=\int f\, d{\bf
r}d{\bf v}$ (see Appendix \ref{sec_cw}). 

{\it Remark:} The Wigner DF $f({\bf r},{\bf v},t)$ is real
but not necessarily positive. Therefore, it does not have the status of
a true DF. One way to overcome this difficulty is to use the
Husimi \cite{husimi} representation which is essentially a smoothed version of
the Wigner quasiprobability distribution.

\subsection{Coarse-grained Wigner equation}
\label{sec_wcg}

As for the classical Vlasov-Poisson
equations (see Refs. \cite{kp,sl,csr,chavmnras,kingen,dubrovnik}
and
Appendix \ref{sec_vrx}), it is relevant to operate a coarse-graining on the
Wigner-Poisson
equations. This is especially justified at sufficiently large scales (i.e. in
the
halo) where quantum effects are weak and the
system essentially behaves like a classical collisionless self-gravitating gas
described by the
Vlasov-Poisson equations. Therefore, it is expected to experience phase
mixing, nonlinear Landau damping, and violent relaxation \cite{bt}. This
coarse-grained description will allow us to take into account the processes of
violent relaxation and gravitational cooling mentioned in Sec. \ref{sec_gc}.

As in the case of
the  classical Vlasov equation, the coarse-grained DF
$\overline{f}({\bf r},{\bf v},t)$ does
not satisfy the Wigner equation because of phase-space correlations (see
Appendix \ref{sec_vrb}).
These phase-space correlations introduce an effective ``collision'' term
in the right hand side of the coarse-grained Wigner equation. For simplicity, we
shall neglect
quantum corrections in this  ``collision'' term and use the same
parametrization as for the classical process of violent
relaxation.\footnote{As we shall see, the effective collision
term accounts for the formation of the halo. In the halo, quantum effects are
erased on the coarse-grained scale (they manifest themselves only on the
fine-grained scale through the presence of granules \cite{hui,bft,bft2,meff}).
By contrast, we keep quantum effects in the left hand side of the coarse-grained
Wigner equation (advection term). This term accounts for the formation of the
soliton in
which quantum effects are dominant.} This parametrization is based on a
heuristic MEPP (see Ref. \cite{csr} and
Appendix \ref{sec_vrb}).
Therefore, we propose a coarse-grained Wigner equation of the form
\begin{eqnarray}
\label{qw3}
\frac{\partial \overline{f}}{\partial t}+{\bf v}\cdot \frac{\partial
\overline{f}}{\partial {\bf
r}}-\frac{im^4}{(2\pi\hbar)^3\hbar}\int e^{im({\bf v}-{\bf v}')\cdot
{\bf y}/\hbar}\nonumber\\
\times \left\lbrack {\Phi}\left ({\bf r}+\frac{{\bf
y}}{2},t\right )-{\Phi}\left ({\bf r}-\frac{{\bf
y}}{2},t\right )\right\rbrack \overline{f}({\bf r},{\bf v}',t)\, d{\bf
y}d{\bf v}'\nonumber\\
=\frac{\partial}{\partial {\bf v}}\cdot \left\lbrack D\left
(\frac{\partial
\overline{f}}{\partial {\bf
v}}+\beta \overline{f}(1-\overline{f}/\eta_0){\bf v}\right
)\right\rbrack,\qquad
\end{eqnarray}
coupled to the Poisson equation
\begin{eqnarray}
\label{qw3c}
\Delta{\Phi}=4\pi G\int \overline{f}\, d{\bf v}.
\end{eqnarray}
The left hand side of Eq. (\ref{qw3}) is the usual Wigner term
(we can write $\Phi$ instead of $\overline{\Phi}$ since it is a smooth field
produced by
$f$ averaged over ${\bf v}$). The  right hand side of Eq. (\ref{qw3}) can be
interpreted as  a fermionic
Kramers (or Fokker-Planck) ``collision'' term. As a result, the coarse-grained Wigner equation
(\ref{qw3}) is similar to the fermionic Wigner-Kramers equation. The
fermionic nature of the collision term arises from the Lynden-Bell
exclusion principle ($\overline{f}\le \eta_0$) introduced in the context of
collisionless stellar systems \cite{lb}. Here, $\eta_0$ denotes the maximum 
value of the initial DF.\footnote{If DM is
made of fermions, we have $\eta_0\sim m^4/h^3$ (see footnote 34 in \cite{clm2}).
Therefore, degeneracy effects in the sense of Lynden-Bell are important. They
lead to a quantum core in the form of a fermion ball. By contrast, the
Heisenberg uncertainty principle (quantum potential) is negligible, or small,
for
fermions. For condensed bosons, this is the opposite. Indeed, $\eta_0$ is in
general very large ($\eta_0\gg m^4/h^3$) so that
degeneracy effects in the sense of Lynden-Bell are usually negligible
($\overline{f}\ll\eta_0$). By
contrast, the
Heisenberg uncertainty principle (quantum potential) is important
for bosons. It leads to a quantum core in the form of a  soliton. Here,
for the sake of generality, we shall take into account all these effects
although some of them may be negligible depending on the situation.} The
Lynden-Bell
exclusion principle is similar to the
Pauli exclusion principle in quantum mechanics but with, of course, a
completely different interpretation (see Appendix \ref{sec_vra}).

The first term
in the fermionic 
Kramers collision operator is a
diffusion and the second term is a friction. The friction may be
interpreted  as a form of nonlinear Landau
damping since it is associated with a collisionless relaxation. The friction coefficient and the diffusion coefficient are linked by the Einstein relation
\begin{eqnarray}
\label{qw3b}
\xi=\beta D,
\end{eqnarray}
where $\beta=1/T_{\rm eff}$
is an inverse effective temperature (see
Appendix \ref{sec_vrb}). The Einstein relation (\ref{qw3b}) expresses the fluctuation-dissipation
theorem. The diffusion coefficient $D$ is not given by the MEPP but it
can be calculated by developing a quasilinear theory of the process of violent
collisionless relaxation like in \cite{kp,sl,chavmnras,kingen,dubrovnik}. This
theory is valid in a regime of ``gentle'' relaxation. In
the present context, it leads to the
Wigner-Landau equation
\begin{eqnarray}
\label{qw3l}
\frac{\partial \overline{f}}{\partial t}+{\bf v}\cdot \frac{\partial
\overline{f}}{\partial {\bf
r}}-\frac{im^4}{(2\pi\hbar)^3\hbar}\int e^{im({\bf v}-{\bf v}')\cdot
{\bf y}/\hbar}\qquad\nonumber\\
\times \left\lbrack {\Phi}\left ({\bf r}+\frac{{\bf
y}}{2},t\right )-{\Phi}\left ({\bf r}-\frac{{\bf
y}}{2},t\right )\right\rbrack \overline{f}({\bf r},{\bf v}',t)\, d{\bf
y}d{\bf v}'\qquad\nonumber\\
=\frac{\partial}{\partial
v_i}\int d{\bf v}' K_{ij}\left\lbrace \overline{f}' \left (1-
\frac{\overline{f}'}{\eta_0}\right )\frac{\partial \overline{f}}{\partial v_j}-
 \overline{f} \left
(1-
\frac{\overline{f}}{\eta_0}\right )\frac{\partial \overline{f}'}{\partial
v'_j}\right\rbrace,\nonumber\\
\end{eqnarray}
\begin{eqnarray}
\label{cle3}
K_{ij}=2\pi
G^2\eta_0\epsilon_r^3\epsilon_v^3\ln\Lambda\frac{u^2\delta_{ij}-u_iu_j}{u^3},
\end{eqnarray}
where ${\bf u}={\bf
v}'-{\bf v}$ is the relative velocity and $\ln\Lambda=\ln(R/\epsilon_r)$ the
Coulomb logarithm. The Wigner-Kramers equation (\ref{qw3}) is recovered
from the Wigner-Landau
equation (\ref{qw3l}) in a thermal bath approximation, and the diffusion
coefficient can be explicitly calculated like in Ref. \cite{chavmnras}. It
is then explicitly demonstrated that the
process of collisionless violent relaxation takes place on a few dynamical
times.\footnote{The
fact that the coarse-grained DF $\overline{f}({\bf r},{\bf v},t)$ relaxes
towards the Lynden-Bell DF on a few
dynamical times is interpreted by Kadomtsev and Pogutse \cite{kp} in terms of
``collisions'' between  macroparticles with a large effective mass $m_{\rm
eff}\sim \eta_0 \epsilon_r^3\epsilon_v^3$ (these macroparticles are
fundamentally different from the quasiparticles
introduced by \cite{hui} in relation to the collisional evolution of FDM
halos on a secular timescale). In the foregoing equations,
$\epsilon_r(t)$ and $\epsilon_v(t)$ denote the position and velocity correlation
scales of the fluctuations $\delta f$ (see Appendix \ref{sec_vrb}).  These
correlation scales are expected to
decrease with time. If this decay is sufficiently rapid, it can slow down and
even stop the
collisionless relaxation before the Lynden-Bell DF is
reached. This kinetic blocking can account for incomplete relaxation and 
solve the infinite mass problem of the Lynden-Bell DF
(see \cite{csr,chavmnras,dubrovnik} and more specifically \cite{incomplete} for
additional discussion
about the concept of incomplete relaxation). This
may explain why DM halos have a NFW or Burkert profile instead of a Lynden-Bell
profile
(see footnote 7).}

For classical particles ($\hbar=0$),
the fermionic Kramers collision
term
accounts for the relaxation of the coarse-grained DF towards
the Lynden-Bell distribution
\begin{eqnarray}
\label{qw4}
\overline{f}_{\rm LB}({\bf r},{\bf v})=\frac{\eta_0}{1+e^{\beta (v^2/2+\Phi({\bf
r}))-\alpha}},
\end{eqnarray}
where $\epsilon=v^2/2+\Phi({\bf r})$ is the individual energy of the particles by unit of mass. The Lynden-Bell distribution, which is similar to the Fermi-Dirac distribution, is a particular steady state of the Vlasov-Poisson equations (\ref{vlasov1}) and (\ref{vlasov2}). It represents the most probable state of the system at statistical equilibrium (see Appendix \ref{sec_vra}). It is obtained by  maximizing the Lynden-Bell entropy at fixed mass and energy. This is also the only equilibrium
state of the coarse-grained Vlasov equation (\ref{mepp10}). It cancels individually the advection (Vlasov) term and the collision (fermionic Kramers) term. Now, for quantum systems
($\hbar\neq 0$), the Lynden-Bell distribution (\ref{qw4}) is not an exact steady
state of the Wigner equation (\ref{qw2}) or of the coarse-grained Wigner equation (\ref{qw3}) because
it does not cancel the advection (Wigner) term
exactly.\footnote{Quantum effects (Heisenberg uncertainty principle) lead to a
small
modification of the Lynden-Bell DF of the
order $O(\hbar^2)$ in
the same manner that they lead to a small modification of the Boltzmann
distribution for a system at thermal equilibrium \cite{wigner}.}  However, in
the present
work, we shall disregard this difficulty. Indeed, we expect
that the Lynden-Bell distribution provides a reasonable description of
the halo where quantum effects are weak. In that case, it cancels the
advection (Wigner) term approximately. Quantum effects are important in
the core (soliton) and they are taken into account in  the advection (Wigner)
term whose cancellation is equivalent to the condition of quantum
hydrostatic equilibrium from Eq. (\ref{gc2}). By contrast, in the core, the
Lynden-Bell
distribution is usually subdominant. The distinction
between a quantum core and an approximately isothermal halo (in the sense of
Lynden-Bell) will become clear in the hydrodynamic representation of the
coarse-grained Wigner equation developed below. The importance of violent relaxation in
establishing an isothermal-like halo in the BECDM model was stressed in
\cite{chavtotal,modeldm}.

\subsection{Truncated Lynden-Bell distribution}
\label{sec_king}

As is well-known, the Lynden-Bell distribution (\ref{qw4}) coupled to the Poisson equation (\ref{qw3c}) generates configurations with an infinite mass.\footnote{Mathematically, this is because the Lynden-Bell
distribution reduces to the Boltzmann distribution (\ref{mep16}) at large
distances (where the system is dilute) so the density $\rho$ decreases as $r^{-2}$ like for the self-gravitating isothermal
sphere \cite{bt}.} Therefore, strictly speaking, the coarse-grained distribution
$\overline{f}({\bf r},{\bf v},t)$ does not relax towards the Lynden-Bell
distribution (\ref{qw4}) since the mass is necessarily finite. This is related
to the problem of incomplete violent relaxation \cite{lb,csr} and to the fact
that the Lynden-Bell distribution does not take into account 
the escape of high energy particles. Nevertheless, the Lynden-Bell distribution
is expected to be approximately valid for tightly bound particles with
sufficiently negative energies ($\epsilon<0$).

We can improve the Lynden-Bell
distribution by taking into account the escape of unbound particles with
positive energy ($\epsilon>0$) or by taking into account tidal effects from
neighboring systems. In particular, from the fermionic Vlasov-Kramers equation
(\ref{mepp10}), one can derive a truncated Lynden-Bell distribution of the form
\cite{chavmnras}
\begin{eqnarray}
\label{qw7}
\overline{f}=A\frac{e^{-\beta(\epsilon-\epsilon_m)}-1}{1+\frac{A}{\eta_0}e^{
-\beta(\epsilon-\epsilon_m)}}\qquad (\epsilon\le \epsilon_m),
\end{eqnarray}
\begin{eqnarray}
\label{qw8}
\overline{f}=0\qquad (\epsilon\ge \epsilon_m).
\end{eqnarray}
This distribution vanishes  above a 
certain escape energy $\epsilon_m$ which is equal to zero for isolated systems
and which is strictly negative for tidally truncated systems.  The truncated
Lynden-Bell distribution (\ref{qw7}) is similar to the fermionic King model
\cite{stella,chavmnras,clm2}. In the nondegenerate (dilute) limit, it becomes
similar to the classical King model
\begin{eqnarray}
\label{qw7b}
\overline{f}=A\left\lbrack e^{-\beta(\epsilon-\epsilon_m)}-1\right\rbrack \qquad (\epsilon\le \epsilon_m),
\end{eqnarray}
\begin{eqnarray}
\label{qw8b}
\overline{f}=0\qquad (\epsilon\ge \epsilon_m),
\end{eqnarray}
which was
introduced in relation to globular clusters evolving under the
effect of
two-body encounters \cite{king}. In the present context, the
thermalization of the
system
is due to Lynden-Bell's type of violent collisionless
relaxation and the fermionic nature of the DF is related to
Lynden-Bell's exclusion principle arising from the Vlasov equation.

The truncated Lynden-Bell distribution (or fermionic King model) has a finite mass and a maximum density in phase space
which prevents gravitational collapse. It has been studied in detail in
\cite{clm2}. The corresponding configurations have a core-halo
structure with
a degenerate core similar to a ``fermion ball'' (a polytrope of index $n=3/2$)
and an isothermal halo. Because of the truncation, the isothermal halo does not
extend to infinity. The density drops to zero at a finite radius identified
with
the
tidal radius.

{\it Remark:} In the context of BECDM halos, Lin {\it et al.}
\cite{lin}, by developing the model from Eqs. (\ref{inter1}) and
(\ref{inter2}) and
using results from numerical simulations, observed that the
virialized state produced by gravitational cooling, in addition of
containing a solitonic core (arising from the bosonic nature of the particles),
has a DF consistent with the truncated Lynden-Bell
distribution (\ref{qw7}) and (\ref{qw8}). This corroborates our
previous qualitative arguments \cite{chavtotal} according to which the
(truncated) Lynden-Bell distribution provides a good description of the
``atmosphere'' of DM halos surrounding the solitonic core.

\subsection{Quantum Jeans equations}
\label{sec_qjeans}

Taking the hydrodynamic moments of the 
coarse-grained
Wigner equation
(\ref{qw3}), we
obtain the following equations\footnote{For simplicity, we assume here that $D$
is constant.}
\begin{equation}
\label{qj1}
\frac{\partial\rho}{\partial t}+\nabla\cdot (\rho {\bf u})=0,
\end{equation}
\begin{equation}
\label{qj2}
\frac{\partial {\bf u}}{\partial t}+({\bf u}\cdot \nabla){\bf
u}=-\frac{1}{\rho}\partial_jP_{ij}-\nabla\Phi-\frac{1}{\rho}D\beta \int
\overline{f}(1-\overline{f}/\eta_0){\bf v}\, d{\bf v},
\end{equation}
where we have introduced the local density
\begin{equation}
\label{qj3}
\rho=\int \overline{f}\, d{\bf v},
\end{equation}
the local velocity 
\begin{equation}
\label{qj4}
{\bf u}=\frac{1}{\rho}\int \overline{f} {\bf v}\, d{\bf v},
\end{equation}
and the pressure tensor
\begin{equation}
\label{qj5}
P_{ij}=\int \overline{f} ({\bf v}-{\bf u})_i  ({\bf v}-{\bf
u})_j\, d{\bf v}.
\end{equation}
Equations (\ref{qj1}) and (\ref{qj2}) are called the quantum damped Jeans
equations.\footnote{More generally, we can build up an infinite hierarchy of
such equations by taking the successive  moments of the DF.} They
coincide with the classical
damped Jeans equations derived from the coarse-grained Vlasov equation
(see Ref. \cite{csr} and the Appendix of \cite{moczSV}). Indeed, $\hbar$ does not explicitly appear in these
equations. Explicit factors of $\hbar$ enter only in the higher
moment equations of the hierarchy. We note that
these hydrodynamic
equations are not closed since the pressure tensor (\ref{qj5}) depends on the
coarse-grained DF $\overline{f}({\bf r},{\bf
v},t)$ which is not explicitly known in general. In the following, we 
propose a heuristic manner to close these equations by combining the results
obtained in two extreme limits of our formalism corresponding to $D=0$ and
$\hbar=0$ respectively.

\subsubsection{Hydrodynamic representation of the fine-grained Wigner equation}
\label{sec_qja}

If we consider
the fine-grained Wigner equation (\ref{qw2}), there is no collision term
($D=\xi=0$) and we
obtain the quantum Jeans
equations
\begin{equation}
\label{qj6}
\frac{\partial\rho}{\partial t}+\nabla\cdot (\rho {\bf u})=0,
\end{equation}
\begin{equation}
\label{qj7}
\frac{\partial {\bf u}}{\partial t}+({\bf u}\cdot \nabla){\bf
u}=-\frac{1}{\rho}\partial_jP_{ij}-\nabla\Phi.
\end{equation}
They coincide with the classical Jeans equations derived from the fine-grained Vlasov
equation (see Ref. \cite{bt} and Appendix \ref{sec_jeans}). Indeed, $\hbar$
does not explicitly appear in these equations. In the classical case
($\hbar=0$), these equations are not closed \cite{bt}. However, in the quantum
case ($\hbar\neq 0$), they can
be closed! Indeed, since the Wigner equation (\ref{qw2})  is equivalent to the
Schr\"odinger  equation  (\ref{h1}), the quantum Jeans equations (\ref{qj6})
and (\ref{qj7}) obtained from the Wigner equation must coincide with the quantum
Euler equations (\ref{mad4}) and   (\ref{mad6}) obtained from
the Schr\"odinger equation. This implies that the pressure
tensor $P_{ij}$ in Eq. (\ref{qj7}) is exactly given by Eq. (\ref{mad10}), i.e.,
\begin{equation}
\label{qj9}
P_{ij}=P_{ij}^{Q}.
\end{equation}
This result can also be obtained by a direct calculation,
substituting Eq.
(\ref{qw1}) into Eq. (\ref{qj5}) and using Eqs. (\ref{mad1})-(\ref{mad3}).
In the same manner, one can show that the density defined by Eq. (\ref{qj3}) and
the velocity defined by Eq. (\ref{qj4}) are equivalent to Eqs.
(\ref{mad1})-(\ref{mad3}). These calculations are detailed in
Appendix \ref{sec_cw}.

\subsubsection{Hydrodynamic representation of the coarse-grained
Vlasov equation}
\label{sec_qjb}

If we consider the process of violent relaxation but ignore quantum effects
($\hbar=0$), we
are led back to the situation studied in Ref. \cite{csr} (see also the Appendix of Ref. \cite{moczSV}) based on the coarse-grained Vlasov equation. This leads to the classical damped
Jeans equations that are equivalent to Eqs. (\ref{qj1}) and (\ref{qj2}) (see the comment made after Eq. (\ref{qj5})). These equations are not closed. In Ref. \cite{csr}, it was
proposed to
compute the pressure tensor (\ref{qj5}) by making a LTE approximation for the DF
based on the 
Lynden-Bell statistics
\begin{eqnarray}
\label{qj10}
\overline{f}_{\rm LTE}({\bf r},{\bf v},t)=\frac{\eta_0}{1+e^{\beta
({\bf v}-{\bf u}({\bf r},t))^2/2-\alpha({\bf r},t)}},
\end{eqnarray}
where ${\bf u}({\bf r},t)$ is the local velocity (\ref{qj4}) and $\alpha({\bf
r},t)$ is a local chemical potential which can be related to the local density
$\rho({\bf
r},t)$ by substituting Eq. (\ref{qj10}) into Eq. (\ref{qj3}).
If we now compute the pressure tensor (\ref{qj5}) with the DF (\ref{qj10}), we
obtain
\begin{equation}
\label{qj11}
P_{ij}=P_{\rm LB}(\rho)\delta_{ij},
\end{equation}
where $P_{\rm LB}(\rho)$ is the Lynden-Bell equation of state which is similar
to the Fermi-Dirac equation of state (see Appendix \ref{sec_vra}). On the other hand, in the last term of Eq. (\ref{qj2}), we shall make for simplicity the approximation
\begin{equation}
\label{qj12}
\int
\overline{f}(1-\overline{f}/\eta_0){\bf v}\, d{\bf v}\simeq \rho {\bf u},
\end{equation}
which amounts to neglecting degeneracy effects in
the friction term. This is a relevant approximation in the dilute 
halo which is nondegenerate or weakly degenerate (in the sense of Lynden-Bell).
With the closure from Eq. (\ref{qj11}) and the approximation from Eq.
(\ref{qj12}),
the damped Jeans equation (\ref{qj2}) becomes
\begin{equation}
\label{qj13}
\frac{\partial {\bf u}}{\partial t}+({\bf u}\cdot \nabla){\bf
u}=-\frac{1}{\rho}\nabla P_{\rm LB}-\nabla\Phi-\xi{\bf u},
\end{equation}
where $P_{\rm LB}(\rho)$ is the Lynden-Bell pressure and $\xi$ is the
friction coefficient given by the Einstein relation
(\ref{qw3b}).\footnote{The Lynden-Bell pressure and the
friction term are in some sense related to the correlation function
$\overline{\delta\rho\nabla\delta\Phi}$ that emerges from the coarse-graining of
the quantum Euler-Poisson equations (\ref{mad4}) and (\ref{mad7}), where
$\delta\rho$ and $\delta\Phi$
denote the fluctuations about the coarse-grained (smooth) fields. However, the
proper description of these correlations requires the analysis of Appendix
\ref{sec_vrb} in phase space.} Eq. (\ref{qj13}) is called the damped Euler
equation. The Lynden-Bell equation
of
state is given in parametric form by Eqs. (\ref{mep9}) and (\ref{mep10}). It has the
same form as the Fermi-Dirac equation of state in quantum mechanics except that
$gm^4/h^3$ (where $g$ is the multiplicity of the quantum states) is replaced by
$\eta_0$, the maximum value of the 
DF. The Lynden-Bell equation of state  is
essentially an isothermal equation of state with a modification at
high densities taking into account the Lynden-Bell exclusion principle which is
similar to the Pauli
exclusion principle in quantum mechanics.
In the nondegenerate limit, valid at low densities, it reduces to the
classical isothermal
equation of state
\begin{equation}
\label{qj16}
P_{\rm LB}^{\rm th}=\rho T_{\rm eff}
\end{equation}
with an effective temperature $T_{\rm eff}$. In the
completely degenerate limit, valid at high densities, we get a pressure of zero-point energy
\begin{equation}
\label{qj17}
P_{\rm LB}^{(0)}=\frac{1}{5}\left (\frac{3}{4\pi\eta_0}\right )^{2/3}\rho^{5/3},
\end{equation}
corresponding to a polytrope of index $\gamma=5/3$ and polytropic constant
$K=({1}/{5})\left ({3}/{4\pi\eta_0}\right )^{2/3}$
like in the theory of nonrelativistic white dwarf stars \cite{chandra}. The
complete 
Lynden-Bell equation of state can be conveniently approximated by
\begin{equation}
\label{qj18}
P_{\rm LB}=P_{\rm LB}^{\rm th}+P_{\rm LB}^{(0)}=\rho T_{\rm
eff}+\frac{1}{5}\left (\frac{3}{4\pi\eta_0}\right )^{2/3}\rho^{5/3},
\end{equation}
where the first term accounts for the effective temperature and the second term
accounts for the Lynden-Bell exclusion principle. When coupled to gravity the Lynden-Bell
equation of state leads to configurations with a core-halo structure. They are
made of a ``fermionic'' core (in the sense of Lynden-Bell) and an isothermal
halo. These core-halo structures have been computed in
\cite{csmnras} in relation to
the Lynden-Bell theory of violent relaxation but they also appear in numerous
works  on
fermionic DM halos in which the DM particle is a fermion like a massive
neutrino (see the Introduction
of Ref. \cite{gr1} and Sec. V.A of \cite{modeldm} for an exhaustive list of
references).

\subsubsection{Hydrodynamic representation
of the coarse-grained Wigner equation}
\label{sec_lbb}

We now consider the coarse-grained Wigner equation (\ref{qw3}) and propose to
close the quantum damped Jeans equations  (\ref{qj1}) and (\ref{qj2}) by
simply superposing the results obtained previously. Therefore, we propose to
approximate the pressure tensor of Eq. (\ref{qj5}) by
\begin{equation}
\label{qj20}
P_{ij}=P_{ij}^{Q}+P_{\rm LB}(\rho)\delta_{ij},
\end{equation}
where $P_{ij}^{Q}$ is the quantum pressure tensor (see Sec. \ref{sec_qja}) and $P_{\rm
LB}(\rho)$ is the Lynden-Bell pressure (see Sec. \ref{sec_qjb}).

{\it Remark:} In principle, there is a correction of
order $O(\hbar^2)$ in
the Lynden-Bell pressure due
to the
effect of the Heisenberg uncertainty principle (see
footnote 22) and the presence of the granules in the halo. However, we shall
neglect this small correction.

\section{Heuristic equations parameterizing the complex dynamics of BECDM halos}
\label{sec_pcd}

\subsection{Hydrodynamic equations}
\label{sec_ch}

Combining the previous results, we find that the hydrodynamic equations
parameterizing the complex dynamics of BECDM halos in our model are
\begin{equation}
\label{ch1}
\frac{\partial\rho}{\partial t}+\nabla\cdot (\rho {\bf u})=0,
\end{equation}
\begin{equation}
\label{ch2}
\frac{\partial {\bf u}}{\partial t}+({\bf u}\cdot \nabla){\bf
u}=-\frac{1}{\rho}\nabla P_{\rm LB}-\frac{1}{\rho}\nabla
P_{\rm int}-\frac{1}{m}\nabla Q-\nabla\Phi-\xi{\bf
u},
\end{equation}
\begin{eqnarray}
\label{ch3}
\Delta\Phi=4\pi G\rho.
\end{eqnarray}
For the sake of generality, we have
considered the case of self-interacting BECs and we have
added the pressure
\begin{equation}
\label{qj21}
P_{\rm int}(\rho)=\frac{2\pi a_s\hbar^2}{m^3}\rho^2
\end{equation}
due to their self-interaction (see, e.g., Ref.
\cite{prd1}).\footnote{According to Eq. (\ref{he7}), this
amounts to making the replacement $\Phi\rightarrow \Phi+4\pi a_s\hbar^2\rho/m^3$
in Eq. (\ref{qj13}).} This is a  polytropic equation of state of
index $\gamma=2$ and polytropic constant $K={2\pi a_s\hbar^2}/{m^3}$. When
$P_{\rm LB}=0$ and
$\xi=0$, we recover the hydrodynamic equations associated with the standard GPP
equations (\ref{he7}) and (\ref{he8}). However, the process of violent
relaxation generates an additional pressure $P_{\rm LB}$ and a friction
$\xi$.

If we introduce the total pressure $P=P_{\rm LB}+P_{\rm int}$, the quantum damped Euler equation (\ref{ch2}) can be rewritten as
\begin{equation}
\label{ze1}
\frac{\partial {\bf u}}{\partial t}+({\bf u}\cdot \nabla){\bf
u}=-\frac{1}{\rho}\nabla P-\frac{1}{m}\nabla
Q-\nabla\Phi-\xi{\bf
u}.
\end{equation}
On the other hand, if  we use the approximate expression  of the 
Lynden-Bell equation of state from Eq. (\ref{qj18}), we obtain
\begin{eqnarray}
\label{ze2}
\frac{\partial {\bf u}}{\partial t}+({\bf u}\cdot \nabla){\bf
u}=-\frac{1}{\rho}\nabla P_{\rm LB}^{(0)}-\frac{1}{\rho}\nabla
P_{\rm int}-\frac{1}{\rho}\nabla P_{\rm LB}^{\rm
th}\nonumber\\
-\frac{1}{m}\nabla Q-\nabla\Phi-\xi{\bf
u}.
\end{eqnarray}
Since $P_{\rm LB}^{(0)}$ and $P_{\rm int}$ are both polytropic equations of state (of index $\gamma=5/3$ and $\gamma=2$ respectively), and since the pressure is additive, it is useful to consider
the ``structural'' hydrodynamic equation
\begin{equation}
\label{ze3}
\frac{\partial {\bf u}}{\partial t}+({\bf u}\cdot \nabla){\bf
u}=-\frac{1}{\rho}\nabla P_{\rm poly}-\frac{1}{\rho}\nabla P_{\rm
th}-\frac{1}{m}\nabla
Q-\nabla\Phi-\xi{\bf
u},
\end{equation}
involving a general polytropic equation of state
\begin{eqnarray}
\label{ze4}
P_{\rm poly}=K\rho^{\gamma}
\end{eqnarray}
and an isothermal equation of state
\begin{eqnarray}
\label{ze5}
P_{\rm th}=\rho T_{\rm eff}.
\end{eqnarray}
We can then incorporate several polytropic equations of state in this model by simply summing the pressures.

{\it Remark:} In the strong friction limit $\xi\rightarrow +\infty$, we can
neglect the inertial term (l.h.s.) in the damped quantum Euler equation
(\ref{ch2}) and substitute the resulting equation into the continuity equation
(\ref{ch1}) thereby obtaining the quantum
Smoluchowski-Poisson equations \cite{pre11}
\begin{equation}
\label{ch7}
\xi\frac{\partial\rho}{\partial t}=\nabla\cdot \left (\nabla
P_{\rm int}+\nabla P_{\rm LB}+\frac{\rho}{m}\nabla
Q+\rho\nabla\Phi\right ),
\end{equation}
\begin{eqnarray}
\label{ch8}
\Delta\Phi=4\pi G\rho.
\end{eqnarray}
If we neglect the quantum potential, in the so-called
Thomas-Fermi (TF) approximation, we recover the classical
Smoluchowski-Poisson equations
\begin{equation}
\label{ch9}
\xi\frac{\partial\rho}{\partial t}=\nabla\cdot \left (\nabla
P_{\rm int}+\nabla P_{\rm LB}+\rho\nabla\Phi\right ),
\end{equation}
\begin{eqnarray}
\label{ch10}
\Delta\Phi=4\pi G\rho
\end{eqnarray}
that have been  exhaustively studied in \cite{spzero} and references therein.
These equations were
introduced in 
relation to a rather academic model of self-gravitating Brownian particles
\cite{crs}. The
present study suggests that they may have some applications in the context of 
DM (see Ref. \cite{crrs} for an illustration of the formation
of a DM halo with a core-halo structure in the framework of these equations).
We can also obtain an equation intermediate
between the quantum Euler
equation and the quantum Smoluchowski equation. It has a form \cite{chavtotal}
\begin{equation}
\label{gp12b}
\frac{\partial^2\rho}{\partial t^2}+\xi\frac{\partial\rho}{\partial
t}=\nabla\cdot \left (\nabla
P_{\rm int}+\nabla P_{\rm LB}+\frac{\rho}{m}\nabla
Q+\rho\nabla\Phi\right )
\end{equation}
similar to the telegrapher's equation.

\subsection{Wave equation}
\label{sec_we}

We can now use the Madelung transformation of Sec. \ref{sec_mad} backwards 
in order to derive the generalized wave equation corresponding to the
generalized hydrodynamic equations
(\ref{ch1}) and (\ref{ch2}).

If we first neglect the terms corresponding to violent relaxation ($\xi=P_{\rm
LB}=0$), the hydrodynamic equations
(\ref{ch1}) and (\ref{ch2}) reduce to
\begin{equation}
\label{nch1}
\frac{\partial\rho}{\partial t}+\nabla\cdot (\rho {\bf u})=0,
\end{equation}
\begin{equation}
\label{nch2}
\frac{\partial {\bf u}}{\partial t}+({\bf u}\cdot \nabla){\bf
u}=-\frac{1}{\rho}\nabla
P_{\rm int}-\frac{1}{m}\nabla Q-\nabla\Phi.
\end{equation}
They correspond to a  GP equation of the form \cite{prd1}
\begin{eqnarray}
\label{we1}
i\hbar \frac{\partial\psi}{\partial
t}=-\frac{\hbar^2}{2m}\Delta\psi
+m\left\lbrack
\Phi+h_{\rm int}(|\psi|^2)\right\rbrack\psi,
\end{eqnarray}
where the effective potential $h_{\rm int}(|\psi|^2)$, which can be interpreted as an enthalpy, takes into account the self-interaction of the
bosons. For a general barotropic equation of state
$P_{\rm int}(\rho)$, it is determined by the relation \cite{prd1,chavtotal}
\begin{eqnarray}
\label{we2}
h'_{\rm int}(\rho)=\frac{P'_{\rm int}(\rho)}{\rho}.
\end{eqnarray}
For a general polytropic equation of state of the form of Eq. (\ref{ze4}) we
have
\begin{eqnarray}
\label{we4}
h_{\rm poly}(\rho)=\frac{K\gamma}{\gamma-1}|\psi|^{2(\gamma-1)}.
\end{eqnarray}
For the standard BEC with the polytropic equation of state (\ref{qj21}), we obtain
\begin{eqnarray}
\label{we8}
h_{\rm int}(|\psi|^2)=\frac{4\pi a_s\hbar^2}{m^3}|\psi|^2,
\end{eqnarray}
leading to the standard GP equation (\ref{he7}).

If we now account for violent relaxation, we obtain the generalized GPP
equations\footnote{The Lynden-Bell effective potential
(enthalpy) and the
friction term are in some sense related to the correlation function
$\overline{\delta\Phi\delta\psi}$ that emerges from the coarse-graining of the
Schr\"odinger-Poisson equations (\ref{h1}) and (\ref{h2}), where $\delta\psi$
and $\delta\Phi$
denote the fluctuations about the coarse-grained (smooth) fields.} 
\begin{eqnarray}
\label{we9}
i\hbar \frac{\partial\psi}{\partial
t}=-\frac{\hbar^2}{2m}\Delta\psi
+m\left\lbrack
\Phi+h_{\rm int}(|\psi|^2)+h_{\rm LB}(|\psi|^2)\right\rbrack\psi\nonumber\\
-i\frac{\hbar}{2}\xi\left\lbrack \ln\left
(\frac{\psi}{\psi^*}\right
)-\left\langle \ln\left (\frac{\psi}{\psi^*}\right
)\right\rangle\right\rbrack\psi,\quad
\end{eqnarray}
\begin{eqnarray}
\label{we9b}
\Delta\Phi=4\pi G|\psi|^2.
\end{eqnarray}
Eq. (\ref{we9}) is the wave equation associated with Eqs.
(\ref{ch1}) and (\ref{ch2}). As compared to Eq. (\ref{we1}) there are two new
terms. The Lynden-Bell enthalpy $h_{\rm LB}(|\psi|^2)$ and
the friction
term $\xi$. The Lynden-Bell enthalpy is determined by the Lynden-Bell
equation of
state $P_{\rm LB}(\rho)$ through the relation
\begin{eqnarray}
\label{we10}
h'_{\rm LB}(\rho)=\frac{P'_{\rm LB}(\rho)}{\rho}.
\end{eqnarray}
Apparently, it is not possible to give a simple explicit expression of $h_{\rm
LB}$. If we
introduce the total enthalpy $h=h_{\rm LB}+h_{\rm
int}$, the generalized GP equation (\ref{we9}) can be written as
\begin{eqnarray}
\label{we17b}
i\hbar \frac{\partial\psi}{\partial
t}=-\frac{\hbar^2}{2m}\Delta\psi+
m\left\lbrack
\Phi+h(|\psi|^2)\right\rbrack\psi\nonumber\\
-i\frac{\hbar}{2}\xi\left\lbrack \ln\left
(\frac{\psi}{\psi^*}\right
)-\left\langle \ln\left (\frac{\psi}{\psi^*}\right
)\right\rangle\right\rbrack\psi.
\end{eqnarray}
This is the wave equation associated with Eqs. (\ref{ch1}) and (\ref{ze1}). It
is of the form of the generalized GPP equations introduced and studied in
\cite{chavtotal} (see also Appendix \ref{sec_gwe}). If we use the approximate
expression
of the Lynden-Bell equation of state from Eq. (\ref{qj18}), we get
\begin{eqnarray}
\label{we11}
h_{\rm LB}(|\psi|^2)=h_{\rm LB}^{(0)}(|\psi|^2)+h_{\rm LB}^{\rm th}(|\psi|^2)
\end{eqnarray}
with
\begin{eqnarray}
\label{we12}
h_{\rm LB}^{(0)}(|\psi|^2)=\frac{1}{2}\left (\frac{3}{4\pi\eta_0}\right
)^{2/3}|\psi|^{4/3}
\end{eqnarray}
and
\begin{eqnarray}
\label{we13}
h_{\rm LB}^{\rm th}(|\psi|^2)=T_{\rm eff}\ln(|\psi|^2).
\end{eqnarray}
In that case, the generalized GP equation (\ref{we9}) can be written as
\begin{eqnarray}
\label{we14}
i\hbar \frac{\partial\psi}{\partial
t}=-\frac{\hbar^2}{2m}\Delta\psi+mT_{\rm eff}\ln(|\psi|^2)\psi\nonumber\\
+m\left\lbrack
\Phi+h_{\rm int}(|\psi|^2)+h_{\rm
LB}^{(0)}(|\psi|^2)\right\rbrack\psi\nonumber\\
-i\frac{\hbar}{2}\xi\left\lbrack \ln\left
(\frac{\psi}{\psi^*}\right
)-\left\langle \ln\left (\frac{\psi}{\psi^*}\right
)\right\rangle\right\rbrack\psi,
\end{eqnarray}
where the effective thermal term has been made explicit. This is the wave
equation
associated with Eq. (\ref{ze2}).  Equation (\ref{we14}) may be
viewed as a coarse-grained GP
equation parameterizing the processes of violent relaxation and gravitational
cooling. For the standard BEC, we get
\begin{eqnarray}
\label{we16}
i\hbar \frac{\partial\psi}{\partial
t}=-\frac{\hbar^2}{2m}\Delta\psi+mT_{\rm eff}\ln(|\psi|^2)\psi\nonumber\\
+\frac{4\pi a_s\hbar^2}{m^2}|\psi|^{2}\psi+\frac{1}{2}\left
(\frac{3}{4\pi\eta_0}\right
)^{2/3}m|\psi|^{4/3} \psi
+m\Phi\psi\nonumber\\
-i\frac{\hbar}{2}\xi\left\lbrack \ln\left
(\frac{\psi}{\psi^*}\right
)-\left\langle \ln\left (\frac{\psi}{\psi^*}\right
)\right\rangle\right\rbrack\psi,
\end{eqnarray}
where all the terms have been made explicit. The structural wave equation
associated with Eqs. (\ref{ch1}) and (\ref{ze3}) is
\begin{eqnarray}
\label{we17}
i\hbar \frac{\partial\psi}{\partial
t}=-\frac{\hbar^2}{2m}\Delta\psi+ mT_{\rm eff}\ln(|\psi|^2)\psi \nonumber\\
+\frac{K\gamma m}{\gamma-1}|\psi|^{2(\gamma-1)}\psi
+m \Phi\psi\nonumber\\
-i\frac{\hbar}{2}\xi\left\lbrack \ln\left
(\frac{\psi}{\psi^*}\right
)-\left\langle \ln\left (\frac{\psi}{\psi^*}\right
)\right\rangle\right\rbrack\psi.
\end{eqnarray}

{\it Remark:} It is interesting to note that, 
at a formal level, the generalized GP equation (\ref{we17b})
allows us to make a connection between the Schr\"odinger (or GP) equation of
quantum
mechanics ($\xi=0$) and the (generalized) Smoluchowski equation of Brownian
theory
($\xi\rightarrow +\infty$) \cite{chavtotal}. We also note that
the generalized GP equation (\ref{we17b}) including a temperature term and a
friction term can be obtained from a ``unified'' formalism based on a
generalization of the theory of scale relativity to the case of
dissipative systems \cite{epjpnottale,ggppdark}. The temperature and the
friction appear as two manifestations of the same concept.

\section{Equilibrium states}
\label{sec_es}

The generalized GPP equations  (\ref{we9}) and
(\ref{we9b}), or equivalently the hydrodynamic equations (\ref{ch1})-(\ref{ch3}), satisfy an
$H$-theorem for a generalized free energy $F=\Theta_c+\Theta_Q+U_{\rm
int}+U_{\rm LB}+W$ (see \cite{chavtotal} and Appendix \ref{sec_gwe} for the
definition of the different functionals) and relax towards a stable equilibrium
state which minimizes $F$ at fixed mass $M$.\footnote{The
$H$-theorem and the relaxation towards an equilibrium
state are due to the friction term $\xi>0$ which provides a source of
dissipation and implies the irreversibility of the generalized  GPP equations
(\ref{we9}) and (\ref{we9b}). By contrast, the usual  GPP equations
(\ref{he7}) and (\ref{he8}) conserve the energy and are reversible. Their relaxation towards a
quasistationary state is due to gravitational cooling and violent relaxation and
can
be understood only at
a coarse-grained level (see Secs. \ref{sec_gc} and
\ref{sec_inter}). It is in this sense that
the generalized GPP equations
(\ref{we9}) and (\ref{we9b}) provide a parametrization of the GPP equations
(\ref{he7}) and (\ref{he8}) taking into account the processes of
gravitational cooling and violent relaxation.} We can determine this
equilibrium state in different manners.

The equation of quantum
hydrostatic equilibrium, which corresponds to the
steady state of the quantum Euler equation (\ref{ch2}), writes
\begin{eqnarray}
\label{es1}
\nabla P_{\rm LB}+\nabla P_{\rm int}+\frac{\rho}{m}\nabla
Q+\rho\nabla\Phi={\bf 0}.
\end{eqnarray}
This equation describes the balance between the Lynden-Bell pressure, the
pressure
due to the self-interaction of the bosons, the
quantum potential, and the gravitational force. If we introduce 
the total pressure $P=P_{\rm LB}+P_{\rm int}$, we can rewrite
Eq. (\ref{es1}) as \cite{chavtotal}
\begin{eqnarray}
\label{es2}
\frac{\rho}{m}\nabla
Q+\nabla P+\rho\nabla\Phi={\bf 0}.
\end{eqnarray}
This is the equilibrium state of Eq. (\ref{ze1}).
Combined with the Poisson equation (\ref{ch3}) we obtain the fundamental
equation of quantum hydrostatic equilibrium determining the structure of DM
halos \cite{chavtotal}
\begin{equation}
\label{es3}
\frac{\hbar^2}{2m^2}\Delta
\left (\frac{\Delta\sqrt{\rho}}{\sqrt{\rho}}\right )-\nabla\cdot \left
(\frac{\nabla P}{\rho}\right )=4\pi G\rho.
\end{equation}

The
foregoing equations can also be obtained from the generalized GPP
equations (\ref{we9}) and
(\ref{we9b}). Writing $\psi({\bf r},t)=\phi({\bf r})e^{-iEt/\hbar}$ where
$\phi({\bf r})=\sqrt{\rho({\bf r})}$ and $E$ are real, the stationary solutions
of the generalized GPP equations are determined by the eigenvalue problem
\begin{eqnarray}
\label{tw4b}
-\frac{\hbar^2}{2m}\Delta\phi+m\left\lbrack
\Phi+h_{\rm int}(\rho)+h_{\rm LB}(\rho)\right\rbrack\phi=E\phi,
\end{eqnarray}
\begin{eqnarray}
\label{tw4c}
\Delta\Phi=4\pi G \phi^2,
\end{eqnarray}
where $E$ is the eigenenergy. Dividing Eq. (\ref{tw4b}) by $\phi$, we get
\begin{eqnarray}
\label{tw4d}
Q+m\Phi+mh_{\rm int}(\rho)+mh_{\rm LB}(\rho)=E.
\end{eqnarray}
Taking the gradient of this equation and 
using Eqs. (\ref{we2}) and (\ref{we10}), we recover the condition of quantum
hydrostatic equilibrium from Eq. (\ref{es1}).

Finally, an equilibrium state of the GPP equations  can  be obtained by
extremizing the free energy $F$ at
fixed mass $M$. 
Writing the variational principle as
\begin{equation}
\label{w35qr}
\delta F-\frac{\mu}{m}\delta M=0,
\end{equation}
where $\mu/m$ is a Lagrange multiplier (global chemical potential) taking into
account the mass constraint, we get 
\begin{equation}
\label{hye}
Q+m\Phi+mh_{\rm int}(\rho)+mh_{\rm LB}(\rho)=\mu,
\end{equation}
which is equivalent to Eq. (\ref{tw4d}) with $E=\mu$. This
establishes that the
eigenenergy is equal to the global chemical potential. Furthermore, one can
show that an equilibrium state of the generalized GPP equations is stable if,
and only if, it is
a minimum of free energy at fixed mass ($\delta^2F>0$ for all
perturbations that conserve mass). These results are a consequence of the
$H$-theorem.

\subsection{Core-halo structure}

We now show that the equilibrium states of the GPP equations (\ref{we9})
and (\ref{we9b}) have a core-halo structure. If we use the approximate
expression of the Lynden-Bell equation of state from
Eq.
(\ref{qj18}), we can rewrite Eq.
(\ref{es1}) as
\begin{eqnarray}
\label{es4}
\frac{\rho}{m}\nabla
Q+\nabla P_{\rm LB}^{(0)}+\nabla P_{\rm int}+\nabla P_{\rm LB}^{\rm
th}+\rho\nabla\Phi={\bf 0}.
\end{eqnarray}
It corresponds to a structural  equation of the form
\begin{eqnarray}
\label{es5}
\frac{\rho}{m}\nabla
Q+\nabla P_{\rm poly}+\nabla P_{\rm th}+\rho\nabla\Phi={\bf 0},
\end{eqnarray}
which is the equilibrium state of Eq. (\ref{ze3}). It involves a
polytropic 
equation of state and an isothermal equation of state. The
total pressure is  \cite{modeldm,ggppdark}
\begin{eqnarray}
\label{es5w}
P=K\rho^{\gamma}+\rho T_{\rm eff}.
\end{eqnarray} 
Combining Eq.
(\ref{es5})
with the Poisson equation (\ref{ch3}) we obtain the fundamental
differential equation \cite{modeldm,ggppdark}
 \begin{equation}
\label{es6}
\frac{\hbar^2}{
2m^2}\Delta
\left (\frac{\Delta\sqrt{\rho}}{\sqrt{\rho}}\right
)-\frac{K\gamma}{\gamma-1}\Delta\rho^{\gamma-1}-T_{\rm
eff}\Delta\ln\rho
=4\pi
G\rho.
\end{equation}
For noninteracting BECs
it reduces to
 \begin{equation}
\label{es6a}
\frac{\hbar^2}{
2m^2}\Delta
\left (\frac{\Delta\sqrt{\rho}}{\sqrt{\rho}}\right
)-T_{\rm
eff}\Delta\ln\rho
=4\pi
G\rho.
\end{equation}
In the TF approximation, we get
 \begin{equation}
\label{es6b}
-\frac{K\gamma}{\gamma-1}\Delta\rho^{\gamma-1}-T_{\rm
eff}\Delta\ln\rho
=4\pi
G\rho.
\end{equation}
If we define
\begin{equation}
\label{es7}
\rho=\rho_0
e^{-\psi},\qquad  \xi=\left (\frac{4\pi G\rho_0}{T_{\rm eff}}\right )^{1/2}r,
\end{equation}
\begin{equation}
\label{es8}
 \chi=\frac{K\gamma \rho_0^{\gamma-1}}{T_{\rm eff}},\qquad \epsilon=\frac{2\pi
G\rho_0\hbar^2}{m^2 T_{\rm eff}^2},
\end{equation}
where $\rho_0$ is the central density, we find that Eq. (\ref{es6}) takes the
form of a
generalized Emden equation
\cite{modeldm,ggppdark}
\begin{equation}
\label{es9}
\epsilon \Delta\left (e^{\psi/2}\Delta
e^{-\psi/2}\right )+\Delta\psi+\chi\nabla\cdot\left \lbrack
e^{-(\gamma-1)\psi}\nabla\psi\right
\rbrack=e^{-\psi}.
\end{equation}
For noninteracting BECs
it reduces to
\begin{equation}
\label{es9ni}
\epsilon \Delta\left (e^{\psi/2}\Delta
e^{-\psi/2}\right )+\Delta\psi=e^{-\psi}.
\end{equation}
In the TF approximation, we get
\begin{equation}
\label{es9tf}
\Delta\psi+\chi\nabla\cdot\left \lbrack
e^{-(\gamma-1)\psi}\nabla\psi\right
\rbrack=e^{-\psi}.
\end{equation}
Alternatively, if we define
\begin{equation}
\label{es10}
\rho=\rho_0\theta^n,\qquad \xi=\left\lbrack \frac{4\pi
G}{K(n+1)\rho_{0}^{1/n-1}}\right\rbrack^{1/2}r,
\end{equation}
\begin{equation}
\label{es12}
\sigma\equiv \frac{\epsilon}{\chi^2}=\frac{2\pi
G\hbar^2}{K^2\gamma^2m^2\rho_0^{2\gamma-3}},
\end{equation}
we find that Eq. (\ref{es6}) takes
the form of a
generalized Lane-Emden equation \cite{modeldm,ggppdark}
\begin{equation}
\label{es11}
-\frac{\sigma}{n^2}\Delta\left
(\frac{\Delta\theta^{n/2}}{\theta^{n/2}}\right )+\frac{1}{\chi}
\Delta\ln\theta+\Delta\theta=-\theta^n.
\end{equation}
In the TF approximation, we get
\begin{equation}
\label{es11tf}
\frac{1}{\chi}
\Delta\ln\theta+\Delta\theta=-\theta^n.
\end{equation}
The above equations describe the
balance between the quantum
potential taking into account the Heisenberg uncertainty principle, the pressure due to the
self-interaction of the bosons, the pressure due to the Lynden-Bell exclusion
principle,
the pressure due to effective thermal effects, and
the self-gravity.  The solutions have a core-halo structure with
a quantum core and an isothermal halo (see Ref. \cite{modeldm}
for explicit calculations of DM halos in the case of a standard self-gravitating
BEC with repulsive self-interaction corresponding to a polytropic index
$\gamma=2$ in the TF approximation).
The quantum core has
a bosonic nature due to the Heisenberg uncertainty principle (soliton) and to
the
self-interaction of the particles leading to a polytropic core of index $n=1$
(boson ball). It has also a fermionic nature in the sense of Lynden-Bell leading to a
polytropic core of index $n=3/2$ (fermion ball) although this
contribution is usually negligible for bosons (see footnote 20). This quantum
core is surrounded by an isothermal halo with an effective temperature $T_{\rm
eff}$. This core-halo structure is
consistent with the structure of large DM halos that are obtained in direct
numerical simulations of
BECDM \cite{ch2,ch3,
schwabe,mocz,moczSV,veltmaat,moczprl,moczmnras,veltmaat2}.

{\it Remark:} Recalling the expressions
of $P_{\rm LB}^{(0)}$ and $P_{\rm int}$ [see Eqs. (\ref{qj17}) and
(\ref{qj21})], we find that the total pressure is
explicitly given by
\begin{eqnarray}
\label{es5wa}
P=\frac{1}{5}\left (\frac{3}{4\pi\eta_0}\right )^{2/3}\rho^{5/3}+\frac{2\pi
a_s\hbar^2}{m^3}\rho^2+\rho T_{\rm eff}.
\end{eqnarray}
On the other hand, in Appendix E of \cite{mcmh} we have shown
that the soliton resulting from the equilibrium between the gravitational
attraction and the quantum repulsion (Heisenberg uncertainty principle) is
similar to a polytrope of index $\gamma=3/2$ (i.e. $n=2$) with an
equation of state
\begin{eqnarray}
\label{es5w1}
P=\left (\frac{2\pi G\hbar^2}{9m^2}\right )^{1/2}\rho^{3/2}.
\end{eqnarray}
Therefore, in order to compute the structure of a soliton surrounded by an
isothermal halo,  instead of solving Eq. (\ref{es6a}) with the quantum term, we
can solve Eq. (\ref{es6b}) without the quantum term but with the pressure from
Eq. (\ref{es5w1}). This leads to a differential equation of the form
 \begin{equation}
\label{es5w3}
-\left (\frac{2\pi G\hbar^2}{m^2}\right
)^{1/2}\Delta\sqrt{\rho}-T_{\rm
eff}\Delta\ln\rho
=4\pi
G\rho,
\end{equation}
corresponding to a total pressure
\begin{eqnarray}
\label{es5w2}
P=\left (\frac{2\pi G\hbar^2}{9m^2}\right )^{1/2}\rho^{3/2}+\rho
T_{\rm eff}.
\end{eqnarray}
We can also take into account the effective fermionic core and
the self-interaction of the bosons by adding the contribution of $P_{\rm
LB}^{(0)}$ and $P_{\rm int}$ in the total pressure. The study of Eq.
(\ref{es5w3}), which is of the general form of Eq. (\ref{es6b}) with
$\gamma=3/2$, is similar to the one performed in \cite{modeldm} for $\gamma=2$.
It will be reported in a future contribution \cite{prep}.

\subsection{Quantum core}

In the core, we can neglect effective thermal effects and take $P_{\rm LB}^{\rm
th}=0$.
In that case,  Eq. (\ref{es4}) reduces to
\begin{eqnarray}
\label{es13}
\frac{\rho}{m}\nabla Q+\nabla P_{\rm LB}^{(0)}+\nabla P_{\rm
int}+\rho\nabla\Phi={\bf 0}.
\end{eqnarray}
It corresponds to a structural  equation of the form
\begin{eqnarray}
\label{es14}
\frac{\rho}{m}\nabla Q+\nabla P_{\rm poly}+\rho\nabla\Phi={\bf
0}.
\end{eqnarray}
Combined with the Poisson equation (\ref{ch3}), we get
 \begin{eqnarray}
\label{es15}
\frac{\hbar^2}{
2m^2}\Delta
\left (\frac{\Delta\sqrt{\rho}}{\sqrt{\rho}}\right
)-\frac{K\gamma}{\gamma-1}\Delta\rho^{\gamma-1}=4\pi
G\rho.
\end{eqnarray}
For noninteracting BECs it reduces to
 \begin{eqnarray}
\label{es15ni}
\frac{\hbar^2}{
2m^2}\Delta
\left (\frac{\Delta\sqrt{\rho}}{\sqrt{\rho}}\right
)=4\pi
G\rho.
\end{eqnarray}
In the TF approximation, we get
 \begin{eqnarray}
\label{es17}
-\frac{K\gamma}{\gamma-1}\Delta\rho^
{\gamma-1}=4\pi
G\rho.
\end{eqnarray}
With the change of
variables from Eq. (\ref{es10}), Eq. (\ref{es15}) takes the form of a
quantum Lane-Emden equation
\begin{equation}
\label{es16}
-\frac{\sigma}{n^2}\Delta\left
(\frac{\Delta\theta^{n/2}}{\theta^{n/2}}\right
)+\Delta\theta=-\theta^n.
\end{equation}
In the TF limit, we recover the ordinary Lane-Emden equation \cite{chandra}
\begin{equation}
\label{es18}
\Delta\theta=-\theta^n.
\end{equation}
The equilibrium of the core
is due to the balance between the quantum potential, the
self-interaction of the bosons,  the pressure due to the Lynden-Bell exclusion principle,
and the gravitational attraction.
Equation (\ref{es15}) with $\gamma=2$ corresponding to a standard BEC has been
solved analytically
(using a Gaussian ansatz) in
\cite{prd1} and numerically in \cite{prd2} for an arbitrary
(repulsive, attractive or vanishing) self-interaction.
It describes a compact quantum core.  Because of
quantum effects (or because of the Lynden-Bell exclusion principle), the central density is finite instead of diverging as in the
CDM model. Therefore, quantum mechanics is able to solve the core-cusp problem.

\subsection{Isothermal halo}

In the halo, we can neglect quantum effects and take
$Q=P_{\rm LB}^{(0)}=0$ and $P_{\rm int}=0$. In that case, Eq. (\ref{es4})
reduces to
\begin{eqnarray}
\label{es19}
\nabla P_{\rm LB}^{\rm th}+\rho\nabla\Phi={\bf 0}.
\end{eqnarray}
Combined with the Poisson equation (\ref{ch3}), we get
 \begin{eqnarray}
\label{es20}
-T_{\rm
eff}\Delta\ln\rho
=4\pi
G\rho,
\end{eqnarray}
which is equivalent to the ordinary Emden equation \cite{chandra}
\begin{equation}
\label{es21}
\Delta\psi=e^{-\psi}.
\end{equation}
The equilibrium of the halo
is due to the balance between the effective thermal pressure and the gravitational
attraction. The  Boltzmann-Poisson equation (\ref{es20}), or the Emden equation (\ref{es21}), has no
simple analytical solution and must be solved
numerically (see, e.g., \cite{modeldm}). However, its asymptotic behavior
is known
analytically \cite{chandra}. The
density of a self-gravitating isothermal
halo decreases as $\rho(r)\sim  T_{\rm eff}/(2\pi G  r^{2})$ for
$r\rightarrow
+\infty$, corresponding to an accumulated mass $M(r)\sim 2 T_{\rm eff} r/G$
increasing
linearly with $r$. This leads to flat rotation curves
\begin{eqnarray}
\label{es22}
v^2(r)=\frac{GM(r)}{r}\rightarrow v_{\infty}^2=2 T_{\rm eff},
\end{eqnarray}
in agreement with the observations \cite{bt}.

\subsection{Conclusion}

In conclusion, the physical meaning of the generalized GPP equations (\ref{we9})
and (\ref{we9b}) is clear. The damping term forces the
system to relax towards a stable equilibrium state with a  core-halo
structure. The friction term and the thermal term provide a parametrization of
gravitational cooling and violent relaxation. The quantum core is able to solve
the
core-cusp
problem and the isothermal halo accounts for the flat rotation
curves of
the galaxies. This core-halo structure is in agreement with the phenomenology of
BECDM halos.

\section{Canonical and microcanonical ensembles}
\label{sec_cmm}

In the previous sections, we have developed a canonical ensemble 
description in which the effective temperature $T_{\rm eff}$ is fixed. The
corresponding thermodynamic potential is the free energy (see Appendix
\ref{sec_gwe})
\begin{eqnarray}
\label{cmm1}
F=\Theta_c+\Theta_Q+U_{\rm int}+U_{\rm LB}^{(0)}+W+U_{\rm LB}^{\rm th}
\end{eqnarray}
where
\begin{eqnarray}
\label{cmm1b}
\Theta_c=\int\rho  \frac{{\bf u}^2}{2}\, d{\bf r},
\end{eqnarray}
\begin{equation}
\label{cmm1c}
\Theta_Q=\frac{1}{m}\int \rho Q\, d{\bf r},
\end{equation}
\begin{equation}
\label{cmm1d}
U_{\rm int}=\frac{2\pi a_s\hbar^2}{m^3}\int \rho^2\, d{\bf r},
\end{equation}
\begin{equation}
\label{cmm1e}
U_{\rm LB}^{(0)}=\frac{3}{10}\left (\frac{3}{4\pi\eta_0}\right )^{2/3}\int \rho^{5/3}\, d{\bf r},
\end{equation}
\begin{eqnarray}
\label{cmm1f}
W=\frac{1}{2}\int\rho\Phi\, d{\bf r},
\end{eqnarray}
\begin{equation}
\label{cmm1g}
U_{\rm LB}^{\rm th}=T_{\rm eff}\int \rho(\ln\rho-1)\, d{\bf r},
\end{equation}
are the classical kinetic energy, the quantum kinetic energy, the internal
energy of self-interaction, the internal energy associated with the Lynden-Bell
pressure of zero-point energy, the gravitational energy, and the internal energy
associated with the Lynden-Bell thermal pressure.\footnote{This is also the
Lynden-Bell entropy multiplied by $-T_{\rm eff}$ (see Appendix
\ref{sec_wce}). As a result, we have $F=E_{\rm tot}-T_{\rm eff}S$.} The
generalized GPP equations (\ref{we9b}) and
(\ref{we14}) satisfy an $H$-theorem for the free energy (\ref{cmm1}) and relax
towards an
equilibrium state which minimizes $F$ at fixed mass $M$ (see Appendix
\ref{sec_wce}). This equilibrium state is canonically stable. However, for
self-gravitating systems, the statistical ensembles are inequivalent
\cite{paddy,katzrevue,ijmpb,campa}. One can show that an equilibrium state that
is
canonically stable is necessarily microcanonically stable but the reciprocal is
wrong \cite{cc}. There exist equilibrium states that are microcanonically
stable while being canonically unstable. Therefore, we may miss important
equilibrium states by using a canonical 
description instead of a microcanonical one. For example, it was shown in Ref.
\cite{mcmh} that the core-halo configurations of BECDM halos with a quantum
core and an isothermal halo, similar to those observed in numerical simulations,
are canonically unstable while being generically microcanonically
stable.\footnote{See \cite{modeldm,mcmh} for a detailed discussion.}
These core-halo configurations have a negative specific heat. We know that
systems with negative specific heats are unstable in the canonical ensemble
while they may be stable in the microcanonical ensemble. Therefore, the
generalized GPP equations  (\ref{we9b}) and (\ref{we14}) with a constant
temperature $T_{\rm eff}$  do {\it not} relax towards the important core-halo
configurations observed in direct numerical simulations since they are unstable
in the canonical ensemble. This is a drawback of this model.

In addition to these considerations of thermodynamical stability, there is no
reason why the effective temperature $T_{\rm eff}$ of the halo should be
constant. On the contrary, it should adapt itself so as to  conserve the total
energy. Indeed, according to the discussion of Sec. \ref{sec_gc}, the mass
$M-M_c$ and the  energy $E_{\rm tot}-E_c$ that are not contained in the core
should be
stored in the halo.

It is therefore important to develop a microcanonical model of BECDM halos
where the total energy is fixed. This can be simply achieved by letting the
effective temperature $T_{\rm eff}(t)$ depend on time so as to conserve the
total energy
\begin{eqnarray}
\label{jme2}
E_{\rm tot}=\Theta_c+\Theta_Q+U_{\rm int}+U_{\rm LB}^{(0)}+W+\Theta_{\rm
LB}^{\rm th},
\end{eqnarray}
where
\begin{equation}
\label{me4}
\Theta_{\rm LB}^{\rm th}=\frac{3}{2}M T_{\rm eff}
\end{equation}
is the Lynden-Bell thermal energy (see Appendix \ref{sec_wmce}). In
line with the discussion of Sec. \ref{sec_es}, the total energy $E_{\rm
tot}=E_c+E_h$ can
be decomposed into the energy of the core $E_c\simeq
(\Theta_c)_c+\Theta_Q+U_{\rm int}+U_{\rm LB}^{(0)}+W_c$ dominated by quantum
effects and the energy of the halo $E_h\simeq (\Theta_c)_h+\Theta_{\rm LB}^{\rm
th}+W_h$ dominated by thermal effects. One can show that the  generalized GPP
equations  (\ref{we9b}) and
(\ref{we14}) with the time-dependent temperature $T_{\rm eff}(t)$ satisfy an $H$-theorem for the Lynden-Bell entropy
\begin{eqnarray}
\label{me2n}
S=-\int\rho (\ln\rho-1)\, d{\bf r}+\frac{3}{2}M\ln T_{\rm eff},
\end{eqnarray}
and  relax towards an equilibrium state which maximizes $S$ at fixed mass $M$
and energy $E_{\rm tot}$ (see Appendix \ref{sec_wmce}). This equilibrium state
is
microcanonically stable. As a result, the  generalized GPP equations 
(\ref{we9b}) and (\ref{we14}) with a time-dependent temperature $T_{\rm eff}(t)$
determined by Eqs. (\ref{jme2}) and (\ref{me4}) with $E_{\rm tot}$ fixed, relax
towards a core-halo structure, similar to the one observed in direct
numerical simulations of BECDM, since it is stable in the microcanonical
ensemble. Furthermore, we showed in Ref. \cite{mcmh} that the core mass-halo
mass relation $M_c(M_h)$ obtained by maximizing the entropy at fixed mass and
energy reproduces the relation observed in direct numerical simulations. This
confirms that our effective thermodynamical description is relevant to describe
BECDM halos. At equilibrium, the virial theorem writes (see
Appendix
\ref{sec_wmce})
\begin{eqnarray}
\label{vircg}
2(\Theta+\Theta_{\rm LB})+W=0,
\end{eqnarray}
where $E_{\rm tot}=\Theta+\Theta_{\rm LB}+W$ is constant (we
consider 
noninteracting systems $U_{\rm int}=0$ for simplicity). This relation is in
agreement with Eq. (\ref{vireq}) with $\langle W\rangle\simeq
W$ and 
$\langle\Theta\rangle=\Theta+\Theta_{\rm LB}$, where $\Theta_{\rm
LB}=\Theta_{\rm LB}^{\rm th}+\Theta_{\rm LB}^{(0)}$ is the Lynden-Bell
energy taking into account fine-grained correlations.

{\it Remark:} For simplicity, we have assumed that the 
temperature $T_{\rm eff}(t)$ is uniform. We can get a more general model where
the temperature  $T({\bf r},t)$ is spatially inhomogeneous (see Ref.
\cite{csr} and Appendix \ref{sec_int}). Its evolution equation can be obtained
from the
second moment of the Wigner-Kramers equation. This leads to a system of three
hydrodynamic equations of the form of Eqs. (\ref{int1})-(\ref{int3}) or their
generalization given
in \cite{csr}.

\section{Fermionic DM}
\label{sec_fermidm}

Although the previous formalism has been developed for
self-gravitating bosons in the form of BECs, it can also be applied to
self-gravitating fermions with only minor modifications. In the case of
fermions, one can ignore the quantum potential in a first
approximation.\footnote{This
is because, in DM models, fermions have a much larger mass than bosons
(see, e.g., \cite{mcmh}).} A system of collisionless self-gravitating fermions
is then
described by the classical Vlasov-Poisson equations (\ref{vlasov1}) and
(\ref{vlasov2}). These equations exhibit a process of violent relaxation leading
to the Lynden-Bell DF (\ref{mep5}). As we have seen, the Lynden-Bell DF is
similar to the Fermi-Dirac DF. Actually, in the case of fermions, the maximum
phase space density $\eta_0$ fixed by the Lynden-Bell exclusion principle is of
the same order as the quantum bound $m^4/h^3$ fixed by the Pauli exclusion
principle (see footnote 20). Therefore, in that case, the
Lynden-Bell DF truly
coincides with the Fermi-Dirac DF. As a result, the statistical theory of
violent relaxation is able to justify the establishment of the Fermi-Dirac DF in
a fermionic DM halo on a very short timescale, of the order of a few dynamical
times, without the need of collisions
which require much more time to develop (see footnote 8) \cite{clm2}. The
resulting
fermionic DM halos have a core-halo structure made of a quantum core (fermion
ball) surrounded by an isothermal halo. Such configurations have been studied by
many authors (see Sec. V.A of \cite{modeldm} for an exhaustive list of
references). However, they have an infinite mass. To solve this problem, a
kinetic
theory of
collisionless violent relaxation on the
coarse-grained scale has been developed in Refs.
\cite{kp,sl,csr,chavmnras,kingen,dubrovnik}. It leads to a fermionic
Vlasov-Kramers or fermionic Vlasov-Landau equation. This kinetic equation
relaxes towards
the truncated Lynden-Bell DF (\ref{qw7}) and (\ref{qw8}) \cite{chavmnras}. This
DF has a finite mass (see also footnote 21) and is stabilized against
gravitational collapse by the
Pauli or Lynden-Bell exclusion principle. This leads to the fermionic King
model of DM halos studied
in \cite{clm2,rarnew}.

In a more general approach, we can take the quantum potential
into account. In that case, the starting point of the analysis is the
Hartree-Fock equations (\ref{ah3}) and (\ref{ah4}) which can be viewed as 
multistate
Schr\"odinger-Poisson equations. These equations take the Heisenberg
uncertainty principle and the Pauli exclusion principle into account. They are
equivalent to
the quantum Euler-Poisson equations (\ref{amad23})-(\ref{amad25}) which are
similar to Eqs.
(\ref{mad4})-(\ref{mad8}) except that they include
an additional pressure term [see Eq. (\ref{amad20})] arising from the  Pauli
exclusion
principle (this term comes from multistate fluctuations). These equations
are also equivalent to the Wigner equation (\ref{qw2}). We can then introduce a
coarse-graining and
proceed as in Secs. \ref{sec_dfaw}-\ref{sec_cmm}. The only difference is that
we have to take into account the Pauli exclusion
principle that
prevents two fermions (with equal spin) to occupy the same quantum state. In
the statistical approach, the Pauli exclusion principle creates a
Fermi-Dirac pressure $P_{\rm FD}$.\footnote{As discussed above, in the case
of fermions, the Fermi-Dirac pressure  $P_{\rm FD}$ is equivalent to the
Lynden-Bell pressure $P_{\rm LB}$. However, for the sake of clarity (and
generality) we shall treat these two terms separately.} This pressure can be
decomposed into a thermal pressure $P_{\rm th}$ and a pressure of zero
point energy $P_{\rm FD}^{(0)}$ corresponding to a completely degenerate Fermi
gas at $T=0$. 
The Fermi pressure plays a role similar to the pressure  $P_{\rm int}$  arising
from the
self-interaction of the bosons. We can also take into account the
Slater correction which is equivalent to a pressure $P_{\rm
Slater}$ (see Appendix \ref{sec_multistate}).\footnote{This correction is
usually negligible in the case of DM halos.} Since the thermal cloud is
taken into account in
$P_{\rm th}$, we just have to add the Fermi pressure $P_{\rm FD}^{(0)}$ and the
Slater pressure $P_{\rm Slater}$ in the formalism of Secs.
\ref{sec_dfaw}-\ref{sec_cmm}. Below, we briefly write the basic equations that
result from this formalism.

The equation of state of the Fermi gas at $T=0$ is \cite{chandra}
\begin{equation}
\label{fermidm1}
P_{\rm FD}^{(0)}(\rho)=\frac{1}{20}\left (\frac{3}{\pi}\right
)^{2/3}\frac{h^2}{m^{8/3}}\rho^{5/3}.
\end{equation}
Using the results of Appendix \ref{sec_gwe}, we obtain the enthalpy
\begin{equation}
\label{fermidm3}
h_{\rm FD}^{(0)}(\rho)=\frac{1}{8}\left (\frac{3}{\pi}\right
)^{2/3}\frac{h^2}{m^{8/3}}\rho^{2/3}
\end{equation}
and the potential
\begin{equation}
\label{fermidm4}
V_{\rm FD}^{(0)}(\rho)=\frac{3}{40}\left (\frac{3}{\pi}\right
)^{2/3}\frac{h^2}{m^{8/3}}\rho^{5/3}.
\end{equation}
The Slater equation of state is (see Appendix \ref{sec_multistate})
\begin{equation}
\label{fermidm1b}
P_{\rm Slater}(\rho)=\frac{C_S}{4m}\rho^{4/3}.
\end{equation}
Using the results of Appendix \ref{sec_gwe}, we obtain the enthalpy
\begin{equation}
\label{fermidm3b}
h_{\rm Slater}(\rho)=\frac{C_S}{m}\rho^{1/3}
\end{equation}
and the potential
\begin{equation}
\label{fermidm4b}
V_{\rm Slater}(\rho)=\frac{3C_S}{4m}\rho^{4/3}.
\end{equation}
Collecting  these results, we obtain a generalized wave equation of the form
\begin{eqnarray}
\label{fermidm5}
i\hbar \frac{\partial\psi}{\partial
t}=-\frac{\hbar^2}{2m}\Delta\psi+mT_{\rm eff}\ln(|\psi|^2)\psi\nonumber\\
+\frac{1}{8}\left (\frac{3}{\pi}\right
)^{2/3}\frac{h^2}{m^{5/3}}|\psi|^{4/3}\psi+C_S |\psi|^{2/3}\psi\nonumber\\
+\frac{1}{2}\left
(\frac{3}{4\pi\eta_0}\right
)^{2/3}m|\psi|^{4/3} \psi
+m\Phi\psi\nonumber\\
-i\frac{\hbar}{2}\xi\left\lbrack \ln\left
(\frac{\psi}{\psi^*}\right
)-\left\langle \ln\left (\frac{\psi}{\psi^*}\right
)\right\rangle\right\rbrack\psi.
\end{eqnarray}
The $|\psi|^{2/3}$ nonlinearity corresponds to the Slater exchange energy term
and the $|\psi|^{4/3}$  nonlinearity accounts for the Pauli exclusion
principle at $T=0$ like in the TF theory of the atoms \cite{thomas,fermi}.
The von Weizs\"acker correction to the TF theory is taken into account in
the kinetic term $-(\hbar^2/2m)\Delta\psi$.
The wave equation (\ref{fermidm5}) with $\xi=T_{\rm
eff}=1/\eta_0=0$ is expected to display a process of violent relaxation leading
to a fermion ball at $T=0$ surrounded by a halo of scalar radiation. This is 
similar to the process of gravitational cooling experienced by the  GPP
equations for bosons (the fermion ball is the counterpart of the bosonic
condensate). The coarse-grained equation (\ref{fermidm5}) with the friction 
term and the Lynden-Bell terms (exclusion principle and effective temperature)
retained parameterizes this process of violent
relaxation. Note that this equation also takes into account the Heisenberg
uncertainty principle through the kinetic term. This provides an additional
source of small-scale
regularization, in addition to the Pauli (or Lynden-Bell) exclusion principle,
preventing
gravitational collapse.

{\it Remark:} For the sake of completeness, we briefly consider the case of a
relativistic Fermi gas. In the ultrarelativistic limit, the equation of state of
the Fermi
gas at $T=0$ is \cite{chandra}
\begin{equation}
\label{fermidm6}
P(\rho)=\frac{1}{8}\left (\frac{3}{\pi}\right )^{1/3}\frac{h
c}{m^{4/3}}\rho^{4/3}.
\end{equation}
Using the results of Appendix \ref{sec_gwe}, we obtain the enthalpy
\begin{equation}
\label{fermidm8}
h(\rho)=\frac{1}{2}\left (\frac{3}{\pi}\right )^{1/3}\frac{h
c}{m^{4/3}}\rho^{1/3}
\end{equation}
and the potential
\begin{equation}
\label{fermidm9}
V(\rho)=\frac{3}{8}\left (\frac{3}{\pi}\right )^{1/3}\frac{h
c}{m^{4/3}}\rho^{4/3}.
\end{equation}
This leads to a generalized wave equation of the form
\begin{eqnarray}
\label{fermidm10}
i\hbar \frac{\partial\psi}{\partial
t}=-\frac{\hbar^2}{2m}\Delta\psi+mT_{\rm eff}\ln(|\psi|^2)\psi\nonumber\\
+ \frac{1}{2}\left (\frac{3}{\pi}\right )^{1/3}\frac{h c}{m^{1/3}} |\psi|^{2/3}
\psi 
+m\Phi\psi\nonumber\\
-i\frac{\hbar}{2}\xi\left\lbrack \ln\left
(\frac{\psi}{\psi^*}\right
)-\left\langle \ln\left (\frac{\psi}{\psi^*}\right
)\right\rangle\right\rbrack\psi.
\end{eqnarray}

\section{Conclusion}
\label{sec_con}

In this paper, we have tried to justify with more precise arguments the
generalized GPP equations (\ref{we9}) and (\ref{we9b}) introduced heuristically
in \cite{chavtotal,modeldm}. These equations aim at providing a
parametrization of
the complicated processes of gravitational cooling and violent relaxation
experienced by a self-gravitating BEC described by the ordinary GPP equations
(\ref{he7}) and (\ref{he8}). First, we have recalled that the Schr\"odinger
equation (\ref{h1}) for the wave function is equivalent to the Wigner equation
(\ref{qw2}) for the DF. Then, we have introduced a
coarse-grained Wigner equation (\ref{qw3}) by analogy with the coarse-grained
Vlasov equation introduced in connection to Lynden-Bell's theory of violent
relaxation (see \cite{moczSV} for the
Schr\"odinger-Poisson--Vlasov-Poisson correspondence). This equation is
consistent with a MEPP. From
the coarse-grained Wigner equation, proceeding as in \cite{csr}, we have derived
a set of quantum hydrodynamic
equations (\ref{ch1})-(\ref{ch3}) which include a quantum potential, a
Lynden-Bell pressure, and a friction force. The Lynden-Bell pressure can be
decomposed into a polytropic equation of state   and an isothermal equation of
state [see Eq. (\ref{qj18})]. The isothermal equation of state is valid at low
or mid densities. It takes into account effective thermal effects which describe
scalar radiation. The polytropic equation of state is valid at high densities.
It takes into account degeneracy effects in the sense of Lynden-Bell. Finally,
we have used the inverse Madelung transformation to derive the corresponding
wave equations (\ref{we9}) and (\ref{we9b}). These generalized GPP equations
include an effective potential associated with the Lynden-Bell pressure and a
damping term. The Lynden-Bell potential can be written as the sum of a power-law
potential describing degeneracy effects and a logarithmic potential describing
effective thermal effects [see Eq. (\ref{we11})]. The self-interaction of the
bosons can be taken into account as usual by introducing an additional potential
in the GPP equations [see Eq. (\ref{we8})]. This leads to Eq. (\ref{we16}). We
have explained the importance of
developing a microcanonical description of the process of violent relaxation
instead of a canonical one. This can be achieved in our formalism by letting the
effective temperature depend on time (and possibly position) so as to conserve
the total energy [see
Eqs. (\ref{jme2}) and (\ref{me4}) or Eqs. (\ref{int1})-(\ref{int3})].

The generalized GPP equations satisfy an $H$-theorem for the Lynden-Bell entropy
and relax towards a stable equilibrium  state (virialized state) which is a
maximum entropy state at fixed mass and energy. This quasistationary state,
which can be viewed as the most probable state of the system, has a core-halo
structure. It is made of a quantum core surrounded by an isothermal halo. The
core results from the balance between the gravitational attraction, the
repulsion due the Heisenberg uncertainty principle, the self-interaction of the
bosons, and the 
repulsion due to the Lynden-Bell exclusion principle.\footnote{The Lynden-Bell
exclusion principle is usually negligible for self-gravitating  bosons (for
which
$\eta_0\gg m^4/h^3$) but it may be relevant for self-gravitating
fermions (for which $\eta_0\sim gm^4/h^3$).} The halo results
from
the balance between the gravitational attraction and the Lynden-Bell effective
thermal pressure. The quantum core solves the core-cusp problem. The isothermal
halo accounts for the flat rotation curves of the galaxies. Since this core-halo
structure is consistent with a MEP,
the core mass can be
obtained by maximizing the entropy $S(M_c)$ at fixed halo mass $M_h$ and halo
energy $E_h$. Interestingly, we have shown in \cite{mcmh} that this maximization
problem is equivalent to the ``velocity dispersion tracing'' relation and
returns the core mass-halo mass relation $M_c(M_h)$ observed in direct
numerical simulations of the GPP equations.  Therefore, our effective
thermodynamic approach is consistent with the structure of BECDM halos.

We have proposed that a similar generalized wave equation may be relevant 
for self-gravitating fermions [see Eqs. (\ref{fermidm5}) and (\ref{fermidm10})].
In that case, the pressure due to the Pauli exclusion principle replaces the
pressure due to the self-interaction of the
bosons.\footnote{One can also account for  the self-interaction
of the fermions and introduce a self-interaction pressure in addition to the
pressure due to the Pauli exclusion principle.} The corresponding wave
equation relaxes towards a stable equilibrium state with a core-halo structure
made of a quantum core and an isothermal halo. This is similar to the
equilibrium state of the generalized GPP equations except that the  bosonic
condensate is replaced by a fermion ball. Explicit density profiles with a
core-halo structure have been computed from our model in the case of
BECs with repulsive self-interaction in the TF approximation \cite{modeldm}.
They are in good
agreement with density profiles of DM halos obtained from observations or from
direct numerical simulations. Similar results will be reported for
noninteracting bosons and fermions in forthcoming
contributions.

\appendix

\section{Properties of the Wigner distribution}
\label{sec_cw}

In this Appendix, we check that the density $\rho$, velocity ${\bf u}$ and
pressure tensor $P_{ij}$ obtained from the Wigner DF
(\ref{qw1}) coincide with the expressions obtained from the
Schr\"odinger equation (\ref{h1}).

To that purpose, let us first establish useful identities. Using Eqs.
(\ref{mad2}) and
(\ref{mad3}), we get
\begin{eqnarray}
\label{cw1}
\rho=|\psi|^2=\psi\psi^*
\end{eqnarray}
and
\begin{eqnarray}
\label{cw2}
{\bf u}=\frac{\nabla
S}{m}=\frac{\hbar}{2im\rho}(\psi^*\nabla\psi-\psi\nabla\psi^*).
\end{eqnarray}
We also note that
\begin{eqnarray}
\label{cw3}
 \int d{\bf
v}\, e^{im{\bf v}\cdot {\bf y}/\hbar} =(2\pi)^3  \delta\left
(\frac{m {\bf y}}{\hbar}\right )=\frac{(2\pi\hbar)^3}{m^3}\delta({\bf y}).
\end{eqnarray}
We are now ready to compute the first moments of the Wigner DF
\begin{eqnarray}
\label{cw5}
f({\bf r},{\bf v},t)=\frac{m^3}{(2\pi\hbar )^3} \int d{\bf
y}\,  e^{im{\bf v}\cdot {\bf y}/\hbar} \nonumber\\
\times\psi^*\left ({\bf
r}+\frac{{\bf y}}{2},t\right )\psi\left ({\bf r}-\frac{{\bf
y}}{2},t\right ).
\end{eqnarray}

Integrating Eq. (\ref{cw5}) over ${\bf v}$ and using Eq. (\ref{cw3}), we
obtain
\begin{eqnarray}
\label{cw6}
\rho\equiv\int f\, d{\bf v}=|\psi|^2,
\end{eqnarray}
which coincide with Eq. (\ref{cw1}).

Multiplying Eq. (\ref{cw5}) by ${\bf v}$ and integrating over ${\bf v}$, we get
\begin{eqnarray}
\label{cw7}
\int f {\bf v}\, d{\bf v}=\frac{m^3}{(2\pi\hbar )^3} \int d{\bf v}d{\bf
y}\, {\bf v} e^{im{\bf v}\cdot {\bf y}/\hbar} \nonumber\\
\times\psi^*\left ({\bf
r}+\frac{{\bf y}}{2},t\right )\psi\left ({\bf r}-\frac{{\bf
y}}{2},t\right )\nonumber\\
=\frac{m^3}{(2\pi\hbar )^3} \int d{\bf v}d{\bf
y}\, \frac{\partial}{\partial {\bf y}}\left (e^{im{\bf v}\cdot {\bf
y}/\hbar}\right )\frac{\hbar}{im} \nonumber\\
\times\psi^*\left ({\bf
r}+\frac{{\bf y}}{2},t\right )\psi\left ({\bf r}-\frac{{\bf
y}}{2},t\right )\nonumber\\
=-\frac{m^3}{(2\pi\hbar )^3} \int d{\bf v}d{\bf
y}\, e^{im{\bf v}\cdot {\bf
y}/\hbar}\frac{\hbar}{im} \nonumber\\
\times \frac{\partial}{\partial {\bf y}}\left\lbrack \psi^*\left ({\bf
r}+\frac{{\bf y}}{2},t\right )\psi\left ({\bf r}-\frac{{\bf
y}}{2},t\right )\right\rbrack,
\end{eqnarray}
where we have used an integration by parts to obtain the last equality.
Calculating the derivative of the term in brackets, and using Eq. (\ref{cw3}),
we get
\begin{eqnarray}
\label{cw8}
\rho {\bf u}\equiv \int f {\bf v}\, d{\bf v}=\frac{\hbar}{2im}\left
(\psi^*\nabla\psi-\psi\nabla\psi^*\right ),
\end{eqnarray}
which coincide with Eq. (\ref{cw2}).

Multiplying Eq. (\ref{cw5}) by $v_iv_j$, integrating over ${\bf v}$, and
proceeding like in Eq. (\ref{cw7}), we
obtain
\begin{eqnarray}
\label{cw9}
\int f v_iv_j\, d{\bf v}=\frac{m^3}{(2\pi\hbar )^3} \int d{\bf v}d{\bf
y}\, v_iv_j e^{im{\bf v}\cdot {\bf y}/\hbar} \nonumber\\
\times\psi^*\left ({\bf
r}+\frac{{\bf y}}{2},t\right )\psi\left ({\bf r}-\frac{{\bf
y}}{2},t\right )\nonumber\\
=\frac{m^3}{(2\pi\hbar )^3} \int d{\bf v}d{\bf
y}\, \frac{\partial^2}{\partial y_i\partial y_j}\left (e^{im{\bf v}\cdot {\bf
y}/\hbar}\right )\left (\frac{\hbar}{im}\right )^2 \nonumber\\
\times\psi^*\left ({\bf
r}+\frac{{\bf y}}{2},t\right )\psi\left ({\bf r}-\frac{{\bf
y}}{2},t\right )\nonumber\\
=\frac{m^3}{(2\pi\hbar )^3} \int d{\bf v}d{\bf
y}\, e^{im{\bf v}\cdot {\bf
y}/\hbar}\left (\frac{\hbar}{im}\right )^2 \nonumber\\
\times \frac{\partial^2}{\partial y_i\partial y_j}\left\lbrack \psi^*\left ({\bf
r}+\frac{{\bf y}}{2},t\right )\psi\left ({\bf r}-\frac{{\bf
y}}{2},t\right )\right\rbrack.
\end{eqnarray}
Calculating the second derivative of the term in brackets, and using
Eq. (\ref{cw3}), we get
\begin{eqnarray}
\label{cw10}
\int f v_iv_j\, d{\bf v}=-\frac{\hbar^2}{4m^2}\biggl (
\psi^*\frac{\partial^2\psi}{\partial x_i\partial
x_j}-\frac{\partial\psi^*}{\partial x_j}\frac{\partial\psi}{\partial
x_i}\nonumber\\
-\frac{\partial\psi}{\partial x_j}\frac{\partial\psi^*}{\partial
x_i}+\psi\frac{\partial^2\psi^*}{\partial x_i\partial
x_j}\biggr ).
\end{eqnarray}
The pressure tensor can be written as
\begin{equation}
\label{cw11}
P_{ij}\equiv \int f ({\bf v}-{\bf u})_i  ({\bf v}-{\bf
u})_j\, d{\bf v}=\int f v_iv_j\, d{\bf v}-\rho u_iu_j.
\end{equation}
Using Eqs. (\ref{cw2}) and (\ref{cw10}), we obtain after simplification
\begin{eqnarray}
\label{cw12}
P_{ij}=\frac{\hbar^2}{4m^2}\biggl
(\frac{\psi^*}{\psi}\frac{\partial\psi}{\partial
x_i}\frac{\partial\psi}{\partial
x_j}
+\frac{\psi}{\psi^*}\frac{\partial\psi^*}{\partial
x_i}\frac{\partial\psi^*}{\partial
x_j}\nonumber\\
-\psi^*\frac{\partial^2\psi}{\partial x_i\partial
x_j}-\psi\frac{\partial^2\psi^*}{\partial x_i\partial
x_j}\biggr ).
\end{eqnarray}
The quantum pressure tensor obtained from the
Schr\"odinger equation (\ref{h1}) can be written as [see Eq. (\ref{mad10})]
\begin{equation}
\label{cw13}
P_{ij}^{Q}=\frac{\hbar^2}{4m^2}\left (\frac{1}{\rho}
\partial_i\rho\partial_j\rho-\partial_i\partial_j\rho\right ).
\end{equation}
Using Eq. (\ref{cw1}), we get after simplification
\begin{eqnarray}
\label{cw14}
P_{ij}^Q=\frac{\hbar^2}{4m^2}\biggl
(\frac{\psi^*}{\psi}\frac{\partial\psi}{\partial
x_i}\frac{\partial\psi}{\partial
x_j}
+\frac{\psi}{\psi^*}\frac{\partial\psi^*}{\partial
x_i}\frac{\partial\psi^*}{\partial
x_j}\nonumber\\
-\psi^*\frac{\partial^2\psi}{\partial x_i\partial
x_j}-\psi\frac{\partial^2\psi^*}{\partial x_i\partial
x_j}\biggr ).
\end{eqnarray}
Comparing Eqs. (\ref{cw12}) and (\ref{cw14}) we see that
\begin{equation}
\label{cw15}
P_{ij}=P_{ij}^{Q}.
\end{equation}

Let us check that the expression of the energy in the Wigner
representation is the same as in the Schr\"odinger representation. In terms of
the Wigner DF, the total energy is given by
\begin{equation}
\label{etot1}
E=\int f \frac{v^2}{2}\, d{\bf r}d{\bf v}+\frac{1}{2}\int \rho\Phi\, d{\bf
r}.
\end{equation}
According to Eqs. (\ref{cw11}) and (\ref{cw15}), it can be rewritten as
\begin{equation}
\label{etot2}
E=\int \rho \frac{{\bf u}^2}{2}\, d{\bf r}+\frac{1}{2}\int P_{ii}^{Q}\,
d{\bf r}+\frac{1}{2}\int \rho\Phi\, d{\bf
r}.
\end{equation}
Using Eq. (G.22) of \cite{chavtotal} establishing that $\int P_{ii}^Q\, d{\bf
r}=2\Theta_Q$, we find that
\begin{equation}
\label{etot3}
E=\int \rho \frac{{\bf u}^2}{2}\, d{\bf r}+\int \rho
\frac{Q}{m}\, d{\bf r}+\frac{1}{2}\int \rho\Phi\, d{\bf
r},
\end{equation}
which is in agreement with Eq. (\ref{hme2}) obtained from the  Schr\"odinger
equation.
From this equivalence, we can directly conclude that the Wigner-Poisson
equations conserve the energy.

Finally, we derive the Wigner equation
(\ref{qw2}). Taking the time derivative
of the Wigner DF
(\ref{cw5}), we get 
\begin{eqnarray}
\label{cw16}
\frac{\partial f}{\partial t}=\frac{m^3}{(2\pi\hbar )^3} \int
d{\bf
y}\,  e^{im{\bf v}\cdot {\bf y}/\hbar} \nonumber\\
\times \Biggl\lbrack \psi^*\left ({\bf
r}+\frac{{\bf y}}{2},t\right )\frac{\partial \psi}{\partial t}\left ({\bf
r}-\frac{{\bf
y}}{2},t\right )\nonumber\\
+\frac{\partial\psi^*}{\partial t}\left ({\bf
r}+\frac{{\bf y}}{2},t\right )\psi\left ({\bf
r}-\frac{{\bf
y}}{2},t\right )\Biggr\rbrack.
\end{eqnarray}
The next step is to substitute the Schr\"odinger equation (\ref{h1}) into the
term in brackets of Eq. (\ref{cw16}). The contribution of the potential term in
the Schr\"odinger equation is
\begin{eqnarray}
\label{cw17}
\left (\frac{\partial f}{\partial t}\right )_{\rm pot}=\frac{i
m^4}{(2\pi)^3\hbar^4}\int
d{\bf
y}\,  e^{im{\bf v}\cdot {\bf y}/\hbar} \nonumber\\
\times \left\lbrack \Phi\left ({\bf
r}+\frac{{\bf y}}{2},t\right )- \Phi\left ({\bf
r}-\frac{{\bf y}}{2},t\right )\right\rbrack \nonumber\\
\times\psi^*\left ({\bf
r}+\frac{{\bf
y}}{2},t\right )\psi\left ({\bf
r}-\frac{{\bf y}}{2},t\right )\nonumber\\
=\frac{im^4}{(2\pi)^3\hbar^4}\int
d{\bf
y}d{\bf v}'\,  e^{im({\bf v}-{\bf v}')\cdot {\bf y}/\hbar} \nonumber\\
\times \left\lbrack \Phi\left ({\bf
r}+\frac{{\bf y}}{2},t\right )- \Phi\left ({\bf
r}-\frac{{\bf y}}{2},t\right )\right\rbrack f({\bf r},{\bf v}',t).
\end{eqnarray}
The second equality can be checked directly by substituting Eq. (\ref{cw5}),
integrating
over ${\bf v}'$, and using identity (\ref{cw3}).
The contribution of the kinetic term  in
the Schr\"odinger equation is
\begin{eqnarray}
\label{cw18}
\left (\frac{\partial f}{\partial t}\right )_{\rm kin}^{(1)}=\frac{i
m^2}{2(2\pi)^3\hbar^2}\int
d{\bf
y}\,  e^{im{\bf v}\cdot {\bf y}/\hbar} \nonumber\\
\times\psi^*\left ({\bf
r}+\frac{{\bf y}}{2},t\right )
\Delta_{\bf r}\psi\left ({\bf
r}-\frac{{\bf y}}{2},t\right ) \nonumber\\
=-\frac{i
m^2}{(2\pi)^3\hbar^2}\int
d{\bf
y}\,  e^{im{\bf v}\cdot {\bf y}/\hbar} \nonumber\\
\times\psi^*\left ({\bf
r}+\frac{{\bf y}}{2},t\right )
\nabla_{\bf y}\cdot \nabla_{\bf r}\psi\left ({\bf
r}-\frac{{\bf y}}{2},t\right ).
\end{eqnarray}
Integrating by parts, we get
\begin{eqnarray}
\label{cw19}
\left (\frac{\partial f}{\partial t}\right )_{\rm kin}^{(1)}=
\frac{i
m^2}{2(2\pi)^3\hbar^2}\int
d{\bf
y}\,  e^{im{\bf v}\cdot {\bf y}/\hbar} \nonumber\\
\times \nabla\psi^*\left ({\bf
r}+\frac{{\bf y}}{2},t\right )\cdot \nabla\psi\left ({\bf
r}-\frac{{\bf y}}{2},t\right ) \nonumber\\
-\frac{
m^3}{(2\pi)^3\hbar^3}{\bf v} \cdot\int
d{\bf
y}\, e^{im{\bf v}\cdot {\bf y}/\hbar} \nonumber\\
\times \psi^*\left ({\bf
r}+\frac{{\bf y}}{2},t\right ) \nabla\psi\left ({\bf
r}-\frac{{\bf y}}{2},t\right ).
\end{eqnarray}
Making the same operations with the kinetic term  of
the complex conjugate Schr\"odinger equation  and
adding the two results, we obtain
\begin{eqnarray}
\label{cw20}
\left (\frac{\partial f}{\partial t}\right )_{\rm kin}=
-\frac{
m^3}{(2\pi\hbar)^3}{\bf v}\cdot \int
d{\bf
y}\,  e^{im{\bf v}\cdot {\bf y}/\hbar} \nonumber\\
\times \psi^*\left ({\bf
r}+\frac{{\bf y}}{2},t\right ) \nabla\psi\left ({\bf
r}-\frac{{\bf y}}{2},t\right ) \nonumber\\
-\frac{
m^3}{(2\pi\hbar)^3}{\bf v} \cdot\int
d{\bf
y}\, e^{im{\bf v}\cdot {\bf y}/\hbar} \nonumber\\
\times \psi\left ({\bf
r}-\frac{{\bf y}}{2},t\right )\nabla\psi^*\left ({\bf
r}+\frac{{\bf y}}{2},t\right )=-{\bf v}\cdot \frac{\partial f}{\partial {\bf
r}}.
\end{eqnarray}
The second equality can be checked directly by substituting Eq. (\ref{cw5}).
Summing
Eqs. (\ref{cw17}) and (\ref{cw20}), we obtain the Wigner equation (\ref{qw2}).

{\it Remark:} When $\hbar\rightarrow 0$, we can make the approximation
$\Phi({\bf r}\pm{\bf y}/2,t)\simeq \Phi({\bf r},t)\pm \nabla\Phi({\bf r},t)\cdot
{\bf y}/2$ in the integral of the Wigner equation (\ref{qw2}) and we obtain 
\begin{eqnarray}
\label{qw2zero}
\frac{\partial f}{\partial t}&+&{\bf v}\cdot \frac{\partial f}{\partial {\bf
r}}\nonumber\\
&=&\frac{im^4}{(2\pi\hbar)^3\hbar}\nabla\Phi \cdot\int e^{im({\bf
v}-{\bf v}')\cdot
{\bf y}/\hbar}
 {\bf y} f({\bf r},{\bf v}',t)\, d{\bf
y}d{\bf v}'\nonumber\\
&=&\frac{m^3}{(2\pi\hbar)^3}\nabla\Phi
\cdot\frac{\partial}{\partial {\bf v}}\int e^{im({\bf
v}-{\bf v}')\cdot
{\bf y}/\hbar} f({\bf r},{\bf v}',t)\, d{\bf
y}d{\bf v}'\nonumber\\
&=&\nabla\Phi
\cdot\frac{\partial}{\partial {\bf v}}\int \delta({\bf
v}-{\bf v}') f({\bf r},{\bf v}',t)\, d{\bf v}'\nonumber\\
&=&\nabla\Phi
\cdot\frac{\partial f}{\partial {\bf v}}.
\end{eqnarray}
This returns the Vlasov equation (\ref{vlasov1}). The
quantum correction to the Vlasov equation (i.e. the difference between the
Vlasov and the Wigner equation) is of order $\hbar^2$. Similarly, let us assume
that orbit crossing has not
yet occurred and let us take the limit $\hbar\rightarrow 0$ in the Wigner DF
(\ref{qw1}). Substituting Eq. (\ref{mad1}) into Eq. (\ref{qw1}), making the
approximations $\rho({\bf r}\pm{\bf
y}/2,t)\simeq \rho({\bf r},t)$ and $S({\bf r}\pm{\bf
y}/2,t)\simeq S({\bf r},t)\pm \nabla S({\bf r},t)\cdot {\bf y}/2$, and using
Eq. (\ref{mad3}), we get
\begin{eqnarray}
\label{qw1zero}
f({\bf r},{\bf v},t)&=&\frac{m^3}{(2\pi\hbar )^3} \rho \int d{\bf
y}\,  e^{im({\bf v}-{\bf u})\cdot {\bf y}/\hbar}\nonumber\\
&=&\rho({\bf r},t) \delta ( {\bf v}-{\bf u}({\bf r},t)).
\end{eqnarray}
This returns the single-speed solution from Eq. (\ref{e1}).
For finite $\hbar/m$, the DF $f$ is proportional to a Gaussian in
${\bf v}$ of thickness $\hbar/m$.

\section{Generalized GPP equations}
\label{sec_gwe}

In this Appendix, we recall the basic properties of the generalized GPP
equations introduced in \cite{chavtotal}. We also discuss the difference between
the canonical and microcanonical formulation.

\subsection{Basic properties}
\label{sec_bp}

Let us consider the generalized GPP equations
\begin{eqnarray}
\label{w1}
i\hbar \frac{\partial\psi}{\partial
t}=-\frac{\hbar^2}{2m}\Delta\psi+
m\left\lbrack
\Phi+\frac{d V}{d|\psi|^2}\right\rbrack\psi\nonumber\\
+mT_{\rm eff}\ln(|\psi|^2)\psi-i\frac{\hbar}{2}\xi\left\lbrack \ln\left
(\frac{\psi}{\psi^*}\right
)-\left\langle \ln\left (\frac{\psi}{\psi^*}\right
)\right\rangle\right\rbrack\psi,\nonumber\\
\end{eqnarray}
\begin{equation}
\label{w2}
\Delta\Phi=4\pi G |\psi|^2,
\end{equation}
with a potential $V(|\psi|^2)$. Introducing the enthalpy
(see Appendix \ref{sec_gicg})
\begin{equation}
\label{w3}
h(\rho)=V'(\rho),
\end{equation}
we can rewrite the generalized GP equation (\ref{w1}) as
\begin{eqnarray}
\label{w4}
i\hbar \frac{\partial\psi}{\partial
t}=-\frac{\hbar^2}{2m}\Delta\psi+
m\left\lbrack
\Phi+h(|\psi|^2)\right\rbrack\psi\nonumber\\
+mT_{\rm eff}\ln(|\psi|^2)\psi-i\frac{\hbar}{2}\xi\left\lbrack \ln\left
(\frac{\psi}{\psi^*}\right
)-\left\langle \ln\left (\frac{\psi}{\psi^*}\right
)\right\rangle\right\rbrack\psi.\nonumber\\
\end{eqnarray}

The stationary solutions of the generalized GPP equations (\ref{w1}) and 
(\ref{w2}) are of the form
\begin{equation}
\label{w3b}
\psi({\bf r},t)=\phi({\bf r})e^{-iEt/\hbar},
\end{equation}
where $\phi({\bf r})=\sqrt{\rho({\bf r})}$ and $E$
(eigenenergy) are real. They are determined by the eigenvalue problem
\begin{eqnarray}
\label{w4b}
-\frac{\hbar^2}{2m}\Delta\phi+m\left\lbrack
\Phi+h(\rho)\right\rbrack\phi+mT_{\rm eff}\ln(\rho)\phi
=E\phi,
\end{eqnarray}
\begin{eqnarray}
\label{w4c}
\Delta\Phi=4\pi G \phi^2.
\end{eqnarray}
Dividing Eq. (\ref{w4b}) by $\phi$, we get
\begin{eqnarray}
\label{w4d}
Q+m\Phi+mh(\rho)+mT_{\rm eff}\ln\rho=E,
\end{eqnarray}
where $Q$ is the quantum potential defined by Eq. (\ref{mad8}).

Using the Madelung \cite{madelung} transformation (see Sec. \ref{sec_mad}), the
hydrodynamic 
equations corresponding to the generalized GPP equations (\ref{w1}) and
(\ref{w2}) are 
\begin{equation}
\label{w5}
\frac{\partial\rho}{\partial t}+\nabla\cdot (\rho {\bf u})=0,
\end{equation}
\begin{equation}
\label{w5b}
\frac{\partial S}{\partial t}+\frac{1}{2m}(\nabla
S)^2+m\lbrack \Phi+T_{\rm eff}\ln\rho+h(\rho)\rbrack+Q+\xi (S-\langle
S\rangle)=0,
\end{equation}
\begin{equation}
\label{w6}
\frac{\partial {\bf u}}{\partial t}+({\bf u}\cdot \nabla){\bf
u}=-\frac{1}{m}\nabla Q-\frac{1}{\rho}\nabla
P-\nabla\Phi-\frac{1}{\rho}\nabla
P_{\rm th}-\xi {\bf u},
\end{equation}
\begin{equation}
\label{w7}
\Delta\Phi=4\pi G \rho,
\end{equation}
where the barotropic equation of state
$P(\rho)$ is determined by the relation (see
Appendix \ref{sec_gicg})
\begin{eqnarray}
\label{w8}
h'(\rho)=\frac{P'(\rho)}{\rho}.
\end{eqnarray}
For a general polytropic equation of state of the form
\begin{eqnarray}
\label{w9}
P_{\rm poly}=K\rho^{\gamma},
\end{eqnarray}
the enthalpy and the potential are given by
\begin{eqnarray}
\label{w10}
h_{\rm poly}(\rho)=\frac{K\gamma}{\gamma-1}\rho^{\gamma-1},\quad V_{\rm poly}(\rho)=\frac{K}{\gamma-1}\rho^{\gamma}.
\end{eqnarray}
In particular, for the standard self-interacting BEC described by the polytropic equation of state
\begin{equation}
\label{w10a}
P_{\rm int}(\rho)=\frac{2\pi a_s\hbar^2}{m^3}\rho^2,
\end{equation}
we have
\begin{eqnarray}
\label{w11}
h_{\rm int}(\rho)=\frac{4\pi a_s\hbar^2}{m^3}\rho,\quad V_{\rm int}(\rho)=\frac{2\pi a_s\hbar^2}{m^3}\rho^2.
\end{eqnarray}
For the Lynden-Bell pressure of zero-point energy
\begin{equation}
\label{w11a}
P_{\rm LB}^{(0)}=\frac{1}{5}\left (\frac{3}{4\pi\eta_0}\right )^{2/3}\rho^{5/3},
\end{equation}
we have
\begin{eqnarray}
\label{w11b}
h_{\rm LB}^{(0)}(\rho)=\frac{1}{2}\left (\frac{3}{4\pi\eta_0}\right
)^{2/3}\rho^{2/3},
\end{eqnarray}
\begin{eqnarray}
\label{w11c}
V_{\rm LB}^{(0)}(\rho)=\frac{3}{10}\left (\frac{3}{4\pi\eta_0}\right
)^{2/3}\rho^{5/3}.
\end{eqnarray}
For the Lynden-Bell isothermal equation of state
\begin{eqnarray}
\label{w12}
P_{\rm th}=\rho T_{\rm eff},
\end{eqnarray}
the enthalpy and the potential are given by
\begin{eqnarray}
\label{w13}
h_{\rm th}(\rho)=T_{\rm eff}\ln\rho,\quad V_{\rm th}(\rho)=T_{\rm eff}\rho(\ln\rho-1).
\end{eqnarray}

An equilibrium state of the quantum damped Euler equations (\ref{w5})-(\ref{w7}) satisfies the quantum equation of hydrostatic equilibrium
\begin{equation}
\label{w13c}
\frac{\rho}{m}\nabla Q+\nabla P+\rho\nabla\Phi+\nabla P_{\rm th}={\bf 0}.
\end{equation}
Dividing Eq. (\ref{w13c}) by $\rho$, using Eq. (\ref{w8}) and integrating, we
obtain Eq. (\ref{w4d}), where $E$ appears as a constant of integration.

\subsection{Microcanonical model}
\label{sec_wmce}

We introduce the mass
\begin{eqnarray}
\label{w13b}
M=\int \rho\, d{\bf r}
\end{eqnarray}
and the energy 
\begin{eqnarray}
\label{w14}
E_{\rm tot}=\Theta_c+\Theta_Q+U+W+\Theta_{\rm th},
\end{eqnarray}
which includes the  classical kinetic energy
\begin{eqnarray}
\label{w15}
\Theta_c=\int\rho  \frac{{\bf u}^2}{2}\, d{\bf r},
\end{eqnarray}
the quantum kinetic energy
\begin{equation}
\label{w16}
\Theta_Q=\frac{1}{m}\int \rho Q\, d{\bf r},
\end{equation}
the internal energy
\begin{equation}
\label{w17}
U=\int V(\rho)\, d{\bf r},
\end{equation}
the gravitational energy
\begin{eqnarray}
\label{w18}
W=\frac{1}{2}\int\rho\Phi\, d{\bf r},
\end{eqnarray}
and the thermal energy
\begin{equation}
\label{w19}
\Theta_{\rm th}=\frac{3}{2}M T_{\rm eff}.
\end{equation}
We also introduce the entropy (up to an additive constant)
\begin{equation}
\label{w20}
S=-\int\rho (\ln\rho-1)\, d{\bf r}+\frac{3}{2}M\ln T_{\rm eff}.
\end{equation}
The functionals (\ref{w15})-(\ref{w18}) are justified in \cite{chavtotal} and
the functionals (\ref{w19}) and (\ref{w20}) are justified in Appendix
\ref{sec_ol}.

In the microcanonical situation, the total energy $E_{\rm tot}$ is fixed and the
temperature $T_{\rm eff}(t)$ evolves in time so as to satisfy the constraint
from Eq. (\ref{w14}) yielding
\begin{equation}
\label{w21}
\frac{3}{2}M T_{\rm eff}(t)=E_{\rm tot}-\Theta_c(t)-\Theta_Q(t)-U(t)-W(t).
\end{equation}
The generalized GPP equations
(\ref{w1}) and (\ref{w2}) with the time-dependent temperature $T_{\rm eff}(t)$
given by Eq. (\ref{w21})\footnote{In terms of the wavefunction,
the effective temperature is given by
\begin{eqnarray}
\label{w21wave}
\frac{3}{2}M T_{\rm eff}(t)&=&E_{\rm tot}-\frac{\hbar^2}{2m^2}\int
|\nabla\psi|^2\, d{\bf r}\nonumber\\
&-&\int V(|\psi|^2)\, d{\bf r}-\frac{1}{2}\int
|\psi|^2\Phi\, d{\bf r}.
\end{eqnarray}
} conserve the energy (by construction) and satisfy an
$H$-theorem for the entropy (\ref{w20}). This can be proven as follows. From Eq.
(\ref{w20}), we have
\begin{eqnarray}
\label{w22}
\dot S=-\int\ln\rho \frac{\partial\rho}{\partial t}\, d{\bf r}+\frac{3}{2}M\frac{\dot T_{\rm eff}}{T_{\rm eff}}.
\end{eqnarray}
Using the equation of continuity (\ref{w5}) and integrating by parts, we get
\begin{equation}
\label{w23}
\dot S=-\int {\bf u}\cdot \nabla\rho\, d{\bf r}+\frac{3}{2}M\frac{\dot T_{\rm eff}}{T_{\rm eff}}.
\end{equation}
On the other hand, taking the time derivative of Eq. (\ref{w21}), and using Eqs.
(\ref{w15})-(\ref{w18}), we obtain
\begin{equation}
\label{w24}
\frac{3}{2}M \dot T_{\rm eff}=-\int \left (\frac{{\bf u}^2}{2}+\frac{Q}{m}+h+\Phi\right ) \frac{\partial\rho}{\partial t}\, d{\bf r}-\int \rho {\bf u}\cdot \frac{\partial {\bf u}}{\partial t}\, d{\bf r},
\end{equation}
where we have used the identities of Appendix C of \cite{chavtotal}. Using 
the equation of continuity (\ref{w5}), the damped quantum Euler equation
(\ref{w6}), and recalling the identity of vector analysis $({\bf u}\cdot
\nabla){\bf u}=\nabla({\bf u}^2/2)-{\bf u}\times (\nabla\times {\bf u})$ and the
fact that the flow is irrotational ($\nabla\times {\bf u}={\bf 0}$) since ${\bf
u}=\nabla S/m$, we get after simplification (using straightforward integrations
by parts)
\begin{equation}
\label{w25}
\frac{3}{2}M \dot T_{\rm eff}=\int {\bf u}\cdot \nabla P_{\rm th}\, d{\bf r}+\xi\int \rho {\bf u}^2\, d{\bf r}.
\end{equation}
Combining Eqs. (\ref{w23}) and (\ref{w25}), and using Eq. (\ref{w12}), we
finally obtain
\begin{equation}
\label{w26}
\dot S=\frac{\xi}{T_{\rm eff}}\int \rho {\bf u}^2\, d{\bf r}\ge 0.
\end{equation}
Therefore, the entropy increases monotonically.  At equilibrium (where $\dot
S=0$), we get ${\bf u}={\bf 0}$ leading to  the quantum equation of hydrostatic
equilibrium (\ref{w13c}).

We can easily show that the generalized GPP equations in the microcanonical
ensemble relax towards an equilibrium state that maximizes the entropy at fixed
mass and energy. An extremum of entropy at fixed mass and energy is determined
by the variational principle
\begin{equation}
\label{w27}
\delta S+\alpha\delta M=0,
\end{equation}
where $\alpha=\mu/mT_{\rm eff}$ is a Lagrange multiplier taking into account the
conservation of mass (the conservation of energy is taken into account in Eq.
(\ref{w21})). Using
\begin{equation}
\label{w28}
\delta S=-\int\ln\rho\, \delta\rho\, d{\bf r}+\frac{3M}{2T_{\rm eff}}\delta T_{\rm eff}
\end{equation}
and
\begin{equation}
\label{w29}
\frac{3}{2}M \delta T_{\rm eff}=-\int \left (\frac{Q}{m}+h+\Phi\right )\delta\rho\, d{\bf r},
\end{equation}
we obtain
\begin{equation}
\label{w30}
mT_{\rm eff}\ln\rho+Q+mh+m\Phi=\mu,
\end{equation}
which is equivalent to Eq. (\ref{w4d}) with $\mu=E$. This shows that 
the eigenenergy $E$ coincides with the chemical potential $\mu$.  Taking the
gradient of Eq. (\ref{w30}) and using Eq. (\ref{w8}), we recover the 
equation of quantum hydrostatic equilibrium (\ref{w13c}). Therefore, an
equilibrium
state of the generalized GPP equations is an extremum of entropy at fixed mass
and energy. By proceeding as explained at the end of Appendix \ref{sec_vrb} we
can show that only
entropy maxima (not
minima or saddle points) at fixed mass and energy are dynamically stable
\cite{prep}. From these results, we conclude that the generalized GPP
equations with a time-dependent effective temperature relax towards an
equilibrium state which maximizes the entropy at fixed mass and energy:
\begin{equation}
\max\ \lbrace {S}\, |\,  M, E_{\rm tot} \,\, {\rm fixed} \rbrace.
\label{maxmce}
\end{equation}

On the other hand, proceeding as in Appendix G of \cite{chavtotal}, we obtain
the damped
quantum  virial theorem
\begin{equation}
\label{vir1}
\frac{1}{2}\ddot I+\frac{1}{2}\xi\dot I=2(\Theta_c+\Theta_{\rm th}+\Theta_Q)+3\int P\, d{\bf r}+W,
\end{equation}
where $I=\int \rho r^2\, d{\bf r}$ is the moment of inertia.  According to Eq.
(\ref{w14}), we have $\Theta_c+\Theta_{\rm th}+\Theta_Q=E_{\rm tot}-U-W$ so we
can rewrite
Eq. (\ref{vir1}) as\footnote{The structure of this equation
shows that a system described by the generalized GPP equations undergoes damped
oscillations towards a virialized state. These damped oscillations are
characteristic of the process of gravitational cooling and violent relaxation
(see Sec. \ref{sec_gc}). In this sense, the generalized GPP equations provide a
parmetrization of the ordinary GPP equations.}
\begin{equation}
\label{vir2}
\frac{1}{2}\ddot I+\frac{1}{2}\xi\dot I=2E_{\rm tot}-2U-W+3\int P\, d{\bf r}.
\end{equation}
For a polytropic equation of state, the internal energy satisfies the identity
\begin{equation}
\label{vir3}
U_{\rm poly}=\frac{1}{\gamma-1}\int P_{\rm poly}\, d{\bf r},
\end{equation}
and the damped quantum   virial theorem becomes
\begin{equation}
\label{vir4}
\frac{1}{2}\ddot I+\frac{1}{2}\xi\dot I=2E_{\rm tot}+(3\gamma-5)U_{\rm poly}-W.
\end{equation}
We note that the term involving the internal energy vanishes for the particular index $\gamma=5/3$ which corresponds to the Lynden-Bell pressure of zero-point energy (\ref{w11a}).

{\it Remark:} In the strong friction limit $\xi\rightarrow +\infty$, we can
neglect the inertial term (l.h.s.) in the damped quantum Euler equation (\ref{w6}) and get
\begin{equation}
\label{w30aa}
{\bf u}=-\frac{1}{\xi\rho} \left (T_{\rm eff}\nabla\rho+\nabla P+\frac{\rho}{m}\nabla
Q+\rho\nabla\Phi\right ).
\end{equation}
Substituting this relation into the continuity equation (\ref{w5}) we  obtain the quantum
Smoluchowski-Poisson equations 
\begin{equation}
\label{w30a}
\xi\frac{\partial\rho}{\partial t}=\nabla\cdot \left (T_{\rm eff}\nabla\rho+\nabla P+\frac{\rho}{m}\nabla
Q+\rho\nabla\Phi\right ),
\end{equation}
\begin{eqnarray}
\label{w30b}
\Delta\Phi=4\pi G\rho.
\end{eqnarray}
Since $|{\bf u}|=O(1/\xi)$ in the strong friction limit, we can neglect the
classical kinetic energy $\Theta_c$  in Eq. (\ref{w21}). Therefore, the
evolution of the effective temperature is given by
\begin{equation}
\label{w30c}
\frac{3}{2}M T_{\rm eff}(t)=E_{\rm tot}-\Theta_Q(t)-U(t)-W(t).
\end{equation}
The $H$-theorem takes the form
\begin{equation}
\label{w30d}
\dot S=\frac{1}{\xi T_{\rm eff}}\int \frac{1}{\rho}\left (T_{\rm eff}\nabla\rho+\nabla P+\frac{\rho}{m}\nabla
Q+\rho\nabla\Phi\right )^2\, d{\bf r}\ge 0,
\end{equation}
and we have the same general properties as those discussed after Eq. (\ref{w26}).
On the other hand, the overdamped quantum virial theorem is given by
\begin{equation}
\label{vir5}
\frac{1}{2}\xi\dot I=2(\Theta_{\rm th}+\Theta_Q)+3\int P\, d{\bf r}+W.
\end{equation}
Using $\Theta_{\rm th}+\Theta_Q=E_{\rm tot}-U-W$, we can rewrite Eq.
(\ref{vir5}) as
\begin{equation}
\label{vir6}
\frac{1}{2}\xi\dot I=2E_{\rm tot}-2U-W+3\int P\, d{\bf r}.
\end{equation}
For a polytropic equation of state, using Eq. (\ref{vir3}), the overdamped
quantum virial theorem becomes
\begin{equation}
\label{vir7}
\frac{1}{2}\xi\dot I=2E_{\rm tot}+(3\gamma-5)U_{\rm poly}-W.
\end{equation}

\subsection{Canonical model}
\label{sec_wce}

In the canonical ensemble, the temperature $T_{\rm eff}$ is fixed and the
appropriate thermodynamic potential is the free energy
\begin{eqnarray}
\label{w31}
F=E_{\rm tot}-T_{\rm eff}S.
\end{eqnarray}
Using Eqs. (\ref{w14}) and (\ref{w20}), we obtain (up to an additive constant)
\begin{eqnarray}
\label{w32}
F=\Theta_c+\Theta_Q+U+W+U_{\rm th},
\end{eqnarray}
where
\begin{equation}
\label{w33}
U_{\rm th}=T_{\rm eff}\int\rho (\ln\rho-1)\, d{\bf r}
\end{equation}
is the internal energy associated with the thermal pressure. This returns the
free energy introduced in \cite{chavtotal}.

The generalized GPP equations (\ref{w1}) and (\ref{w2}) with 
a fixed temperature $T_{\rm eff}$  satisfy an $H$-theorem for the free energy
(\ref{w32}). Indeed, using calculations similar to the previous ones (see also
Appendix D of \cite{chavtotal}) we get
\begin{equation}
\label{w34}
\dot F=-\xi\int \rho {\bf u}^2\, d{\bf r}\le 0.
\end{equation}
Therefore, the free energy decreases monotonically. At equilibrium (where $\dot
F=0$), we get ${\bf u}={\bf 0}$ leading to  the  equation of quantum hydrostatic
equilibrium (\ref{w13c}). 

We can easily show that the generalized GPP equations in the canonical ensemble
relax towards an equilibrium state that minimizes the free energy at fixed mass.
An extremum of free energy at fixed mass is determined by the variational
principle
\begin{equation}
\label{w35}
\delta F-\frac{\mu}{m}\delta M=0,
\end{equation}
where $\mu/m$ is a Lagrange multiplier taking into account the conservation of
mass. Using calculations similar to the previous ones (see also section 3.4 of
\cite{chavtotal}) we get Eq. (\ref{w30}) which is equivalent to Eq. (\ref{w4d})
with
$\mu=E$ or to the quantum equation of hydrostatic equilibrium (\ref{w13c}).
Therefore, an equilibrium state of the generalized GPP equations
is an extremum of free energy at fixed mass. By proceeding as explained at the
end of Appendix \ref{sec_vrb}  we can show
that only minima (not
maxima or saddle points) of free energy at
fixed mass are dynamically stable \cite{prep}. From these results, we
conclude that
the generalized GPP equations with a constant effective temperature relax
towards an equilibrium state which minimizes the free energy at fixed mass:
\begin{equation}
\min\ \lbrace {F}\, |\,  M \,\, {\rm fixed} \rbrace.
\label{mince}
\end{equation}

On the other hand, the  damped quantum  virial theorem is given by Eq.
(\ref{vir1}) or, equivalently, by
\begin{equation}
\label{vir8}
\frac{1}{2}\ddot I+\frac{1}{2}\xi\dot I=2(\Theta_c+\Theta_Q)+3M T_{\rm eff}+3\int P\, d{\bf r}+W,
\end{equation}
where $T_{\rm eff}$ is constant.

{\it Remark:} In the strong friction limit $\xi\rightarrow +\infty$, we get the quantum Smoluchowski-Poisson equations (\ref{w30a}) and (\ref{w30b}) where $T_{\rm eff}$ is constant. The $H$-theorem now writes
\begin{equation}
\label{w36}
\dot F=-\frac{1}{\xi}\int \frac{1}{\rho}\left (T_{\rm eff}\nabla\rho+\nabla P+\frac{\rho}{m}\nabla
Q+\rho\nabla\Phi\right )^2\, d{\bf r}\le 0.
\end{equation}
On the other hand, the overdamped quantum virial theorem is given by Eq.
(\ref{vir5}) or,
equivalently, by
\begin{equation}
\label{vir9}
\frac{1}{2}\xi\dot I=2\Theta_Q+3M T_{\rm eff}+3\int P\, d{\bf r}+W,
\end{equation}
where $T_{\rm eff}$ is constant.

\subsection{Numerical algorithm}
\label{sec_na}

A stationary solution  of the GPP equations 
\begin{eqnarray}
\label{w1s}
i\hbar \frac{\partial\psi}{\partial
t}=-\frac{\hbar^2}{2m}\Delta\psi+
m\left\lbrack
\Phi+\frac{d V}{d|\psi|^2}\right\rbrack\psi,
\end{eqnarray}
\begin{equation}
\label{w2s}
\Delta\Phi=4\pi G |\psi|^2,
\end{equation}
is of the form $\psi({\bf r},t)=\phi({\bf r})e^{-iEt/\hbar}$,
where $\phi({\bf r})=\sqrt{\rho({\bf r})}$ and $E$
(eigenenergy) are real. They are determined by the eigenvalue problem
\begin{eqnarray}
\label{w4bs}
-\frac{\hbar^2}{2m}\Delta\phi+m\left\lbrack
\Phi+h(\rho)\right\rbrack\phi=E\phi,
\end{eqnarray}
\begin{eqnarray}
\label{w4cs}
\Delta\Phi=4\pi G \phi^2.
\end{eqnarray}
Dividing Eq. (\ref{w4bs}) by $\phi$, we get
\begin{eqnarray}
\label{w4dy}
Q+m\Phi+mh(\rho)=E,
\end{eqnarray}
where $Q$ is the quantum potential defined by Eq. (\ref{mad8}).

Using  the Madelung transformation, the ordinary GPP equations
(\ref{w1s}) and (\ref{w2s}) are equivalent to the  hydrodynamic 
equations
\begin{equation}
\label{w5s}
\frac{\partial\rho}{\partial t}+\nabla\cdot (\rho {\bf u})=0,
\end{equation}
\begin{equation}
\label{w5sb}
\frac{\partial S}{\partial t}+\frac{1}{2m}(\nabla
S)^2+m\Phi+mh(\rho)+Q=0,
\end{equation}
\begin{equation}
\label{w6s}
\frac{\partial {\bf u}}{\partial t}+({\bf u}\cdot \nabla){\bf
u}=-\frac{1}{m}\nabla Q-\frac{1}{\rho}\nabla
P-\nabla\Phi,
\end{equation}
\begin{equation}
\label{w7s}
\Delta\Phi=4\pi G \rho.
\end{equation}
A stationary solution of the quantum Euler-Poisson equations
(\ref{w5s})-(\ref{w7s}) satisfies the condition
of quantum hydrostatic equilibrium 
\begin{equation}
\label{w13cs}
\frac{\rho}{m}\nabla Q+\nabla P+\rho\nabla\Phi={\bf 0}.
\end{equation}
This equation is equivalent to Eq. (\ref{w4dy}) as can be seen by taking the
gradient of  Eq. (\ref{w4dy}) and using Eq. (\ref{w8}).

On the other hand, a stationary
solution of the ordinary GPP equations (\ref{w1s}) and (\ref{w2s}) is an
extremum of energy 
\begin{eqnarray}
\label{w32s}
E_{\rm tot}=\Theta_c+\Theta_Q+U+W
\end{eqnarray}
at fixed mass $M$. Indeed, writing
the variational principle as
\begin{equation}
\label{w32sa}
\delta E_{\rm tot}-\frac{\mu}{m}\delta M=0,
\end{equation}
where $\mu/m$ is a Lagrange multiplier (global chemical potential) taking into
account the mass constraint, we get
\begin{equation}
\label{hyesa}
Q+m\Phi+mh(\rho)=\mu,
\end{equation}
which is equivalent to Eq. (\ref{w4dy}) with $E=\mu$, and to Eq. (\ref{w13cs}).
Furthermore, it can be shown that an equilibrium state is dynamically stable if,
and only if, it is a {\it minimum} of
energy  at fixed mass \cite{prd1}.\footnote{These
results are basically due to
the
fact that both $E_{\rm tot}$ and $M$ are
conserved by the GPP equations \cite{holm}. A direct proof of the
equivalence
between the energy principle (minimum of energy at fixed mass) and the condition
of dynamical stability (positivity of the squared pulsation $\omega^2$ of
all modes) for the GPP equations is given in Appendix B of \cite{jeansMR}.}
Therefore, a stable stationary
solution of the GPP equations is determined by the minimization problem
\begin{equation}
\min\ \lbrace {E_{\rm tot}}\, |\,  M \,\, {\rm fixed} \rbrace.
\label{na1}
\end{equation}
We stress
that the ordinary GPP 
equations do {\it not} relax towards the stationary state that minimizes the
energy at fixed mass (ground state) since
these equations are reversible and the energy is conserved.
They rather experience a process of
gravitational cooling and violent relaxation  (on a coarse-grained scale)
towards a quasistationary state with a
core-halo structure (see Sec. \ref{sec_gc}). The characterization of this
core-halo state was the topic of this paper.

Independently from the 
physical problem treated in this paper, it is an interesting mathematical
problem in itself to be able to construct stationary solutions of the ordinary
GPP equations (\ref{w1s}) and (\ref{w2s}). However, it is difficult
in practice to numerically solve the
nonlinear eigenvalue problem defined by Eqs. (\ref{w4bs}) and (\ref{w4cs}) and
make sure that the solution is dynamically stable. Interestingly, the
generalized GPP equations (\ref{w1}) and (\ref{w2})  provide a useful numerical
algorithm to reach that goal.\footnote{Note that the
thermal term $T_{\rm eff}\ln(|\psi|^2)$ can be absorbed in the potential
$V(|\psi|^2)$ so we take $T_{\rm eff}=0$ here.} Indeed, we have shown in
Appendix
\ref{sec_wce} that these equations satisfy
an $H$-theorem and that they relax towards a stationary state that minimizes the
free energy $F$ defined by Eq. (\ref{w32}) at fixed mass $M$. Since $F$
coincides with $E_{\rm tot}$, this
equilibrium state solves the minimization problem (\ref{na1}). Therefore,  it is
a dynamically stable steady state of the ordinary GPP equations (\ref{w1s}) and
(\ref{w2s}). Consequently,
the
generalized GPP equations (\ref{w1}) and (\ref{w2}) provide a useful numerical
algorithm to construct
stable stationary solutions of the ordinary GPP equations (\ref{w1s}) and
(\ref{w2s}). By construction, the
equilibrium solution reached by the generalized GPP equations (which are true
relaxation equations) is guaranteed to be a  stable
stationary
solution of the ordinary GPP equations.

{\it Remark:} Instead of solving the generalized GPP equations (\ref{w1}) and (\ref{w2}), we may equivalently solve the quantum damped Euler equations (\ref{w5})-(\ref{w7}) or the simpler (diffusive) quantum Smoluchowski-Poisson equations (\ref{w30a}) and (\ref{w30b})  which also relax towards a stationary solution that satisfies the minimization problem (\ref{na1}).

\subsection{General identities for a cold gas}
\label{sec_gicg}

For a cold ($T=0$) gas, the first principle of thermodynamics
\begin{equation}
\label{fo1}
d\left (\frac{u}{\rho}\right )=-P d\left (\frac{1}{\rho}\right )+T d\left
(\frac{s}{\rho}\right ),
\end{equation}
where $u$ is the density of internal energy, $s$ the density of entropy, $\rho$
the mass density, and $P$ the pressure reduces to
\begin{equation}
\label{fo2}
d\left (\frac{u}{\rho}\right )=-P d\left (\frac{1}{\rho}\right
)=\frac{P}{\rho^2}d\rho.
\end{equation}
Introducing the enthalpy per particle
\begin{equation}
\label{fo3}
h=\frac{P+u}{\rho},
\end{equation}
we get
\begin{equation}
\label{fo4}
du=h d\rho\qquad {\rm and}\qquad dh=\frac{dP}{\rho}.
\end{equation}
Comparing Eq. (\ref{fo3}) with the Gibbs-Duhem relation at $T=0$:
\begin{equation}
\label{fo4gd}
u=-P+Ts+\frac{\mu}{m} \rho \qquad \Rightarrow \qquad
\frac{\mu}{m}=\frac{P+u}{\rho},
\end{equation}
we  see that the enthalpy $h({\bf r})$ is equal to the local chemical potential
$\mu({\bf r})$ by unit of mass: $h({\bf r})=\mu({\bf r})/m$. On the other hand,
for a barotropic gas for which $P=P(\rho)$, the foregoing equations can be
written as
\begin{equation}
\label{fo5}
P(\rho)=-\frac{d(u/\rho)}{d(1/\rho)}=\rho^2\left \lbrack
\frac{u(\rho)}{\rho}\right \rbrack'=\rho u'(\rho)-u(\rho),
\end{equation}
\begin{equation}
\label{fo5b}
P'(\rho)=\rho u''(\rho),\qquad h(\rho)=\frac{P(\rho)+u(\rho)}{\rho},
\end{equation}
\begin{equation}
\label{fo6}
h(\rho)=u'(\rho),\qquad h'(\rho)=\frac{P'(\rho)}{\rho}.
\end{equation}
\begin{equation}
\label{fo6b}
u(\rho)=\rho\int^{\rho} \frac{P(\rho')}{{\rho'}^2}\, d\rho'.
\end{equation}
Comparing Eq. (\ref{fo6}) with Eqs. (\ref{w3}) and (\ref{w8}), we see that the
potential $V(\rho)$ represents the density of internal energy
\begin{equation}
\label{fo7}
u(\rho)=V(\rho).
\end{equation}
We then have
\begin{equation}
\label{nfo5}
P(\rho)=\rho^2\left \lbrack \frac{V(\rho)}{\rho}\right \rbrack'=\rho
V'(\rho)-V(\rho),
\end{equation}
\begin{equation}
\label{bfo5b}
P'(\rho)=\rho V''(\rho),\qquad h(\rho)=\frac{P(\rho)+V(\rho)}{\rho},
\end{equation}
\begin{equation}
\label{wfo6}
h(\rho)=V'(\rho),\qquad h'(\rho)=\frac{P'(\rho)}{\rho}.
\end{equation}
\begin{equation}
\label{wfo6b}
V(\rho)=\rho\int^{\rho} \frac{P(\rho')}{{\rho'}^2}\,
d\rho'.
\end{equation}
The squared speed of sound is
\begin{equation}
\label{jfo5b}
c_s^2=P'(\rho)=\rho V''(\rho).
\end{equation}

\section{Classical collisionless self-gravitating  systems}
\label{sec_ccsgs}

In this Appendix and in the following one, we discuss certain aspects of the dynamical evolution of
classical collisionless self-gravitating  systems in order to facilitate the
comparison with the dynamical evolution of quantum self-gravitating  systems
treated in the main text.

\subsection{Vlasov equation }
\label{sec_vlasov}

A classical collisionless self-gravitating system (such as a stellar system
or such as CDM) is governed by
the
Vlasov-Poisson
equations 
\begin{eqnarray}
\label{vlasov1}
\frac{\partial f}{\partial t}+{\bf v}\cdot \frac{\partial f}{\partial {\bf
r}}-\nabla\Phi\cdot \frac{\partial f}{\partial {\bf
v}}=0,
\end{eqnarray}
\begin{eqnarray}
\label{vlasov2}
\Delta\Phi=4\pi G\int f\, d{\bf v},
\end{eqnarray}
where $f=f({\bf r},{\bf v},t)$ is the six-dimensional DF. The Vlasov
equation is also known as the
collisionless Boltzmann equation \cite{bt}. It states that, in the absence of
encounters (``collisions'') between the particles, the density (DF) of a
``fluid'' particle is conserved when we
follow its
motion in phase space, i.e., $Df/Dt=0$ where $D/Dt=\partial/\partial
t+{\bf v}\cdot \partial/\partial {\bf r}-\nabla\Phi\cdot\partial/\partial {\bf
v}$ is the material derivative (Stokes operator). As a result,
the Vlasov-Poisson equations conserve the energy $E=(1/2)\int f v^2\, d{\bf
r}d{\bf v}+(1/2)\int \rho\Phi\, d{\bf r}$ and an infinite class of Casimir
integrals of the form $\int h(f)\, d{\bf
r}d{\bf v}$ where $h(f)$ is an arbitrary function of $f$ \cite{lb,csr}.

{\it Remark:} The Vlasov equation can be viewed as the expression of
the Liouville theorem in the individual phase space. Under the circumstance in
which stellar encounters can be ignored, each star can be idealized as an
independent conservative system described by the  Hamiltonian
$H={\bf v}^2/2+\Phi({\bf r},t)$ yielding the equations of motion $d{\bf
r}/dt={\bf v}$, $d{\bf v}/dt=-\nabla\Phi$. The equation of continuity $\partial
f/\partial t+\nabla_6\cdot (f {\bf U}_6)=0$, where $\nabla_6=(\partial_{\bf
r},\partial_{\bf v})$ is a generalized nabla operator and ${\bf U}_6=({\bf
v},-\nabla\Phi)$ a generalized velocity field, and the fact that the flow in
phase space in incompressible, $\nabla_6\cdot {\bf U}_6=0$, leads to the
Liouville equation $\partial f/\partial t+{\bf U}_6\cdot \nabla_6 f=0$, which is
the Vlasov equation.

\subsection{Hydrodynamics of the Vlasov equation: Jeans equations
}
\label{sec_jeans}

From the  Vlasov equation, we can derive a system of
hydrodynamic equations called the Jeans equations.\footnote{Actually, the Vlasov
equation \cite{vlasov} was introduced by Jeans \cite{jeans} (see
\cite{henonvlasov}) and
the Jeans equations were introduced by Maxwell (see \cite{bt}).} By integrating
the Vlasov
equation (\ref{vlasov1}) over velocity, we get the continuity equation
(expressing the local
mass conservation)
\begin{equation}
\label{j1}
\frac{\partial\rho}{\partial t}+\nabla\cdot (\rho {\bf u})=0,
\end{equation}
where we have introduced the local density
\begin{equation}
\label{j2}
\rho=\int f\, d{\bf v}
\end{equation}
and the local velocity
\begin{equation}
\label{j3}
{\bf u}=\langle {\bf v}\rangle=\frac{1}{\rho}\int f {\bf v}\, d{\bf v}.
\end{equation}
Then, multiplying the Vlasov
equation (\ref{vlasov1}) by ${\bf v}$ and integrating over velocity, we obtain
the momentum equation
\begin{equation}
\label{j4}
\frac{\partial }{\partial t}(\rho u_i)+\frac{\partial}{\partial
x_j}\int f v_iv_j\, d{\bf
v}=-\rho\frac{\partial\Phi}{\partial x_i}.
\end{equation}
Introducing the difference ${\bf w}={\bf v}-{\bf u}$ between the velocity
${\bf v}$ of a particle and the mean velocity ${\bf u}({\bf r},t)$, and using
the fact that $\langle {\bf w}\rangle={\bf 0}$, we get
\begin{eqnarray}
\label{j5}
\int f v_iv_j\, d{\bf v}=\rho u_i u_j+\int  f
w_i w_j\, d{\bf
v}.
\end{eqnarray}
Therefore, the momentum equation (\ref{j4}) takes the form
\begin{equation}
\label{j6}
\frac{\partial }{\partial t}(\rho u_i)+\frac{\partial}{\partial
x_j}(\rho u_i u_j)=-\partial_j P_{ij}-\rho  \frac{\partial\Phi}{\partial
x_i},
\end{equation}
where we have introduced the pressure tensor ($-P_{ij}$ is the
stress tensor)
\begin{eqnarray}
\label{j7}
P_{ij}=\rho\langle
w_iw_j\rangle=\int f ({\bf v}-{\bf u})_i  ({\bf v}-{\bf
u})_j\, d{\bf v}.
\end{eqnarray}
It can be written as
\begin{eqnarray}
\label{j7b}
P_{ij}=\rho (\langle v_i v_j\rangle
-  \langle v_i \rangle  \langle v_j\rangle).
\end{eqnarray}
Using the continuity equation
(\ref{j1}), we obtain the identity 
\begin{equation}
\frac{\partial}{\partial t}(\rho {\bf
u})+\nabla(\rho {\bf u}\otimes {\bf u})=\rho\left\lbrack \frac{\partial {\bf
u}}{\partial t}+({\bf
u}\cdot \nabla){\bf
u}\right\rbrack.
\end{equation}
As a result, the momentum equation (\ref{j6}) can be rewritten as
\begin{equation}
\label{j8}
\frac{\partial {\bf u}}{\partial t}+({\bf u}\cdot \nabla){\bf
u}=-\frac{1}{\rho}\partial_jP_{ij}-\nabla\Phi.
\end{equation}
These equations are essentially
those for a compressible fluid which is supported by pressure in the form of a
velocity dispersion. These equations
are not closed because the pressure tensor $P_{ij}$ depends on the DF which is
not explicitly known in general. Actually, we can build up an
infinite hierarchy of equations by
introducing higher and higher moments of the velocity. In general, there is no
simple way to close this hierarchy of Jeans equations except in the single
speed case (see Appendix \ref{sec_euler}). In more general cases, some
approximations must be
introduced.

\subsection{Single-speed solution: pressureless Euler equations}
\label{sec_euler}

The Vlasov-Poisson equations admit a particular solution of the form
\begin{equation}
\label{e1}
f({\bf r},{\bf v},t)=\rho({\bf r},t)\delta({\bf v}-{\bf u}({\bf r},t)).
\end{equation}
This is called the single-speed solution because there is a single velocity
attached to any given point ${\bf r}$ at time $t$. It corresponds to the ``dust
model'' where the pressure is zero because there is no thermal motion (at a
given location all the particles have the same velocity). The
density $\rho({\bf r},t)$ and the
velocity ${\bf
u}({\bf r},t)$ satisfy the pressureless Euler 
equations
\begin{equation}
\label{e2}
\frac{\partial\rho}{\partial t}+\nabla\cdot (\rho {\bf u})=0,
\end{equation}
\begin{equation}
\label{e3}
\frac{\partial {\bf u}}{\partial t}+({\bf u}\cdot \nabla){\bf
u}=-\nabla\Phi,
\end{equation}
\begin{eqnarray}
\label{e4}
\Delta\Phi=4\pi G\rho.
\end{eqnarray}
These equations are exact. They can be deduced from the Jeans equations
(\ref{j1}) and (\ref{j8}) by closing the hierarchy with the condition
$P_{ij}=0$ obtained by substituting Eq. (\ref{e1}) into Eq. (\ref{j7}).
They correspond to very particular initial conditions where the particles at a
given location all have the same velocity. However, there is a well-known
difficulty with the solution (\ref{e1}).
Even if one starts with a DF of the form of Eq. (\ref{e1}) then, after
a finite time, the solution of the Vlasov-Poisson equations becomes multi-stream
because of
particle crossing. This leads to the formation of
caustics (singularities) in the density field at
shell-crossing. Therefore, Eq. (\ref{e1}) ceases to be valid.  This phenomenon
renders the pressureless hydrodynamical
description (\ref{e2})-(\ref{e4}) useless beyond the first time of
crossing when the fast particles cross the slow ones. Therefore, after
shell-crossing, the  pressureless Euler equations are
not defined anymore and we must come
back to the original Vlasov-Poisson equations, or to the Jeans
equations, because we need to account for a velocity dispersion. Indeed, the
velocity field becomes multi-valued even if, initially, it is single-valued.
The  pressureless Euler equations are only valid until
shell
crossing and they fail as soon as orbit crossing
(multistreaming) occurs. 

Different attempts to cure this problem have been proposed in order to continue using a hydrodynamical
model.

(i) A first heuristic possibility to avoid multi-streaming is to introduce a
viscosity term $\nu\Delta {\bf u}$ in the momentum equation (\ref{e3}) yielding
the  Navier-Stokes equation
\begin{equation}
\label{e5}
\frac{\partial {\bf u}}{\partial t}+({\bf u}\cdot \nabla){\bf
u}=-\nabla \Phi+\nu\Delta {\bf u}.
\end{equation}
In order for the diffusion term to have a smoothing effect only in the
regions where particle-crossing is about to occur, the viscosity $\nu$
should be small. More precisely, the limit $\nu\rightarrow 0$ should be
taken, which is different from setting $\nu=0$.\footnote{In an expanding
universe, using the  Zel'dovich
approximation, the pressureless Euler equations can be reduced to the
so-called Burgers equation \cite{vergassola}. On the other hand, the 
Navier-Stokes equation can be reduced to the viscous Burgers equation which
corresponds to the adhesion model \cite{gurbatov} of cosmology. This
model is relevant provided that the viscosity is sufficiently small
($\nu\rightarrow 0$).}

(ii) A second
heuristic possibility is to introduce a pressure term $(-1/\rho)\nabla P$  in
the momentum equation (\ref{e3}) yielding
\begin{equation}
\label{e6}
\frac{\partial \mathbf{u}}{\partial t}
+ (\mathbf{u} \cdot\nabla) \mathbf{u}
= -\nabla \Phi - \frac{1}{\rho} \nabla P,
\end{equation}
where $P$ is the fluid pressure, a local quantity given by a specified equation
of state which takes into account velocity dispersion. This amounts to closing
the hierarchy of Jeans equations with the
isotropy ansatz $P_{ij}=P(\rho) \delta_{ij}$. In this manner, there is no
shell-crossing singularities. The velocity dispersion giving rise to the
pressure can be a consequence of
the multi-streaming or it can be already present in the initial condition. We
need $P\rightarrow 0$ at large scales to recover the CDM model and $P\neq 0$ at
small
scales in order to avoid singularities.\footnote{Sensible expressions
of $P(\rho)$ have been considered by Buchert {\it et
al.} \cite{buchert}. In particular, they singled
out an equation of state of the form $P(\rho)\propto \rho^2$ that leads, under
certain assumptions, to a viscous Burgers-like equation in cosmology.
Interestingly, as discussed in \cite{prd3}, this expression
is similar to the pressure created by self-interacting bosons in the BECDM
model.}

(iii) A third heuristic possibility, proposed by Widrow and Kaiser \cite{wk},
is to replace the Vlasov
equation by  a wave equation having the form of a Schr\"odinger
equation with
an effective Planck constant $\hbar_{\rm eff}$ controlling the spatial
resolution.\footnote{The Schr\"odinger equation is equivalent to the Wigner
equation. In the Schr\"odinger equation, $\psi({\bf r},t)$ encodes both position
and momentum
information in a single position-space function. It is argued that, when
$\hbar_{\rm eff}\rightarrow 0$, the
Vlasov-Poisson equations are recovered and that a finite value of $\hbar_{\rm
eff}$
provides a small-scale regularization of the dynamics. In that case, the
Schr\"odinger-Poisson
system has nothing to do with quantum mechanics since it aims at describing the
evolution of classical collisionless matter under the influence of gravity.}
Using the Madelung transformation, this prescription leads, instead
of  Eq.  (\ref{e3}), to a momentum equation of the form
\begin{equation}
\label{e7}
\frac{\partial {\bf u}}{\partial t}+({\bf u}\cdot \nabla){\bf
u}=-\nabla\Phi-\frac{1}{\rho}\partial_jP_{ij}^{Q},
\end{equation}
with an effective  ``quantum'' pressure tensor
\begin{equation}
\label{e8}
P_{ij}^{Q}=-\frac{\hbar_{\rm eff}^2}{4m^2}\rho\,
\partial_i\partial_j\ln\rho=\frac{\hbar_{\rm eff}^2}{4m^2}\left (\frac{1}{\rho}
\partial_i\rho\partial_j\rho-\partial_i\partial_j\rho\right ).
\end{equation}
Interestingly, the effective quantum pressure tensor $P_{ij}^Q$ bares some
resemblance with the Jeans pressure tensor $P_{ij}$ of
Eq. (\ref{j7b})
with the velocity average operators replaced by density gradients (a nonlocal
quantity). This amounts to closing the hierarchy of Jeans
equations with the condition $P_{ij}=P_{ij}^Q$.\footnote{For
self-gravitating BECs described by the ``true''
Schr\"odinger equation, this
identification is exact (see Sec. \ref{sec_qja}). Inversely, this identification may
suggest an interpretation of the quantum potential $P_{ij}^Q$ in terms of
a classical
kinetic theory.} Therefore, although the system is classical, the procedure of
Widrow and
Kaiser \cite{wk} amounts to  closing the Jeans
equations by introducing an effective quantum
potential of order $O(\hbar_{\rm eff}^2/m^2)$. When
$\hbar_{\rm eff}/m\rightarrow 0$ the
effective quantum pressure  $P_{ij}^Q$ tens to zero at large scales
while $P_{ij}^Q\neq 0$ at small
scales (in line with the fact that
quantum mechanics is negligible at large scales and important at small scales in
BECDM), thereby preventing singularities. This is exactly what is required. The
effective quantum pressure tensor acts as a regularizer of caustics and
singularities
in classical solutions. The
quantum pressure also replaces the role of the viscosity in the adhesion model
(see footnote 42) \cite{sc}. Therefore, the Schr\"odinger
method can handle multiple streams in phase space.

(iv) A fourth possibility is to use the heuristic kinetic theory of violent
relaxation
developed by Chavanis {\it et al.} \cite{csr} (see also the Appendix of
\cite{moczSV}). In that case, the 
pressureless Euler equation (\ref{e3}) is replaced by the
damped Euler equation
\begin{equation}
\label{e10}
\frac{\partial {\bf u}}{\partial t}+({\bf u}\cdot \nabla){\bf
u}=-\nabla\Phi-\frac{1}{\rho}\nabla P_{\rm LB}-\xi{\bf u},
\end{equation}
where $P_{\rm LB}$ is the Lynden-Bell pressure and $\xi$ is a friction
coefficient
accounting for nonlinear Landau damping.

In these different hydrodynamic models, we can go beyond the first time
of crossing. Therefore, these hydrodynamic models are defined for all times. The
solution of these hydrodynamic equations is expected to remain close to the
solution of the
Vlasov-Poisson equations for all times provided
that $\nu\rightarrow 0$, $P\rightarrow 0$, and $\hbar_{\rm eff}/m\rightarrow 0$
in these respective models.
We note that the limits $\nu,P,\hbar_{\rm eff}/m\rightarrow 0$ are crucially
different
from taking  $P=\nu=\hbar_{\rm eff}/m=0$.

{\it Remark:} The Vlasov-Poisson equations (\ref{vlasov1}) and (\ref{vlasov2})
are valid for all times. Similarly, the
Jeans equations (\ref{j1})-(\ref{j8}), which are equivalent to the Vlasov
equation, are valid for all times but they are not closed (we have to consider
the whole hierarchy). By contrast, the pressureless Euler-Poisson equations
(\ref{e2})-(\ref{e4}) are
valid only until the first time of crossing. Before that time they coincide with
the Vlasov-Poisson equations and after that time they break down. The modified
Euler-Poisson equations (\ref{e5})-(\ref{e10}) are valid for all times. If
$\nu\rightarrow 0$, $P\rightarrow 0$ and $\hbar_{\rm eff}\rightarrow 0$,
they are expected to be close to the Vlasov-Poisson equations. In particular,
Eqs. (\ref{e7})-(\ref{e8}) are equivalent to the Schr\"odinger and Wigner
equations, which are themselves equivalent to the Vlasov equation when
$\hbar_{\rm eff}\rightarrow 0$.

\section{Violent relaxation of classical collisionless self-gravitating systems}
\label{sec_vrx}

The Vlasov-Poisson equations (\ref{vlasov1}) and (\ref{vlasov2}) describing
classical collisionless stellar
systems are known to experience a
process of
violent relaxation \cite{lb} caused by the rapid fluctuations of the strongly varying
gravitational potential at the early stage of galaxy formation. While the
fine-grained DF ${f}({\bf r},{\bf v},t)$ always evolves
in time, forming intermingled filaments  in phase space at smaller and smaller
scales, the coarse-grained DF $\overline{f}({\bf r},{\bf v},t)$, which smoothes
out this intricate
filamentation, rapidly relaxes
 towards a quasistationary state  $\overline{f}_{\rm QSS}({\bf r},{\bf v})$.
This process takes
place on a very short timescale of the order of the dynamical time $t_D$. For
$t>t_D$, the evolution of ${f}({\bf r},{\bf v},t)$ occurs
on scales smaller than the coarse-graining mesh. This complicated
dynamics is associated with phase mixing, violent relaxation and nonlinear
Landau damping. As a result, the coarse-grained DF $\overline{f}({\bf r},{\bf
v},t)$ does
not satisfy the Vlasov equation. Phase-space correlations introduce an effective ``collision'' term
${\cal C}[\overline{f}]$ on the right hand side of the coarse-grained Vlasov
equation. This collision term drives the relaxation of the coarse-grained
DF towards  the quasistationary state. The determination of the quasistationary
state $\overline{f}_{\rm QSS}({\bf
r},{\bf v})$ and of the collision term ${\cal C}[\overline{f}]$ is a problem of
fundamental interest but also of great difficulty because of the very nonlinear
nature of the process. Here, we tackle this problem
through a thermodynamical approach.  We use a MEP
to determine the quasistationary state 
$\overline{f}_{\rm QSS}({\bf r},{\bf v})$ and we use a MEPP to
determine the collision term
${\cal C}(\overline{f})$.

\subsection{Maximum entropy principle }
\label{sec_vra}

In a seminal paper, Lynden-Bell \cite{lb} argued that the coarse-grained DF
$\overline{f}({\bf r},{\bf v},t)$ violently relaxes towards a quasistationary
state 
$\overline{f}_{\rm QSS}({\bf r},{\bf v})$ which maximizes a suitable mixing 
entropy while taking into
account
all the constraints of the Vlasov equation.\footnote{The
Lynden-Bell
entropy is proportional to the logarithm of 
the number of microstates corresponding to a given macrostate. This is a
measure of disorder.
Therefore, the maximization of $S$ under constraints determines the most
probable state of the system, i.e., the macrostate that is the most represented
at the ``microscopic'' level.} In the two-levels approximation
of the theory, where the fine-grained DF ${f}({\bf r},{\bf v},t)$ takes only
two values $f=\eta_0$ and
$f=0$,\footnote{The general case is treated in \cite{lb,csr}.} the
constraints reduce to the conservation of mass and energy and the Lynden-Bell
entropy writes
\begin{equation}
\label{mep1}
S=-\eta_0\int \biggl\lbrace {\overline{f}\over \eta_{0}}\ln
{\overline{f}\over \eta_{0}}+\biggl
(1-{\overline{f}\over \eta_{0}}\biggr )\ln \biggl (1-{\overline{f}\over
\eta_{0}}\biggr )
\biggr\rbrace \, d{\bf r} d{\bf v}.
\end{equation}
It can be obtained from a combinatorial analysis taking into account the
specificities of the Vlasov equation. In particular, the coarse-grained
DF must satisfy the inequality $\overline{f}({\bf r},{\bf v},t)\le \eta_{0}$ arising from the
incompressibility of the flow in phase space and the conservation of the DF on
the fine-grained scale by the Vlasov equation.\footnote{More generally, the
coarse-grained DF must always be smaller than the maximum value of
the fine-grained (or initial) DF.} This constraint is similar to the Pauli
exclusion principle in
quantum mechanics (with another interpretation) and this is why the
Lynden-Bell entropy
(\ref{mep1}) resembles the Fermi-Dirac entropy of quantum mechanics. In this
sense, the
process of violent relaxation is similar in some respects to the collisional relaxation of
self-gravitating fermionic particles. The Lynden-Bell DF
corresponds to a fourth type of statistics corresponding to distinguishable 
particles experiencing an exclusion principle \cite{lb}.

According to the MEP, the  DF of the quasistationary state
 $\overline{f}_{\rm QSS}$ maximizes the Lynden-Bell entropy (\ref{mep1}) at
fixed
mass
\begin{equation}
\label{mep2}
M=\int \rho \, d{\bf r}
\end{equation}
and energy
\begin{equation}
\label{mep3}
E=\int{1\over 2}f v^{2} \, d{\bf  r}
 d{\bf  v}+{1\over 2}\int \rho \Phi \, d{\bf  r}=K+W,
\end{equation}
where $K$ is the kinetic energy and $W$ the potential (gravitational) energy.
Introducing Lagrange multipliers and writing the
variational principle under the form
\begin{equation}
\label{mep4}
\delta S-\beta\delta E+\alpha\delta M=0,
\end{equation}
we find that the extrema of entropy at fixed mass and energy correspond to the
Lynden-Bell DF
\begin{eqnarray}
\label{mep5}
\overline{f}_{\rm LB}({\bf r},{\bf v})=\frac{\eta_0}{1+e^{\beta (v^2/2+\Phi({\bf
r}))-\alpha}},
\end{eqnarray}
where $\epsilon={v^{2}/2}+\Phi({\bf r})$ is the energy of a particle
by unit of mass.  The Lagrange multipliers $\beta=1/T_{\rm eff}$ and
$\alpha=\mu_{\rm eff}/T_{\rm eff}$ are the effective inverse temperature and the
effective chemical
potential (divided by the effective temperature). The Lynden-Bell DF
(\ref{mep5})
which maximizes the Lynden-Bell entropy at fixed mass and energy is the most
probable, or most mixed, state taking into account all the constraints of the
Vlasov equation. The Lynden-Bell DF is similar to the Fermi-Dirac DF in
quantum mechanics provided
that we make the correspondence
\begin{eqnarray}
\label{mep6}
\eta_0=\frac{gm^4}{h^3},
\end{eqnarray}
where $g=2s+1$ is the multiplicity of the quantum states. In particular, the
Lynden-Bell DF (\ref{mep5}) satisfies the
constraint $\overline{f}_{\rm LB}({\bf r},{\bf v})\le \eta_{0}$ which is
similar to the Pauli exclusion
principle in quantum mechanics. We note that the effective temperature $T_{\rm eff}$ has
not the dimension of a temperature. It should rather be
interpreted as a velocity dispersion. However, we shall use this notation which
is more transparent. We also note that the mass $m$ of the particles does not appear in the
Lynden-Bell theory since it is based on the Vlasov equation for collisionless
systems which is independent of the mass of the particles. In this sense, we
can say that the temperature in
Lynden-Bell's theory is proportional to the  mass of the particles \cite{lb}.

From the Lynden-Bell DF (\ref{mep5}), one can determine the density and the
pressure through the equations
\begin{eqnarray}
\label{mep7}
\rho=\int \overline{f}\, d{\bf v}=\int_0^{+\infty} \frac{\eta_0}{1+e^{\beta
(v^2/2+\Phi({\bf
r}))-\alpha}} 4\pi v^2\, dv,
\end{eqnarray}
\begin{equation}
\label{mep8}
P=\frac{1}{3}\int \overline{f} v^2\, d{\bf v}=\frac{1}{3}\int_0^{+\infty}
\frac{\eta_0}{1+e^{\beta
(v^2/2+\Phi({\bf
r}))-\alpha}} \, 4\pi v^4\, dv.
\end{equation}
They can be rewritten as
\begin{eqnarray}
\label{mep9}
\rho({\bf r})=\frac{4\pi\eta_0\sqrt{2}}{\beta^{3/2}}I_{1/2}\left\lbrack
e^{\beta\Phi({\bf r})-\alpha}\right\rbrack,
\end{eqnarray}
\begin{eqnarray}
\label{mep10}
P({\bf r})=\frac{8\pi\eta_0\sqrt{2}}{3\beta^{5/2}}I_{3/2}\left\lbrack
e^{\beta\Phi({\bf r})-\alpha}\right\rbrack,
\end{eqnarray}
where
\begin{eqnarray}
\label{mep11}
I_n(t)=\int_0^{+\infty}\frac{x^n}{1+te^x}\, dx
\end{eqnarray}
denote the Fermi-Dirac integrals. Eliminating formally the gravitational
potential $\Phi({\bf r})$ between Eqs. (\ref{mep9}) and (\ref{mep10}), we see
that the equation of state is barotropic: $P({\bf
r})=P[\rho({\bf r})]$. Eqs. (\ref{mep9}) and (\ref{mep10}) determine the
Lynden-Bell
equation of state $P_{\rm LB}(\rho)$ in parametric form. The Lynden-Bell
equation of state is formally similar to the Fermi-Dirac equation of state.
Using Eq. (\ref{mep8}), the kinetic energy can be
written as
\begin{eqnarray}
\label{kinlb}
K=\frac{3}{2}\int P_{\rm LB}(\rho)\, d{\bf r}.
\end{eqnarray}
On the other hand, the
Lynden-Bell DF
(\ref{mep5}), and more generally any DF of the form
$f=f(\epsilon)$, implies the condition of hydrostatic equilibrium (see, e.g.,
\cite{gr1} and the Remark of Appendix \ref{sec_vb})
\begin{eqnarray}
\label{mep11c}
\nabla P+\rho\nabla\Phi={\bf 0}.
\end{eqnarray}

In the completely degenerate limit $\overline{f}_{\rm LB}\sim \eta_0$, the
Lynden-Bell DF
reduces to a step function
\begin{eqnarray}
\label{mep11b}
f_{\rm LB}({\bf r},{\bf v})=\eta_0
H(\epsilon-\epsilon_{\rm LB}),
\end{eqnarray}
where $H$ is the Heaviside function ($H(x)=1$ if $x<1$ and $H(x)=0$ if $x>1$)
and $\epsilon_{\rm LB}$ is 
the Lynden-Bell energy, which is the counterpart of the Fermi energy. In that
case,
Eqs.
(\ref{mep7}) and (\ref{mep8}) reduce to
\begin{eqnarray}
\label{mep12}
\rho=\int_0^{v_{\rm LB}} \eta_0  4\pi v^2\, dv=4\pi\eta_0 \frac{v_{\rm
LB}^3}{3},
\end{eqnarray}
\begin{equation}
\label{mep13}
P=\frac{1}{3}\int_0^{v_{\rm LB}} \eta_0 v^2 4\pi v^2\, dv=\frac{4\pi\eta_0}{3}
\frac{v_{\rm LB}^5}{5},
\end{equation}
where $v_{\rm LB}({\bf r})=\sqrt{2(\epsilon_{\rm LB}-\Phi({\bf r}))}$ is
the Lynden-Bell velocity. Eqs. (\ref{mep12}) and (\ref{mep13}) lead to the
equation of state
\begin{equation}
\label{mep14}
P=\frac{1}{5}\left (\frac{3}{4\pi\eta_0}\right )^{2/3}\rho^{5/3}.
\end{equation}
This is a polytropic equation of state of index $n=3/2$ like the one
arising in the theory of nonrelativistic white dwarf stars at $T=0$
corresponding to the ground state of the
self-gravitating Fermi gas \cite{chandra}.

In the nondegenerate, or dilute,  limit $\overline{f}_{\rm LB}\ll
\eta_{0}$, the Lynden-Bell entropy becomes similar to the Boltzmann entropy
\begin{equation}
\label{mep15}
S\simeq -\int \overline{f} \left\lbrack \ln
\left (\frac{\overline{f}}{\eta_{0}}\right )-1\right\rbrack \, d{\bf r} d{\bf
v},
\end{equation}
the Lynden-Bell DF becomes similar to the Maxwell-Boltzmann DF
\begin{eqnarray}
\label{mep16}
\overline{f}_{\rm LB}({\bf r},{\bf v})\simeq \eta_0 e^{\alpha-\beta
\lbrack v^2/2+\Phi({\bf
r})\rbrack},
\end{eqnarray}
and the Lynden-Bell equation of state becomes similar to the equation of state
of an
isothermal gas
\begin{eqnarray}
\label{mep17}
P=\rho T_{\rm eff}.
\end{eqnarray}
This equation of state has been studied in relation to isothermal
stars \cite{chandra} and to the statistical mechanics of ``collisional''
self-gravitating systems relaxing under the effect of gravitational encounters
\cite{paddy}. Remarkably, the Lynden-Bell theory of violent
relaxation explains how a collisionless self-gravitating system
can ``thermalize''
on a very short timescale, much shorter that the collisional relaxation time
$t_{\rm relax} \sim (N/\ln N)t_D$, without the need of ``collisions'' \cite{lb}.

So far, we have assumed that the system is isolated so that it
conserves the mass and the energy. This corresponds to  the microcanonical
ensemble. We now consider the canonical ensemble in which the temperature
$T_{\rm eff}=1/\beta$ is fixed instead of the energy. In that case, the
equilibrium state is obtained by minimizing the free energy $F=E-T_{\rm eff}S$
at fixed mass $M$ or, equivalently, by maximizing the Massieu function
$J=S-\beta E$ at fixed mass $M$. The variational principle determining the
extrema of free energy at fixed mass writes
\begin{equation}
\label{mep18}
\delta J+\alpha\delta M=0.
\end{equation}
Since $\beta$ is fixed, this variational principle for the first variations of
the thermodynamical potential is equivalent to Eq. (\ref{mep4}) and it returns
the Lynden-Bell DF (\ref{mep5}). Therefore, the equilibrium states are the same
in the
microcanonical and canonical ensembles (the extrema of entropy at fixed mass and
energy coincide with the extrema of free energy at fixed mass). However, their
thermodynamical stability (related to the sign of the second variations of
the  thermodynamical potential) may be different in the microcanonical and
canonical ensembles. This is the notion of ensembles inequivalence for systems
with long-range interactions \cite{paddy,ijmpb,katzrevue,campa}. An equilibrium
state is
thermodynamically stable in the microcanonical ensemble if it is a maximum of
entropy at fixed mass and energy. This corresponds to
\begin{eqnarray}
\label{mep19}
\delta^2 J=\delta^2 S-\beta \delta^2 E
=-\int\frac{1}{\overline{f}(1-\overline{f}/\eta_0)}\frac{(\delta f)^2}{2}\, d{\bf r}d{\bf v}\nonumber\\
-{1\over 2}\beta\int \delta\rho \delta\Phi \, d{\bf  r}\le 0\qquad
\end{eqnarray}
for all perturbations $\delta f$ that conserve mass and energy at first order.
An equilibrium state is thermodynamically stable in the canonical ensemble if it
is a minimum of free energy at fixed mass. This corresponds to the inequality of
Eq. (\ref{mep19}) for all perturbations $\delta f$ that conserve mass. We note
that canonical stability implies microcanonical stability but the converse is
wrong \cite{cc}. For example, it is shown in \cite{mcmh} that  the core-halo
solution with a negative specific heat is stable in the microcanonical
ensemble while it is unstable in the canonical ensemble.

\subsection{Maximum entropy production principle }
\label{sec_vrb}

We now consider the  dynamical evolution of the coarse-grained
DF $\overline{f}({\bf r},{\bf v},t)$.  Writing
$f=\overline{f}+\delta f$ and $\Phi=\overline{\Phi}+\delta \Phi$, where $\delta
f$ and $\delta\Phi$ denote fluctuations about the coarse-grained fields, and
taking the local average of the Vlasov equation (\ref{vlasov1}), we get
\begin{eqnarray}
\label{mepp0}
\frac{\partial \overline{f}}{\partial t}+{\bf v}\cdot \frac{\partial
\overline{f}}{\partial {\bf
r}}-\nabla {\Phi}\cdot \frac{\partial \overline{f}}{\partial {\bf
v}}=\frac{\partial}{\partial {\bf v}}\cdot \overline{\delta f\nabla\delta\Phi}.
\end{eqnarray}
This equation shows that the correlations of the fluctuations of the
gravitational potential and DF create an effective
``collision'' term ${\cal C}[\overline{f}]$ in the r.h.s. of Eq.
(\ref{mepp0}). In Ref. \cite{csr} we obtained an explicit
expression of this collision term by using heuristic arguments based on a
MEPP.\footnote{Since the process of violent relaxation is very
nonlinear, we cannot in principle use perturbation methods (see, however,
\cite{kp,sl,chavmnras,kingen,dubrovnik} in a regime of ``gentle''
relaxation). We thus capitalize our ignorance and assume that the system evolves
so
as to maximize its rate of entropy production at fixed mass and
energy. This thermodynamic principle is expected to determine the most probable
evolution of the system.} We considered the general
case where
the fine-grained DF may take an arbitrary number of values. Below, we detail
this procedure in the simpler case where the fined-grained DF takes only two
values $f=0$ and $f=\eta_0$ as in the preceding section.

To apply the MEPP, we first write the relaxation equation for
the coarse-grained DF under the form
\begin{eqnarray}
\label{mepp1}
\frac{\partial \overline{f}}{\partial t}+{\bf v}\cdot \frac{\partial
\overline{f}}{\partial {\bf
r}}-\nabla {\Phi}\cdot \frac{\partial \overline{f}}{\partial {\bf
v}}=-\frac{\partial}{\partial {\bf v}}\cdot {\bf J},
\end{eqnarray}
where ${\bf J}$ is the current to be determined. The form of Eq. (\ref{mepp1})
ensures
the conservation of mass provided that ${\bf J}$ decreases sufficiently rapidly
for large $|{\bf v}|$. From Eqs. (\ref{mep1}), (\ref{mep3}) and (\ref{mepp1}),
we get
\begin{eqnarray}
\label{mepp3}
\dot S=-\int \frac{1}{\overline{f}(1-\overline{f}/\eta_0)} {\bf J}\cdot
\frac{\partial \overline{f}}{\partial {\bf v}}\, d{\bf r}d{\bf v},
\end{eqnarray}
\begin{eqnarray}
\label{mepp2}
\dot E=\int {\bf J}\cdot {\bf v}\, d{\bf r}d{\bf v},
\end{eqnarray}
where we have used straightforward  integrations  by parts. Following
the MEPP, we shall determine the optimal current ${\bf J}$ which
maximizes the rate of entropy production (\ref{mepp3}) while satisfying the
conservation of energy $\dot E=0$. For this problem to have a solution, we shall
also impose a limitation on the current $|{\bf J}|$ characterized by a
bound $C({\bf r},{\bf v},t)$ which exists but is not explicitly known, so that
\begin{equation}
\label{mepp4}
{J^{2}\over 2f}\le C({\bf r},{\bf v},t).
\end{equation}
It can be shown by a convexity argument that reaching the bound (\ref{mepp4}) is
always favorable for increasing $\dot S$, so this constraint can be
replaced by an equality. The variational problem can then be solved by
introducing at each time $t$ Lagrange multipliers $\beta(t)$ and $1/{\tilde
D}({\bf
r},{\bf v},t)$ for the two constraints. The condition
\begin{equation}
\label{mepp5}
\delta\dot S-\beta(t)\delta\dot E-\int {1\over {\tilde D}}\delta \biggl
({J^{2}\over
2f}\biggr )\, d{\bf r}d{\bf v}=0
\end{equation}
yields an optimal current of the form
\begin{equation}
\label{mepp6}
{\bf J}=-D\biggl\lbrack {\partial \overline{f}\over\partial {\bf
v}}+\beta(t)\overline{f}(1-\overline{f}/\eta_0){\bf v}\biggr \rbrack,
\end{equation}
where we have set $\tilde D=D(1-\overline{f}/\eta_0)$ to avoid divergences when $\overline{f}\rightarrow \eta_0$.\footnote{We have some freedom on the determination of the diffusion coefficient since it is related to a constraint that is not explicitly known.} The time evolution of the Lagrange multiplier $\beta(t)$ is determined by the
conservation of energy $\dot E=0$, introducing Eq. (\ref{mepp6}) into the
constraint
(\ref{mepp2}). This yields
\begin{equation}
\label{mepp7}
\beta(t)=-{\int D {\partial \overline{f}\over\partial {\bf v}}\cdot {\bf
v}\, d{\bf r}d{\bf v}\over \int D \overline{f}(1-\overline{f}/\eta_0) v^{2}\,
d{\bf r}d{\bf v}}.
\end{equation}
Note that the optimal
current (\ref{mepp6}) can be written as
\begin{equation}
\label{mepp8}
{\bf J}=-{\tilde D} \frac{\partial\alpha}{\partial {\bf v}},
\end{equation}
where
\begin{equation}
\label{mepp9}
\alpha({\bf r},{\bf
v},t)\equiv \ln\left (\frac{\overline{f}/\eta_0}{1-\overline{f}/\eta_0}\right
)+\beta(t)\left \lbrack\frac{v^2}{2}+\Phi({\bf r},t)\right \rbrack
\end{equation}
is a time-dependent chemical potential which is uniform at equilibrium
according to Eq. (\ref{mep5}). The relation from Eq. (\ref{mepp8}) then 
corresponds to the linear thermodynamics of Onsager \cite{onsager31a,onsager31b}
where the currents
are proportional to the gradients of the thermodynamic potentials that are
uniform at statistical equilibrium. Therefore, the MEPP can be viewed as a
variational formulation of Onsager's linear thermodynamics
\cite{nfp}.\footnote{See Ref. \cite{entropy} for a discussion of the 
connection between the MEPP and the minimization of the Onsager-Machlup
\cite{om} functional.}

Introducing the optimal current (\ref{mepp6}) into Eq. (\ref{mepp1}), we obtain
the relaxation  equation
\begin{eqnarray}
\label{mepp10}
&&\frac{\partial \overline{f}}{\partial t}+{\bf v}\cdot \frac{\partial
\overline{f}}{\partial {\bf
r}}-\nabla {\Phi}\cdot \frac{\partial \overline{f}}{\partial {\bf
v}}\nonumber\\
&=&\frac{\partial}{\partial {\bf v}}\left\lbrack D\left
(\frac{\partial
\overline{f}}{\partial {\bf
v}}+\beta(t) \overline{f}(1-\overline{f}/\eta_0){\bf v}\right )\right\rbrack.
\end{eqnarray}
Morphologically, this relaxation equation has the form of a nonlinear Fokker-Planck equation or, more
precisely, the form of a fermionic Kramers equation \cite{gen,nfp}. The
first term is a diffusion term  and the
second term is a friction term. The fermionic factor
$\overline{f}(1-\overline{f}/\eta_0)$ takes into account the Lynden-Bell
exclusion
principle. The function $\beta(t)$
can be considered as a time dependant inverse temperature evolving
with time so as to conserve the energy (microcanonical formulation).\footnote{In
the nondegenerate 
limit $\overline{f}\ll \eta_0$, Eq. (\ref{mepp10}) is similar to the classical
Kramers equation. On the other hand, assuming $D={\rm cst}$, Eq. (\ref{mepp7})
can be written as $\beta(t)=3M/2K(t)=1/\langle T_{\rm kin}\rangle(t)$ where
$\langle T_{\rm kin}\rangle(t)$ is the average kinetic temperature of the system
(see Appendix \ref{sec_ol}).} The
friction coefficient $\xi$ satisfies a generalized Einstein relation: $\xi=D\beta$.
Note that $D$ is not determined by the MEPP since it is related to the unknown
bound $C({\bf r},{\bf v},t)$ in Eq. (\ref{mepp4}).

It is straightforward to check that Eq. (\ref{mepp10}) with the constraint
(\ref{mepp7}) satisfies an $H$-theorem for the Lynden-Bell entropy
(\ref{mep1}). From Eq. (\ref{mepp3}), we can write
\begin{eqnarray}
\label{mepp11}
\dot S=-\int {1\over \overline{f}(1-\overline{f}/\eta_0)}{\bf J}\cdot
\biggl\lbrack
{\partial \overline{f}\over\partial {\bf
v}}+\beta(t)\overline{f}(1-\overline{f}/\eta_0){\bf v}\biggr \rbrack \, d{\bf
r}d{\bf v}\nonumber\\
+\beta(t) \int {\bf J}\cdot {\bf v} \, d{\bf r}d{\bf v}.\qquad
\end{eqnarray}
The last integral vanishes due to the conservation of the energy (see Eq.
(\ref{mepp2}) with $\dot E=0$). Using Eq. (\ref{mepp6}), we obtain
\begin{equation}
\label{mepp12}
\dot S=\int {{\bf J}^{2}\over D\overline{f}(1-\overline{f}/\eta_0)}\, d{\bf
r}d{\bf v},
\end{equation}
which is positive ($\dot S\ge 0$) provided that $D>0$. This proves the
$H$-theorem. At equilibrium, we have $\dot S=0$ which implies
${\bf J}={\bf 0}$. Then, according to Eq. (\ref{mepp6}), we obtain
\begin{equation}
\label{mepp13}
{\partial \overline{f}\over\partial {\bf
v}}+\beta\overline{f}(1-\overline{f}/\eta_0){\bf v}={\bf 0}.
\end{equation}
Integrating this equation with respect to ${\bf v}$, we get
\begin{equation}
\label{mepp14}
\ln\left (\frac{\overline{f}/\eta_0}{1-\overline{f}/\eta_0}\right )+\beta
{v^{2}\over 2}=A({\bf r}),
\end{equation}
where $A({\bf r})$ is a constant of integration that may depend on ${\bf r}$.
Taking the gradient of the foregoing equation with respect to ${\bf r}$, we find that
\begin{equation}
\label{mepp15}
\frac{1}{\overline{f}(1-\overline{f}/\eta_0)}{\partial
\overline{f}\over\partial {\bf
r}}=\nabla A({\bf r}).
\end{equation}
On the other hand, since $\partial_t\overline{f}=0$ and ${\bf J}={\bf 0}$, the advection  term in Eq.
(\ref{mepp1}) cancels out:
\begin{eqnarray}
\label{mepp16}
{\bf v}\cdot \frac{\partial
\overline{f}}{\partial {\bf
r}}-\nabla {\Phi}\cdot \frac{\partial \overline{f}}{\partial {\bf
v}}=0.
\end{eqnarray}
Together with Eqs. (\ref{mepp13}) and (\ref{mepp15}), this implies the relation
\begin{eqnarray}
\label{mepp17}
{\bf v}\cdot (\nabla A+\beta \nabla\Phi)=0.
\end{eqnarray}
This relation must be true for all ${\bf v}$ so that
\begin{eqnarray}
\label{mepp18}
\nabla A+\beta \nabla\Phi={\bf 0},
\end{eqnarray}
which can be integrated into
\begin{eqnarray}
\label{mepp19}
A({\bf
r})=-\beta\Phi({\bf r})+\alpha,
\end{eqnarray}
where $\alpha$ is a constant. Substituting  Eq. (\ref{mepp19}) into Eq.  (\ref{mepp14}), we finally recover the
Lynden-Bell DF (\ref{mep5}) with $\beta=\lim_{t\rightarrow +\infty}\beta(t)$.
Therefore, the
stationary
solutions of the generalized Fokker-Planck
equation (\ref{mepp10}) are the Lynden-Bell DFs which extremize the Lynden-Bell
entropy at
fixed energy and
mass.  In addition, one can show that an equilibrium state is linearly
dynamically stable if, and
only if, it is a  (local) {\it maximum} of $S$ at fixed $M$.\footnote{This
result can also be directly obtained from the
$H$-theorem using Lyapunov's direct method.}  Indeed, considering the linear dynamical
stability of a stationary
solution of Eqs. (\ref{mepp7}) and (\ref{mepp10}), we can derive the general
relation \cite{gen}
\begin{eqnarray}
\label{mepp20}
2\lambda\delta^{2}{J}=\delta^{2}\dot S\ge 0,
\end{eqnarray}
connecting the growth rate $\lambda$ of the perturbation $\delta f\sim
e^{\lambda t}$ to the second order variations of the free energy (Massieu function)
$J=S-\beta E$ and the second order variations of the rate of entropy
production $\delta^{2}\dot S$.  Since the product
$\lambda\delta^{2}{J}$ is positive because $\delta^{2}\dot S\ge 0$ according to
Eq. (\ref{mepp12}), we conclude that a stationary
solution of Eqs. (\ref{mepp7}) and (\ref{mepp10}) is
linearly dynamically stable ($\lambda<0$) if, and only if, it is an entropy
maximum at fixed mass and energy ($\delta^2 J<0$). This
aesthetic formula shows the equivalence between dynamical and
thermodynamical stability for the generalized Fokker-Planck
equation (\ref{mepp10}) with the constraint from Eq. (\ref{mepp7}). Therefore,
this equation can only relax towards maxima of $S$, not towards minima
or saddle points.

A relaxation equation of the form of Eq. (\ref{mepp1}) appropriate to the canonical ensemble can be obtained by
maximizing the rate of free energy dissipation $\dot J=\dot S-\beta\dot E$ with the constraint (\ref{mepp4}). The corresponding variational principle
\begin{equation}
\label{mepp21}
\delta\dot J-\int {1\over {\tilde D}}\delta \biggl ({J^{2}\over 2f}\biggr
)\, d{\bf
r}d{\bf v}=0,
\end{equation}
again yields an optimal current of the form of Eq. (\ref{mepp6}) now involving a
constant
inverse temperature $\beta$. This equation satisfies an $H$-theorem for the
Lynden-Bell free energy. From Eqs. (\ref{mepp3}) and (\ref{mepp2}) we have
\begin{eqnarray}
\label{mepp22}
\dot J&=&\dot S-\beta \dot E\nonumber\\
&=&-\int {1\over \overline{f}(1-\overline{f}/\eta_0)}{\bf J}\cdot
{\partial \overline{f}\over\partial {\bf
v}}  \, d{\bf r}d{\bf v}-\beta \int {\bf J}\cdot {\bf v} \, d{\bf
r}d{\bf v}.\nonumber\\
\end{eqnarray}
Using Eq. (\ref{mepp6}), we obtain
\begin{equation}
\label{mepp23}
\dot J=\int {{\bf J}^{2}\over D\overline{f}(1-\overline{f}/\eta_0)}\, d{\bf
r}d{\bf v},
\end{equation}
which is positive ($\dot J\ge 0$) provided that $D>0$.  Therefore, the free
energy $F$ decreases 
monotonically until an equilibrium state
of the form (\ref{mep5}) is reached ($H$-theorem). In the canonical ensemble, we can
show that $2\lambda\delta^{2}{J}=\delta^{2}\dot J\ge 0$ \cite{nfp}.
Therefore, a stationary solution of the generalized Fokker-Planck equation
(\ref{mepp10}) with constant temperature is linearly stable if, and only if, it
is a (local) minimum of free energy $F$ at fixed mass. Therefore,
this equation can only relax towards minima of $F$, not towards maxima
or saddle points.

{\it Remark:} By using the MEPP (or the linear thermodynamics of Onsager), we
have constructed a kinetic equation for the coarse-grained DF, Eq.
(\ref{mepp10}),  which relaxes towards the Lynden-Bell DF (\ref{mep5}). This
kinetic
equation has the form of a fermionic Kramers equation. We stress that this
thermodynamical approach is phenomenological in nature. It is also possible to
derive a kinetic equation for the coarse-grained DF through a systematic
procedure by developing a quasilinear theory of the Vlasov-Poisson equations
\cite{kp,sl,chavmnras,kingen,dubrovnik}. This kinetic
equation  has the form of a fermionic
Landau
equation. Although it is satisfying to derive this equation
from
a systematic
procedure, its domain of validity remains unclear (it may only describe a
``gentle'' relaxation). However, the fermionic Kramers equation can be obtained
from the
fermionic Landau equation by making a thermal bath approximation
\cite{chavmnras}. In
that case, the diffusion coefficient (which is not determined by the MEPP) can
be explicitly calculated. The fermionic Kramers equation with a time-dependent
temperature and the fermionic Landau equation both satisfy an $H$-theorem for
the Lynden-Bell entropy. We note, however, that the fermionic Landau equation
conserves
the energy locally while the  fermionic Kramers equation with a time-dependent
temperature conserves the energy globally. This difference affects the
hydrodynamic equations derived from these kinetic equations: they
involve either a viscosity (Landau) or a friction (Kramers)
\cite{prep}.\footnote{The  generalized wave equation associated
with the quantum Navier-Stokes equation involving
a viscous term $\nu\Delta {\bf u}$ is given in Sec. 7 and in Appendix L of
\cite{epjpnottale}, and in Appendix L of \cite{chavtotal}.}
We finally note that it is possible to generalize the results of this Appendix
to an arbitrary entropy of the form $S=-\int C(\overline{f})\, d{\bf r}d{\bf
v}$, where $C$ is a convex function \cite{gen}. This generalization may allow
us to solve heuristically the problem of incomplete relaxation
mentioned in Sec. \ref{sec_king} by considering entropic functionals different
from the Lynden-Bell entropy that yield DFs with a finite mass. It can also
allow us to deal with more complex situations than those considered here.
Finally, the MEPP provides a numerical algorithm in the form
of a generalized Fokker-Planck equation that can be used to construct
dynamically stable stationary
solutions of the Vlasov-Poisson equations, a nontrivial problem in general
\cite{gen}.

\subsection{Hydrodynamic equations}
\label{sec_ol}

We now take the hydrodynamic moments of the coarse-grained Vlasov equation
(\ref{mepp10}). In order to make the results fully explicit, we consider the
nondegenerate limit of the theory (the general case is treated in \cite{csr}).
In that case, the coarse-grained Vlasov equation takes the form
\begin{equation}
\label{hyd1}
\frac{\partial \overline{f}}{\partial t}+{\bf v}\cdot \frac{\partial
\overline{f}}{\partial {\bf
r}}-\nabla {\Phi}\cdot \frac{\partial \overline{f}}{\partial {\bf
v}}=\frac{\partial}{\partial {\bf v}}\cdot \left\lbrack D\left
(\frac{\partial
\overline{f}}{\partial {\bf
v}}+\beta(t) \overline{f}{\bf v}\right )\right\rbrack,
\end{equation}
which is similar to the classical Kramers equation. The evolution of the inverse effective temperature $\beta(t)=1/T_{\rm eff}(t)$ is given by
\begin{equation}
\label{hyd2}
\beta(t)=-{\int D {\partial \overline{f}\over\partial {\bf v}}\cdot {\bf
v}\, d{\bf r}d{\bf v}\over \int D \overline{f} v^{2}\,
d{\bf r}d{\bf v}}.
\end{equation}
Assuming that $D$ is constant, and making an integration by parts, we get
\begin{equation}
\label{hyd3}
\beta(t)=\frac{3M}{2K(t)},
\end{equation}
where
\begin{equation}
\label{hyd4}
K(t)=\int f\frac{v^2}{2}\, d{\bf r}d{\bf v}
\end{equation}
is the total kinetic energy. Eq. (\ref{hyd3}) can be rewritten as 
$K(t)=\frac{3}{2}MT_{\rm eff}(t)$. This relation
shows that the effective temperature $T_{\rm eff}(t)=1/\beta(t)$ can be
interpreted as the average kinetic temperature of the system, i.e., $T_{\rm
eff}(t)=\langle T_{\rm kin}({\bf r},t)\rangle$ where $\frac{3}{2}\rho T_{\rm
kin}({\bf r},t)=\int f \frac{v^2}{2}\, d{\bf v}$ is the local kinetic energy.

Taking the hydrodynamic moments of the coarse-grained Vlasov equation
(\ref{hyd1}), we obtain a system of equations similar to the Jeans
equations (see Appendix \ref{sec_jeans}) but including dissipative effects:
\begin{equation}
\label{hyd5}
\frac{\partial\rho}{\partial t}+\nabla\cdot (\rho {\bf u})=0,
\end{equation}
\begin{equation}
\label{hyd6}
\frac{\partial {\bf u}}{\partial t}+({\bf u}\cdot \nabla){\bf
u}=-\frac{1}{\rho}\partial_jP_{ij}-\nabla\Phi-\xi {\bf u},
\end{equation}
where $\xi=D\beta$. They are called the damped
Jeans equations. On the other hand, the effective collision term in Eq. (\ref{hyd1}) provides
a source of relaxation
which allows us to close the hierarchy of moment equations. Indeed, we can
compute the pressure tensor in Eq. (\ref{hyd6})  by making a LTE
approximation
\begin{eqnarray}
\label{hyd7}
\overline{f}_{\rm LTE}({\bf r},{\bf v},t)=\frac{\rho({\bf r},t)}{\lbrack 2\pi T_{\rm eff}(t)\rbrack^{3/2}}e^{-\frac{\lbrack {\bf v}-{\bf u}({\bf r},t)\rbrack^2}{2T_{\rm eff}(t)}}.
\end{eqnarray}
In that case, we get
\begin{eqnarray}
\label{hyd8}
P_{ij}=P_{\rm th}\delta_{ij}\qquad {\rm with}\qquad P_{\rm th}=\rho T_{\rm
eff}(t).
\end{eqnarray}
This leads to
a system of hydrodynamic equations of the form
\begin{equation}
\label{hyd9}
\frac{\partial\rho}{\partial t}+\nabla\cdot (\rho {\bf u})=0,
\end{equation}
\begin{equation}
\label{hyd10}
\frac{\partial {\bf u}}{\partial t}+({\bf u}\cdot \nabla){\bf
u}=-\frac{1}{\rho}\nabla P_{\rm th}-\nabla\Phi-\xi{\bf u},
\end{equation}
\begin{equation}
\label{hyd11}
\Delta\Phi=4\pi G\rho,
\end{equation}
called the damped Euler equations. Using the LTE approximation (\ref{hyd7}), the
total energy (\ref{mep3}) can be written as
\begin{equation}
\label{hyd12}
E=\int \rho \frac{{\bf u}^2}{2}\, d{\bf r}+\frac{3}{2}MT_{\rm eff}+W.
\end{equation}
In the microcanonical ensemble where the energy $E$ is fixed, this equation
determines the evolution of the effective temperature $T_{\rm eff}(t)$. On the
other hand, the Boltzmann-like entropy (\ref{mep15}) can be written (up to an
additive constant) as
\begin{equation}
\label{hyd13}
S=-\int\rho (\ln\rho-1)\, d{\bf r}+\frac{3}{2}M\ln T_{\rm eff}.
\end{equation}
This justifies the expressions of $E$ and $S$ used in Appendix \ref{sec_wmce}.
One can then show (like in Appendix \ref{sec_wmce}) that the hydrodynamic
equations (\ref{hyd9})-(\ref{hyd11}) with a time-dependent effective temperature
$T_{\rm eff}(t)$ determined by Eq. (\ref{hyd12}) satisfy an $H$-theorem for the
entropy $S$ given by Eq. (\ref{hyd13}). Analogously, in the canonical ensemble
where the effective temperature $T_{\rm eff}$ is fixed, one can show (like in
Appendix \ref{sec_wce}) that the hydrodynamic equations
(\ref{hyd9})-(\ref{hyd11}) with a constant effective temperature $T_{\rm eff}$
satisfy an $H$-theorem for the free energy $F=E-T_{\rm eff}S$. One can also
derive explicit
expressions of the virial theorem like in Appendices
\ref{sec_wmce} and \ref{sec_wce} (one simply has to take $P=Q=0$ in the
equations of these Appendices).

{\it Remark:} We note that, in Eq. (\ref{hyd12}), the effective temperature $T_{\rm eff}$ represents the
thermal kinetic energy $\frac{3}{2}M T_{\rm eff}(t)=\int f \frac{w^2}{2}\, d{\bf
r}d{\bf v}$ where ${\bf w}={\bf v}-{\bf u}({\bf r},t)$ is the fluctuating
velocity while the exact relation (\ref{hyd3}) indicates that it should
represent the total kinetic energy $\frac{3}{2}M T_{\rm eff}(t)=\int f
\frac{v^2}{2}\, d{\bf r}d{\bf v}$ including the mean kinetic energy. This is an
artefact of the LTE approximation (\ref{hyd7}) which
involves  a uniform temperature $T_{\rm eff}(t)$ instead of a space-dependent
temperature $T({\bf r},t)$. We can solve
this problem by replacing $T_{\rm eff}(t)$ by $T({\bf r},t)$ in Eq. (\ref{hyd7})
and introducing a hydrodynamic equation for the local temperature $T({\bf r},t)$
(second moment) as in
Ref. \cite{csr} (see Appendix  \ref{sec_int}). However,  the
LTE approximation (\ref{hyd7}) becomes exact close to equilibrium or in the
strong
friction limit $\xi\rightarrow +\infty$ where $|{\bf u}|=O(1/\xi)$. Indeed, in
the strong friction limit, we have
\begin{equation}
\label{tra1}
{\bf u}=-\frac{1}{\xi\rho} \left ( T_{\rm
eff}\nabla\rho+\rho\nabla\Phi\right ).
\end{equation}
Substituting this relation into the continuity equation (\ref{hyd9}) we  obtain
the
Smoluchowski-Poisson equations 
\begin{equation}
\label{tra2}
\xi\frac{\partial\rho}{\partial t}=\nabla\cdot \left (T_{\rm
eff}\nabla\rho+\rho\nabla\Phi\right ),
\end{equation}
\begin{eqnarray}
\label{tra3}
\Delta\Phi=4\pi G\rho.
\end{eqnarray}
Since $|{\bf u}|=O(1/\xi)$, we can neglect
the kinetic energy $\Theta_c$  in Eq. (\ref{hyd12}). Therefore, the
evolution of the effective temperature $T_{\rm eff}(t)$ is given by the
energy constraint
\begin{equation}
\label{w30c2}
E=\frac{3}{2}M T_{\rm eff}+\frac{1}{2}\int\rho\Phi\, d{\bf r}.
\end{equation}
In the canonical ensemble, Eqs. (\ref{tra2}) and (\ref{tra3}) are valid with
fixed $T_{\rm eff}$. One
can then derive the $H$-theorems and the virial theorems like in the Remarks at
the end of Appendices
\ref{sec_wmce} and \ref{sec_wce} (one simply has to take $P=Q=0$ in the
equations of these Appendices). We refer to \cite{crs} (and following papers)
for a detailed study of these equations.

\subsection{Inhomogeneous temperature}
\label{sec_int}

In the previous section, we have assumed that the temperature (velocity
dispersion) is uniform. In a more elaborate model (see Ref. \cite{csr}), we can
take into account an inhomogeneous temperature by considering the second moment
of the coarse-grained Vlasov equation and closing the hierarchy of hydrodynamic
equations with the LTE approximation (\ref{hyd7}) with $T_{\rm eff}(t)$ replaced
by $T({\bf r},t)$. This yields (see Ref. \cite{csr} for details)
\begin{equation}
\label{int1}
\frac{\partial\rho}{\partial t}+\nabla\cdot (\rho {\bf u})=0,
\end{equation}
\begin{equation}
\label{int2}
\frac{\partial {\bf u}}{\partial t}+({\bf u}\cdot \nabla){\bf
u}=-\frac{1}{\rho}\nabla (\rho T)-\nabla\Phi-\xi {\bf u},
\end{equation}
\begin{equation}
\label{int3}
\frac{3}{2}\left (\frac{\partial T}{\partial t}+{\bf u}\cdot \nabla T\right )+T\nabla\cdot {\bf u}=-3\xi(T-T_{\rm eff}(t)),
\end{equation}
where $T_{\rm eff}(t)=1/\beta(t)$ evolves according to Eq. (\ref{hyd3}) so as to
conserve the total energy
\begin{equation}
\label{int4}
E=\int \rho \frac{{\bf u}^2}{2}\, d{\bf r}+\frac{3}{2}\int \rho T\, d{\bf r}+W.
\end{equation}
The pressure is $P=\rho T$ and the thermal energy is $\Theta_{\rm th}=\int f \frac{w^2}{2}\, d{\bf
r}d{\bf v}=\frac{3}{2}\int \rho T\, d{\bf r}$. These equations conserve the mass
and the energy and satisfy an $H$-theorem for the Boltzmann-like entropy
\begin{equation}
\label{int5}
S=-\int\rho (\ln\rho-1)\, d{\bf r}+\frac{3}{2}\int \rho \ln T\, d{\bf r}.
\end{equation}
As a result, they relax towards the equilibrium Boltzmann-like distribution with
a uniform temperature $T=T_{\rm eff}$. The equation for the entropy density
$\sigma=-\ln(\rho T^{-3/2})$ is
\begin{equation}
\label{int6}
\frac{\partial\sigma}{\partial t}+{\bf u}\cdot \nabla \sigma=3\xi\frac{T_{\rm eff}(t)-T}{T}.
\end{equation}
In the absence of dissipation ($\xi=0$) we recover the usual Euler equations
\cite{landaulifshitz} which conserve the entropy. In that case,
$\rho T^{-3/2}$ is constant along the fluid (Lagrangian) trajectories,
corresponding to the adiabatic law $P\rho^{-5/3}={\rm cst}$. The damped virial
theorem is given by (see, e.g., Appendix G of \cite{chavtotal})
\begin{equation}
\label{int7}
\frac{1}{2}\ddot I+\frac{1}{2}\xi\dot I=2\Theta_c+3\int \rho T\, d{\bf r}+W.
\end{equation}
Using the energy conservation equation (\ref{int4}), it can be rewritten as
\begin{equation}
\label{int8}
\frac{1}{2}\ddot I+\frac{1}{2}\xi\dot I=2E-W.
\end{equation}
In the canonical ensemble, $T_{\rm eff}$ is constant and the
hydrodynamic equations satisfy an $H$-theorem for the free energy $F=E-T_{\rm
eff}S$.

\section{The classical limit $\hbar\rightarrow 0$}
\label{sec_cl}

In this Appendix, we discuss how the quantum equations studied in the
present paper pass to the limit $\hbar\rightarrow
0$.\footnote{We note that the Schr\"odinger-Poisson
equations (\ref{h1}) and (\ref{h2}) depend only on $\hbar$ through the ratio
$\hbar/m$. Therefore, the classical limit corresponds to $\hbar/m\rightarrow
0$ (i.e. $\hbar\rightarrow 0$ or $m\rightarrow +\infty$).} Our discussion is
essentially heuristic. A mathematically rigorous treatment of the
classical limit $\hbar\rightarrow 0$ is difficult but would certainly be very
valuable.

\subsection{Classical versus quantum descriptions}

In the  collisionless regime, classical self-gravitating systems are
described by the Vlasov-Poisson equations (\ref{vlasov1}) and (\ref{vlasov2}).
The Vlasov equation is equivalent to the infinite hierarchy of Jeans equations
(see Appendix  \ref{sec_jeans}). This is true for all times and for arbitrary
initial conditions. Let us now assume that we start from a single-speed initial
condition. Then, as long as the DF remains single-speed, the DF is given by Eq.
(\ref{e1}), the pressure tensor vanishes ($P_{ij}=0$), and the Jeans equations
reduce to the 
pressureless Euler equations (\ref{e2}) and (\ref{e3}). This is true
until
shell crossing, after which the 
pressureless Euler equations develop singularities and the
DF becomes multi-streamed. Before shell crossing the Vlasov equation is
equivalent to the 
pressureless Euler equations but after shell crossing
the
pressureless Euler 
equations are not valid anymore. In that case $P_{ij}\neq 0$. Unfortunately, it
is not possible to calculate the pressure tensor exactly so the Jeans equations
are not closed. One has to come back to the Vlasov-Poisson equations. Some
heuristic procedures to compute $P_{ij}$ approximately in order to continue
using a hydrodynamical approach are described in Appendix \ref{sec_euler}.

In the collisionless regime, quantum self-gravitating systems made
of condensed bosons are described by the Schr\"odinger-Poisson equations
(\ref{h1}) and (\ref{h2}). The Schr\"odinger equation is equivalent to the
quantum Euler equations (\ref{mad4})-(\ref{mad6}). These equations are also
equivalent to the Wigner equation (\ref{qw2}), whose exact solution is given by
Eq. (\ref{qw1}), and to the quantum Jeans equations (see Sec. \ref{sec_qja}).
For condensed bosons, $P_{ij}=P_{ij}^Q$, where $P_{ij}^Q$ is given by Eq.
(\ref{mad10}), and the quantum Jeans equations are closed (they reduce to the
quantum Euler equations (\ref{mad4})-(\ref{mad6})). This is true for all
times and for arbitrary initial conditions.

\subsection{Comparison between the Wigner equation and the Vlasov equation}

When $\hbar\rightarrow 0$, the Wigner equation (\ref{qw2}) reduces to the
Vlasov equation (\ref{vlasov1}) which corresponds to $\hbar=0$ (see
Appendix \ref{sec_cw}). On the other hand, 
the Wigner equation is equivalent to the Schr\"odinger equation (\ref{h1}) and
to the quantum Euler equations (\ref{mad4})-(\ref{mad6}). Therefore, the
solution of the Vlasov equation ($\hbar=0$) 
is expected to be well-approximated by the solution of the Wigner, Schr\"odinger
and quantum Euler equations with $\hbar\rightarrow 0$. In particular, $f({\bf
r},{\bf
v},t)\simeq f_W({\bf r},{\bf
v},t)$ where $f_W$ is given by Eq. (\ref{qw1}).  This equivalence is valid
for all times. Therefore, for what concerns the Wigner equation, the
limit $\hbar\rightarrow 0$ is equivalent to $\hbar=0$.

\subsection{Comparison between the quantum Euler equation and the
pressureless Euler equations}

When $\hbar\rightarrow 0$, the quantum Euler equations (\ref{mad4})-(\ref{mad6})
seem to reduce to the 
pressureless Euler equations (\ref{e2}) and (\ref{e3})
which correspond to $P=\hbar=0$. However, this equivalence is only valid before
shell crossing ($t<t_*$).  After shell crossing ($t>t_*$),
the quantum
Euler equations are still valid while the
pressureless Euler equations are  not valid
anymore (see Appendix \ref{sec_euler}). Therefore, for what concerns the
quantum Euler
equations (\ref{mad4})-(\ref{mad6}), the limit $\hbar\rightarrow 0$ is
different from $\hbar=0$. A small but finite value of $\hbar$ allows us to
extend the
solutions of the hydrodynamic equations past $t_*$ for all times.

\subsection{Conclusion}

The Vlasov equation is valid for all times. By contrast,
the
pressureless Euler equations are
valid only for a single speed solution until the first time of crossing. Before that time, they are equivalent to the
Vlasov equation but after that time they break down. The Schr\"odinger equation,
the quantum Euler equations, and the Wigner equation are equivalent and they
are valid for all times. When $\hbar\rightarrow 0$ their solutions are expected
to be close to the solution of the  Vlasov equation.

In conclusion, the Vlasov equation and the Schr\"odinger, quantum Euler,
and Wigner equations with  $\hbar\rightarrow 0$ are superior to the classical
pressureless Euler equations. They take into account velocity
dispersion whereas the classical pressureless fluid description does not. They
can be used to describe multistreaming and caustics in the nonlinear regime,
whereas the classical pressureless fluid equations break down in that regime.
Indeed, the
pressureless Euler equations develop shocks so they are not
well-defined after
the first shock. The velocity dispersion, or the quantum pressure, allow to
regularize the dynamics and solve the problems of the classical pressureless
hydrodynamic description.

\section{The Vlasov-Bohm equation}
\label{sec_vb}

Instead of using the rather complicated Wigner equation (\ref{qw2}), one
could
consider the simpler Vlasov-Bohm equation
\begin{eqnarray}
\label{vb1}
\frac{\partial f}{\partial t}+{\bf v}\cdot \frac{\partial f}{\partial {\bf
r}}-\nabla\Phi\cdot \frac{\partial f}{\partial {\bf
v}}-\frac{1}{m}\nabla Q\cdot \frac{\partial f}{\partial {\bf
v}}=0.
\end{eqnarray}
This equation is similar to the classical Vlasov equation (\ref{vlasov1}) except
that it
includes the quantum potential $Q$ from Eq. (\ref{mad8}).  We stress, however,
that this
equation is heuristic and that it is not expected to give an exact description
of the dynamics. The Vlasov-Bohm-Poisson equations conserve the
energy $E=(1/2)\int f v^2\, d{\bf
r}d{\bf v}+(1/2)\int \rho\Phi\, d{\bf r}+(1/m)\int \rho Q\, d{\bf r}$ and an
infinite class of Casimir
integrals of the form $\int h(f)\, d{\bf
r}d{\bf v}$ where $h(f)$ is an arbitrary function of $f$. Coarse-graining the
Vlasov-Bohm equation like in  Sec. \ref{sec_wcg}, we obtain the
fermionic Vlasov-Bohm-Kramers equation
\begin{eqnarray}
\label{vb2}
\frac{\partial \overline{f}}{\partial t}&+&{\bf v}\cdot \frac{\partial
\overline{f}}{\partial {\bf
r}}-\nabla\Phi\cdot \frac{\partial \overline{f}}{\partial {\bf
v}}-\frac{1}{m}\nabla Q\cdot \frac{\partial \overline{f}}{\partial {\bf
v}}\nonumber\\
&=&\frac{\partial}{\partial {\bf v}}\cdot \left\lbrack D\left
(\frac{\partial
\overline{f}}{\partial {\bf
v}}+\beta \overline{f}(1-\overline{f}/\eta_0){\bf v}\right
)\right\rbrack
\end{eqnarray}
in place of Eq. (\ref{qw3}) [one can also obtain the fermionic
Vlasov-Bohm-Landau equation in place of Eq. (\ref{qw3l})].
This equation relaxes towards an equilibrium DF of the form
\begin{eqnarray}
\label{vb3}
\overline{f}_{\rm LB}({\bf r},{\bf v})=\frac{\eta_0}{1+e^{\beta (v^2/2+\Phi({\bf
r})+Q/m)-\alpha}},
\end{eqnarray}
which can be viewed as a Lynden-Bell DF incorporating the quantum
potential. Since $Q$ depends on the density itself this equation is a
complicated integral equation. It is equivalent to the condition of
quantum hydrostatic equilibrium from Eq. (\ref{es2}) (see the Remark below).

We can then
take the hydrodynamic moments of the coarse-grained Vlasov-Bohm-Kramers equation
(\ref{vb2}) as in Sec. \ref{sec_qjeans}. Instead of Eq. (\ref{qj2}), we
get
\begin{eqnarray}
\label{vb4}
\frac{\partial {\bf u}}{\partial t}&+&({\bf u}\cdot \nabla){\bf
u}=-\frac{1}{\rho}\partial_jP_{ij}-\nabla\Phi\nonumber\\
&-&\frac{1}{m}\nabla
Q-\frac{1}{\rho} D\beta \int
\overline{f}(1-\overline{f}/\eta_0){\bf v}\, d{\bf v},
\end{eqnarray}
in which the quantum potential appears explicity. Closing the hierarchy
of quantum Jeans equations by  making the LTE
approximation from Eq. (\ref{qj10}) to compute the pressure tensor, we obtain
\begin{equation}
\label{vb5}
\frac{\partial {\bf u}}{\partial t}+({\bf u}\cdot \nabla){\bf
u}=-\frac{1}{\rho}\nabla P_{\rm LB}-\frac{1}{m}\nabla
Q-\nabla\Phi-\xi{\bf
u},
\end{equation}
which coincides with Eq. (\ref{ch2}). Therefore, at the level of the
hydrodynamic
equations, and within the assumptions  made in our approach, the Vlasov-Bohm
equation yields the same results as the Wigner equation. The reason is that our
approach neglects collective effects (encapsulated in the dielectric
function) where differences between the Wigner
equation and  the Vlasov-Bohm equation occur. However, this approximation may
be justified during the very nonlinear process of violent relaxation.

{\it Remark:} For spherically symmetric systems, the stationary
solutions of the Vlasov-Bohm equation (\ref{vb1}) are of the form
$f=f(\epsilon_Q)$ with
$\epsilon_Q=v^2/2+\Phi+Q/m$. We can easily show that this relation implies the
condition of quantum hydrostatic equilibrium from Eq. (\ref{es2}). Indeed,
defining the density and the pressure by
$\rho=\int
f\, d{\bf v}$ and $P=\frac{1}{3}\int f v^2\, d{\bf v}$, we get
\begin{eqnarray}
\label{vb6}
\nabla P&=&\frac{1}{3}\int v^2\frac{\partial f}{\partial {\bf r}}\, d{\bf
v}\nonumber\\
&=&\frac{1}{3}\left (\nabla\Phi+\frac{1}{m}\nabla Q\right )\int
v^2 f'(\epsilon_Q)\, d{\bf v}\nonumber\\
&=&\frac{1}{3}\left
(\nabla\Phi+\frac{1}{m}\nabla Q\right )\int \frac{\partial f}{\partial {\bf
v}}\cdot {\bf v}\, d{\bf v}\nonumber\\
&=&-\left (\nabla\Phi+\frac{1}{m}\nabla
Q\right )\int f\, d{\bf v}
\end{eqnarray}
yielding
\begin{eqnarray}
\label{vb7}
\nabla P+\rho\nabla\Phi+\frac{\rho}{m}\nabla
Q={\bf 0}.
\end{eqnarray}
The effective collision term on the right hand side of Eq. (\ref{vb2}) selects
the form of the equilibrium DF $f=f(\epsilon_Q)$ among all the possible
stationary solutions of
the Vlasov-Bohm equation. In the present case, the fermionic Kramers operator
selects the Lynden-Bell DF from Eq. (\ref{vb3}).

\section{Multistate systems}
\label{sec_multistate}

\subsection{Hartree equations}
\label{sec_ah}

We consider a system of $N$ quantum particles in gravitational
interaction. These particles may be fermions or bosons. Fundamentally, they are
described by an $N$-body wave function
$\Psi({\bf r}_1,...,{\bf r}_N,t)$ satisfying the exact  $N$-body Schr\"odinger
equation \cite{parr} 
\begin{eqnarray}
\label{ah1}
i\hbar \frac{\partial\Psi}{\partial
t}=-\frac{\hbar^2}{2m}\Delta\Psi-\sum_{\alpha<\beta}\frac{Gm^2}{|{\bf
r}_\alpha-{\bf r}_\beta|}\Psi.
\end{eqnarray}
When $N$ is large, this equation is completely untractable. The aim of
statistical mechanics and kinetic theory is to obtain simpler models described
by one-body equations. Following Hartree \cite{hartreea,hartreeb} we shall make
a mean
field
approximation and ignore correlations among the particles. We thus assume that
the
$N$-body wave function $\Psi({\bf r}_1,...,{\bf r}_N,t)$  can be written as a
product of $N$
one-body wave functions $\psi_{\alpha}({\bf r},t)$ so that
\begin{eqnarray}
\label{ah2}
\Psi({\bf r}_1,...,{\bf r}_N,t)=\prod_{\alpha=1}^{N} \psi_\alpha({\bf
r}_\alpha,t).
\end{eqnarray}
The  mean field approximation is
known to become exact for systems with long-range interactions (like
self-gravitating systems) when $N\rightarrow +\infty$ in a proper thermodynamic
limit \cite{campa}. In that case, the
evolution of the quantum system is described by
$N$ coupled mean field  Schr\"odinger equations of the
form 
\begin{eqnarray}
\label{ah3}
i\hbar \frac{\partial\psi_{\alpha}}{\partial
t}=-\frac{\hbar^2}{2m}\Delta\psi_{\alpha}+m\Phi\psi_{\alpha},
\end{eqnarray}
\begin{eqnarray}
\label{ah4}
\Delta\Phi=4\pi G\sum_{\alpha=1}^N p_{\alpha} |\psi_{\alpha}|^2,
\end{eqnarray}
where $p_{\alpha}$ is the occupation probability of state $\alpha$ such
that the normalization condition $\sum_{\alpha}p_{\alpha}=1$ is fulfilled. We
shall assume that
$p_{\alpha}$ is constant or that it varies slowly with ${\bf r}$ and $t$. The
total density is
$\rho=\sum_{\alpha} p_{\alpha} |\psi_{\alpha}|^2$. In the mean field
approximation, each particle evolves in a self-consistent gravitational
potential $\Phi$ that is produced  by the particles themselves through the
Poisson equation (\ref{ah4}). Eqs (\ref{ah3}) and
(\ref{ah4}) are called the Hartree (mean field) equations.

From the wavefunctions $\psi_{\alpha}({\bf r},t)$ of the multistate system,  we
can define the Wigner DF by 
\begin{eqnarray}
\label{ah5}
f({\bf r},{\bf v},t)=\sum_{\alpha=1}^{N} \frac{m^3}{(2\pi\hbar )^3}
p_{\alpha} \int d{\bf
y}\,  e^{im{\bf v}\cdot {\bf y}/\hbar} \nonumber\\
\times\psi_{\alpha}^*\left ({\bf
r}+\frac{{\bf y}}{2},t\right )\psi_{\alpha}\left ({\bf r}-\frac{{\bf
y}}{2},t\right ).
\end{eqnarray}
It satisfies the Wigner
equation (\ref{qw2}). The Wigner DF (\ref{ah5}) involves the density matrix
$\rho({\bf r},{\bf
r}',t)=\sum_{\alpha=1}^N p_{\alpha}\psi_{\alpha}({\bf r},t)\psi_{\alpha}^*({\bf
r},t)$ which appears in the von Neumann equation. Actually, the Hartree, Wigner
and von Neumann equations are equivalent.

In the case of fermions, the wave function $\Psi$ is antisymmetric with respect
to the exchange of any two variables ${\bf r}_\alpha$ and ${\bf r}_\beta$. The
antisymmetry condition stems from the Pauli exclusion principle which
establishes that two fermions cannot share the same position so that the
probability density $|\Psi|^2$ vanishes as ${\bf r}_\alpha={\bf r}_\beta$ (for a
rigorous treatment the spin of the fermions must be considered). A system of
fermions is
necessarily in a mixed quantum state in order to respect the Pauli exclusion
principle. 
In the Hartree-Fock \cite{hartree2,fock} theory, the $N$-body wave function is
expressed as 
\begin{eqnarray}
\label{ah6}
\Psi({\bf r}_1,...,{\bf r}_N,t)=\frac{1}{\sqrt{N!}}\, {\rm
det}\lbrack \psi_{\alpha}({\bf r}_{\beta},t)\rbrack.
\end{eqnarray}
In other words, the wave function for many fermions is taken as a Slater
determinant
which is zero for two particles at the same position.
This expression guarantees that the system obeys the Pauli exclusion principle.
This leads
to the Hartree-Fock equations corresponding to the Hartree Eqs. (\ref{ah3}) and
(\ref{ah4}) with an
additional exchange energy term introduced by Fock \cite{fock}. This exchange
energy term
can be simplified by using the Slater \cite{slater} approximation (see also
Dirac
\cite{dirac} and Kohn-Sham \cite{ks}) which amounts to introducing a term $C_S
\rho^{1/3}\psi_{\alpha}$ in the right hand side of Eq.  (\ref{ah3}), where
$C_S$ is a positive constant in the gravitational case (it is negative in the
electrostatic case). The Hartree-Fock equations neglect correlations.
More general equations taking into account correlations
(assuming that the
fermions interact by microscopic forces in addition to
the gravitational force) can be obtained through the density functional theory
\cite{parr} initiated by Kohn and Sham \cite{ks}.

A gas of bosons at $T=0$ forms a BEC in which  all the bosons
are in the same quantum state described by a single wave function $\psi({\bf
r},t)$. In the mean field approximation, the $N$-body wave function is expressed
as 
\begin{eqnarray}
\label{ah8}
\Psi({\bf r}_1,...,{\bf r}_N,t)=\prod_{\alpha=1}^{N} \psi({\bf
r}_\alpha,t).
\end{eqnarray}
In that case, the evolution of the self-gravitating
BEC
is described by the Schr\"odinger-Poisson equations (\ref{h1}) and
(\ref{h2}). This
amounts to taking $\alpha=1$ and $p_{\alpha}=1$ in Eqs. (\ref{ah3})
and
(\ref{ah4}). A system of
bosons at $T=0$ is in a  pure quantum state.

{\it Remark:} Actually, even in the case of BECs, we can have a
quantum superposition of modes. Indeed, the wave function $\psi$ can be
decomposed into eigenfunctions of the Schr\"odinger-Poisson equations involving
the fundamental state ($n=0$) and the excited states ($n>0$)  as in Eq.
(\ref{inter1}). As a result, the multistate equations (\ref{ah3}) and
(\ref{ah4}) are also relevant for bosons with this interpretation (in that case
$p_\alpha$ is the probability of mode $\alpha$ and  $N$ is the number of
modes). The resulting equations can be called the Hartree (mean field) equations
for bosons. In the case of fermions, the Hartree equations (\ref{ah3}) and
(\ref{ah4}) are basically valid with $p_{\alpha}=1/N$ if $\alpha=1,...,N$ and
$p_{\alpha}=0$ otherwise.  However, as explained above,
they can also be viewed as an expansion over the modes $\alpha$ of the system
where $p_{\alpha}$ represents the probability of mode $\alpha$ and $N$
represents the number of
modes.\footnote{If the system reaches a
state of statistical equilibrium through a collisional relaxation, $p_{\alpha}$
should be given by the Bose-Einstein DF for bosons and by the
Fermi-Dirac DF for fermions. However, if the evolution of the system is
collisionless, like in the present situation, $p_{\alpha}$ should be given by
the Lynden-Bell DF, which is similar to the Fermi-Dirac DF, in
all cases (fermions and bosons). In the nondegenerate limit, the Lynden-Bell
DF is similar to the Maxwell-Boltzmann DF.} Below, we generalize
the
hydrodynamic representation of the SP equations to the  case of a multistate
system and show how a pressure term arises in the quantum Euler equations.
This pressure term is similar to the one obtained from the approach of Sec.
\ref{sec_dfaw}.

\subsection{Madelung transformation}
\label{sec_amad}

We can use the Madelung transformation to rewrite the Hartree-Fock
equations (\ref{ah3}) and (\ref{ah4}) under the form of hydrodynamic equations
for each state $\alpha$ (we shall use the Slater approximation $C_S
\rho^{1/3}\psi_{\alpha}$ to evaluate the exchange term for fermions). Let us
write the wave
function as
\begin{equation}
\label{amad1}
\psi_{\alpha}({\bf r},t)=\sqrt{{\rho_{\alpha}({\bf r},t)}} e^{iS_{\alpha}({\bf
r},t)/\hbar},
\end{equation}
where  $\rho_{\alpha}({\bf r},t)$ is the density and $S_{\alpha}({\bf r},t)$ is
the action given by
\begin{equation}
\label{amad2}
\rho_{\alpha}=|\psi_{\alpha}|^2\qquad {\rm and}\qquad
S_{\alpha}=-i\frac{\hbar}{2}\ln\left
(\frac{\psi_{\alpha}}{\psi_{\alpha}^*}\right ).
\end{equation}
Following Madelung, we introduce the velocity field
\begin{equation}
\label{amad3}
{\bf u}_{\alpha}=\frac{\nabla S_{\alpha}}{m}.
\end{equation}
Substituting
Eq. (\ref{amad1}) into Eqs. (\ref{ah3}) and
(\ref{ah4}) and separating the real and
the imaginary parts, we find that the Hartree-Fock equations are equivalent to
hydrodynamic
equations of the form
\begin{equation}
\label{amad4}
\frac{\partial\rho_{\alpha}}{\partial t}+\nabla\cdot (\rho_{\alpha} {\bf
u}_{\alpha})=0,
\end{equation}
\begin{equation}
\label{amad5}
\frac{\partial S_{\alpha}}{\partial t}+\frac{1}{2m}(\nabla
S_{\alpha})^2+m\Phi+C_S\rho^{1/3}+Q_{\alpha}=0,
\end{equation}
\begin{equation}
\label{amad6}
\frac{\partial {\bf u}_{\alpha}}{\partial t}+({\bf u}_{\alpha}\cdot \nabla){\bf
u}_{\alpha}=-\frac{1}{m}\nabla
Q_{\alpha}-\nabla\Phi-\frac{1}{\rho}\nabla P_{\rm Slater},
\end{equation}
\begin{eqnarray}
\label{amad7}
\Delta\Phi=4\pi G\sum_{\alpha=1}^N p_{\alpha}\rho_{\alpha},
\end{eqnarray}
where  
\begin{equation}
\label{amad8}
Q_{\alpha}=-\frac{\hbar^2}{2m}\frac{\Delta\sqrt{\rho_{\alpha}}}{\sqrt{\rho_{
\alpha}}}=-\frac{\hbar^2}{4m}\left\lbrack
\frac{\Delta\rho_{\alpha}}{\rho_{\alpha}}-\frac{1}{2}\frac{(\nabla\rho_{\alpha}
)^2}{ \rho_{\alpha}^2} \right\rbrack
\end{equation}
is the quantum potential and 
\begin{equation}
\label{amad9}
P_{\rm Slater}=\frac{C_S}{4m}\rho^{4/3}
\end{equation}
is the Slater pressure. It corresponds to a polytrope of index $n=3$ with a
polytropic constant $K_S=C_S/4m$. The Slater pressure is positive
($P_{\rm Slater}>0$) meaning that the exchange interaction is effectively
repulsive in the gravitational case (it is attractive in the electrostatic
case).
Using the continuity equation
(\ref{amad4}), we obtain the identity 
\begin{equation}
\label{amad10}
\rho_{\alpha}\left\lbrack \frac{\partial {\bf u}_{\alpha}}{\partial t}+({\bf
u}_{\alpha}\cdot \nabla){\bf
u}_{\alpha}\right\rbrack=\frac{\partial}{\partial t}(\rho_{\alpha} {\bf
u}_{\alpha})+\nabla(\rho_{\alpha} {\bf u}_{\alpha}\otimes {\bf u}_{\alpha}).
\end{equation}
The quantum Euler equation (\ref{amad6}) can then be
rewritten as
\begin{eqnarray}
\label{amad11}
\frac{\partial}{\partial t}(\rho_{\alpha} {\bf u}_{\alpha})+\nabla(\rho_{\alpha}
{\bf u}_{\alpha}\otimes {\bf u}_{\alpha})
=-\frac{\rho_{\alpha}}{\rho}\nabla P_{\rm
Slater}\nonumber\\
-\rho_{\alpha}\nabla\Phi
-\frac{\rho_{\alpha}}{m}\nabla
Q_{\alpha}.
\end{eqnarray}
The foregoing  equations are valid for
each state $\alpha$. 
We now introduce
the total
density 
\begin{equation}
\label{amad12}
\rho=\sum_{\alpha=1}^N p_{\alpha}\rho_{\alpha}
\end{equation}
and the total velocity
\begin{equation}
\label{amad13}
{\bf u}=\frac{1}{\rho}\sum_{\alpha=1}^N p_{\alpha}\rho_{\alpha}{\bf
u}_{\alpha}.
\end{equation}
From Eqs. (\ref{amad4}), (\ref{amad12}) and  (\ref{amad13}), we obtain the
continuity equation 
\begin{equation}
\label{amad14}
\frac{\partial\rho}{\partial t}+\nabla\cdot (\rho {\bf
u})=0.
\end{equation}
From Eqs. (\ref{amad11}), (\ref{amad12}) and  (\ref{amad13}), we get
\begin{eqnarray}
\label{amad15}
\frac{\partial}{\partial t}(\rho{\bf
u})+\sum_{\alpha=1}^N p_{\alpha}\nabla\left\lbrack\rho_{\alpha}
({\bf u}+{\bf w}_{\alpha})\otimes ({\bf u}+{\bf w}_{\alpha})\right\rbrack
=\nonumber\\
-\nabla
P_{\rm
Slater}-\rho\nabla\Phi-\sum_{\alpha=1}^N p_{\alpha}\frac{\rho_{\alpha}}{m}\nabla
Q_{\alpha},
\end{eqnarray}
where ${\bf w}_{\alpha}={\bf
u}_{\alpha}-{\bf u}$. Expanding the advection term
in Eq. (\ref{amad15}) and using the identity
\begin{equation}
\label{amad16}
\sum_{\alpha=1}^Np_{\alpha}\rho_{\alpha}{\bf
w}_{\alpha}=\sum_{\alpha=1}^N p_{\alpha}\rho_{\alpha}({\bf
u}_{\alpha}-{\bf u})=\rho{\bf u}-\rho{\bf u}={\bf 0},
\end{equation}
we are left with
\begin{eqnarray}
\label{amad17}
\frac{\partial}{\partial t}(\rho{\bf
u})+\nabla(\rho{\bf u}\otimes {\bf
u})=-\sum_{\alpha=1}^N p_{\alpha}\nabla(\rho_{\alpha}{\bf w}_{\alpha}\otimes
{\bf
w}_{\alpha})\nonumber\\
-\nabla
P_{\rm
Slater}-\rho\nabla\Phi-\sum_{\alpha=1}^N p_{\alpha}\frac{\rho_{\alpha}}{m}\nabla
Q_{\alpha}.
\end{eqnarray}
Using the continuity equation
(\ref{amad14}), we obtain the identity
\begin{equation}
\label{amad18}
\rho\left\lbrack \frac{\partial {\bf u}}{\partial t}+({\bf
u}\cdot \nabla){\bf
u}\right\rbrack=\frac{\partial}{\partial t}(\rho {\bf
u})+\nabla(\rho {\bf u}\otimes {\bf u}).
\end{equation}
The Euler equation (\ref{amad17}) can then be
rewritten as
\begin{eqnarray}
\label{amad19}
\frac{\partial {\bf u}}{\partial t}+({\bf u}\cdot \nabla){\bf
u}=-\frac{1}{\rho}\partial_j P_{ij}-\frac{1}{\rho}\nabla
P_{\rm
Slater}\nonumber\\
-\frac{1}{\rho m}\sum_{\alpha=1}^N
p_{\alpha}\rho_{\alpha} \nabla
Q_{\alpha}-\nabla\Phi,
\end{eqnarray}
where
\begin{equation}
\label{amad20}
P_{ij}=\sum_{\alpha=1}^N p_{\alpha}\rho_{\alpha} {\bf w}_{\alpha}\otimes {\bf
w}_{\alpha}
\end{equation}
is the pressure tensor arising from the difference between the multistate
velocities ${\bf u}_{\alpha}$ and the total velocity ${\bf u}$. It can also be
written as
\begin{equation}
\label{amad21}
P_{ij}=\sum_{\alpha=1}^N p_{\alpha}\rho_{\alpha} {\bf u}_{\alpha}\otimes {\bf
u}_{\alpha}-\rho {\bf u}\otimes {\bf
u}.
\end{equation}
If we make the approximation
\begin{equation}
\label{amad22}
-\frac{1}{\rho m}\sum_{\alpha=1}^N
p_{\alpha}\rho_{\alpha} \nabla
Q_{\alpha}\simeq -\frac{1}{m}\nabla Q,
\end{equation}
where $Q$ is the quantum potential defined by Eq. (\ref{mad8}), we obtain the
hydrodynamic
equations
\begin{equation}
\label{amad23}
\frac{\partial\rho}{\partial t}+\nabla\cdot (\rho {\bf
u})=0,
\end{equation}
\begin{equation}
\label{amad24}
\frac{\partial {\bf u}}{\partial t}+({\bf u}\cdot \nabla){\bf
u}=-\frac{1}{\rho}\partial_j P_{ij}-\frac{1}{\rho}\nabla
P_{\rm
Slater}-\frac{1}{m}\nabla
Q-\nabla\Phi,
\end{equation}
\begin{eqnarray}
\label{amad25}
\Delta\Phi=4\pi G\rho.
\end{eqnarray}
These equations are not closed since the pressure
tensor $P_{ij}$ depends on the multistate variables. We note that the pressure
tensor defined by Eq. (\ref{amad20}) is similar to the pressure tensor defined
by Eq.
(\ref{qj5}) if we identify
$p_{\alpha}\rho_{\alpha}$ with the coarse-grained DF $\overline{f}({\bf
r},{\bf v})$. We shall consider successively the
case of bosons and fermions.

\subsection{Bosons}
\label{sec_endb}

In the case of bosons  at $T=0$ forming a BEC, we just have one pure state
($\alpha=p_\alpha=1$), and we trivially find that the pressure tensor defined
by Eq. (\ref{amad20}) vanishes
\begin{equation}
\label{madb1}
P_{ij}=0.
\end{equation}
We also have $P_{\rm Slater}=0$ in that case ($C_S=0$). As a result, the
hydrodynamic equations (\ref{amad23})-(\ref{amad25}) reduce to
Eqs. (\ref{mad4})-(\ref{mad7}) as it should. However, if we write $\psi$ as a
superposition of modes [see Eq. (\ref{inter1})], and
use Eqs. (\ref{ah3}) and (\ref{ah4})  with this interpretation (see the Remark
at the end of Appendix \ref{sec_ah}), we can close
Eqs.  (\ref{amad23})-(\ref{amad25}) by using  the Lynden-Bell pressure from Eq.
(\ref{qj11}). This leads to Eqs. (\ref{ch1})-(\ref{ch3}) and, finally, to
Eq. (\ref{we16}). Note, however, that
the friction term, which represents of form of nonlinear Landau damping, does
not
explicitly appear in the present formalism.

\subsection{Fermions}
\label{sec_endf}

In the case of fermions, the pressure tensor $P_{ij}$ takes
into account the Pauli exclusion principle but, without further assumption, it
cannot be explicitly evaluated. Now, if we view $\sum_\alpha$ as
a sum over the modes with $p_\alpha$ being the probability of mode $\alpha$
(see the Remark
at the end of Appendix \ref{sec_ah}), we can close
Eqs.  (\ref{amad23})-(\ref{amad25}) by using  the Lynden-Bell pressure from Eq.
(\ref{qj11}) which, in the case of fermions, coincides with the Fermi-Dirac
pressure. This leads to the hydrodynamic equations
\begin{equation}
\label{madf1}
\frac{\partial\rho}{\partial t}+\nabla\cdot (\rho {\bf
u})=0,
\end{equation}
\begin{equation}
\label{madf2}
\frac{\partial {\bf u}}{\partial t}+({\bf u}\cdot \nabla){\bf
u}=-\frac{1}{\rho}\nabla P_{\rm LB/FD}-\frac{1}{\rho}\nabla P_{\rm
Slater}-\frac{1}{m}\nabla
Q-\nabla\Phi,
\end{equation}
\begin{eqnarray}
\label{madf3}
\Delta\Phi=4\pi G\rho,
\end{eqnarray}
and, finally, to Eq. (\ref{fermidm5}) with the same comment as
above concerning the absence of the friction term. We note that the Lynden-Bell
(or Fermi-Dirac) pressure has a statistical origin (it depends on the
specification of  $p_\alpha$)
while the Slater pressure has a purely quantum nature.

\end{document}